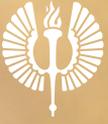

**TURUN
YLIOPISTO**
UNIVERSITY
OF TURKU

# NATURE'S BREWERY
# TO BEDTIME
## The Role of Hops in GABA$_A$ Receptor
## Modulation and Sleep Promotion

Ali Yasser Benkherouf

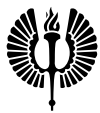

**TURUN
YLIOPISTO**

UNIVERSITY
OF TURKU

# NATURE'S BREWERY
# TO BEDTIME

The Role of Hops in GABA$_A$ Receptor Modulation
and Sleep Promotion

Ali Yasser Benkherouf

## University of Turku

Faculty of Medicine
Institute of Biomedicine
Pharmacology, Drug Development and Therapeutics
Drug Research Doctoral Programme

## Supervised by

Adjunct Professor Mikko Uusi-Oukari, PhD
Institute of Biomedicine
University of Turku
Turku, Finland

Senior Researcher Sanna Soini, PhD
Institute of Biomedicine
University of Turku
Turku, Finland

## Reviewed by

Professor Garry Wong
Faculty of Health Sciences
University of Macau
Macau, China

Adjunct Professor Olli Kärkkäinen
School of Pharmacy
University of Eastern Finland
Kuopio, Finland

## Opponent

Professor Esa Korpi
Department of Pharmacology
Faculty of Medicine
University of Helsinki
Helsinki, Finland

*To my grandmother, whose belief in me remains unwavering, and to the cherished ones no longer with us. As stars pierce the night's veil, so has my family illuminated my path. This work stands in tribute to the trailblazers of this discipline, and to humanity's enduring spirit, always seeking nature's gentle touch.*

UNIVERSITY OF TURKU
Faculty of Medicine
Institute of Biomedicine
Pharmacology, Drug Development and Therapeutics
ALI YASSER BENKHEROUF: Nature's Brewery to Bedtime: The Role of
Hops in GABA$_A$ Receptor Modulation and Sleep Promotion
Doctoral Dissertation, 189 pp.
Drug Research Doctoral Programme
December 2023

ABSTRACT

Insomnia, a prevalent health challenge, often requires pharmacological interventions to improve sleep onset, maintenance, and quality. Benzodiazepines and Z-drugs, like other positive modulators, enhance the inhibitory effects of gamma-aminobutyric acid (GABA) by stabilizing the open conformation of the GABA$_A$ receptor (GABA$_A$R) chloride ion channels, facilitating the transition to sleep. However, prolonged use raises concerns, including dependence and cognitive issues. Among herbal alternatives, *Humulus lupulus* (hops) is gaining attention due to its role as a natural relaxant, sleep aid, and brewing component. Neuroactive phytochemicals in hops may modulate GABA$_A$Rs differently from benzodiazepines. This research uncovers these hop constituents and potential therapeutic mechanisms.

The α-acid humulone and hop prenylflavonoids (PFs), including xanthohumol/ isoxanthohumol, 6/8-prenylnaringenin, enhanced GABA-induced displacement of [$^3$H]EBOB, a GABA$_A$R function marker, in the low micromolar range. These potent effects were flumazenil-insensitive and α6β3δ subtype-selective. Molecular docking at the α1β2γ2 isoform identified the extracellular α+/β− interface as the PF binding site. An additional 6-prenylnaringenin site was recognized at the extracellular α+/γ2− interface, aligning with its inhibition of [$^3$H]flunitrazepam and [$^3$H]Ro 15-4513 binding. Given humulone's prominence and relatively high potency, its activity was confirmed electrophysiologically, where it enhanced GABA-evoked currents in the sedation-mediating α1β3γ2 subtype. In mice, humulone reduced locomotor activity, shortened sleep onset induced by pentobarbital, and prolonged sleep duration induced by either pentobarbital or ethanol. Moreover, [$^3$H]EBOB binding assays showed synergies between humulone and ethanol, and additive interactions with PFs, suggesting enhanced alcohol intoxication in hop-rich beers.

In summary, we revealed positive modulators of GABA$_A$Rs that act independently of the classical benzodiazepine site. 6-prenylnaringenin also acts as a silent modulator with the potential to block benzodiazepine responses. Humulone plays a pivotal role in the sedative and sleep-promoting properties of hops. These findings offer novel mechanistic insights into hop neuroactive constituents and potential avenues for sleep aid optimization.

KEYWORDS: GABA(A) receptors, allosteric modulation, pharmacodynamics, *Humulus lupulus*, flavonoids, humulone, benzodiazepines, sleep, alcohol

TURUN YLIOPISTO
Lääketieteellinen tiedekunta
Biolääketieteen laitos
Farmakologia, lääkekehitys ja lääkehoito
ALI YASSER BENKHEROUF: Luonnon panimosta apua nukahtamiseen:
Humalan vaikutuksesta GABA~A~-reseptoriin ja unen edistämiseen
Väitöskirja, 189 s.
Lääketutkimuksen tohtoriohjelma
Joulukuu 2023

TIIVISTELMÄ

Pitkäaikainen unettomuus on terveysriski, joka voi vaatia lääkehoitoa parantamaan unen alkamista, ylläpitoa ja laatua. Bentsodiatsepiinit ja muut positiiviset säätelijät, vahvista-vat gamma-aminovoihapon (GABA) estäviä vaikutuksia vakauttamalla GABA~A~-resep-torin (GABA~A~R) kloridikanavan avointa muotoa ja helpottavat nukahtamista. Kuitenkin pitkäaikaisen käytön seurauksena voi syntyä haittoja, kuten riippuvuutta ja kognitiivisia ongelmia. Humalalla (*Humulus lupulus*) on pitkät perinteet luonnollisena rauhoittavana ja unilääkkeenä sekä merkittävä rooli panimoteollisuudessa. Humalan neuroaktiiviset fytokemikaalit saattavat säädellä GABA~A~R:a eri tavalla kuin bentsodiatsepiinit. Tämä tutkimus pyrki selvittämään näitä humalan yhdisteitä ja niiden mahdollisia terapeuttisia mekanismeja.

Humalan prenyyliflavonoidit ksantohumoli, isoksantohumoli, 6- ja 8-prenyyli-narin-geniini sekä α-happo humuloni tehostivat GABA~A~R-toimintaa. Nämä alhais-ten mikro-molaaristen pitoisuuksien vaikutukset olivat selektiivisempiä α6β3δ-alatyypille ja epä-herkkiä flumatseniilille. Sitoutumisennuste molekulaarisen tela-koinnin avulla α1β2γ2 reseptorialatyypissä paljasti solunulkoisen α+/β—alayksik-körajapinnan prenyyliflavo-noidien vahvistamisen sitoutumiskohtana. Lisäpaikka 6-prenyylinaringeniinille tunnis-tettiin solunulkoisessa α+/γ2−-rajapinnassa, mikä sopii sen aiheuttamaan [³H]flunitratse-paamin ja [³H]Ro 15-4513:n sitoutumisen estoon samassa kohdassa. Ottaen huomioon humulonin suuren pitoisuuden humalassa tutkimme sen vaikutusta GABA~A~R:in toimintaan sähköfysiologisesti. Humuloni tehosti GABAn aikaansaamia virtoja sedaatiota välittävässä α1β3γ2-alatyypissä. Se inhiboi hiirten spontaania liikeak-tiivisuutta ja lisäsi pentobarbitaalin ja etanolin aiheuttaman unen kestoa. [³H]EBOB-sitoutumiskokeet paljastivat yhteisvaikutuksia humulonin, prenyyliflavonoidien ja eta-nolin välillä, mikä viittaa alkoholin vaikutusten tehostamiseen humalapitoisissa oluissa.

Löysimme humalasta GABA~A~R:ien positiivisia säätelijöitä, jotka vaikuttivat ilman klassista bentsodiatsepiinien sitoutumispaikkaa. 6-prenyylinaringeniini toimi myös hil-jaisena säätelijänä, jolla on potentiaalia estää bentsodiatsepiinivasteita. Humulonilla on keskeinen rooli humalan sedatiivisissa ja unta edistävissä ominai-suuksissa. Nämä löydökset tarjoavat uusia mekanistisia näkemyksiä humalan neuro-aktiivisista yhdisteistä sekä aihioita olemassa olevien unilääkkeiden paranteluun.

AVAINSANAT: GABA(A) reseptorit, allosteerinen modulaatio, farmakodyna-miikka, *Humulus lupulus*, flavonoidit, humuloni, bentsodiatsepiinit, uni, alkoholi

# Table of Contents

# Abbreviations

| | |
|---|---|
| 2M3B | 2-methyl-3-buten-2-ol |
| 6PN | 6-prenylnaringenin |
| 8PN | 8-prenylnaringenin |
| ANOVA | Analysis of variance |
| BBB | Blood-Brain Barrier |
| cDNA | Complementary DNA |
| CNS | Central nervous system |
| cRNA | Complementary RNA |
| Cryo-EM | Cryogenic transmission electron microscopy |
| DRN | Dorsal raphe nucleus |
| EBOB | Ethynylbicycloorthobenzoate |
| $EC_{50}$ | Concentration producing half-maximal enhancement |
| ECD | Extracellular domain |
| EEG | Electroencephalogram |
| GABA | Gamma-aminobutyric acid |
| $GABA_AR$ | Gamma-aminobutyric acid type A receptor |
| HEK293 | Human embryonic kidney 293 cells |
| HPLC | High-performance liquid chromatography |
| $IC_{50}$ | Concentration producing half-maximal inhibition |
| ICD | Intracellular domain |
| i.p. | Intraperitoneal |
| IXN | Isoxanthohumol |
| $K_D$ | Equilibrium dissociation constant |
| $K_i$ | Inhibition constant |
| LC | locus coeruleus |
| LGIC | Ligand-gated ion channels |
| NAM | Negative allosteric modulator |
| NREM | Non-rapid eye movement |
| PAM | Positive allosteric modulator |
| p.o. | Per Os (by mouth) |
| REM | Rapid eye movement |

| Ro 15-4513 | Ethyl-8-azido-5,6-dihydro-5-methyl-6-oxo-4H-imidazo-1,4-benzodiazepine-3-carboxylate |
| --- | --- |
| SAM | Silent allosteric modulator |
| SWA | Slow-wave activity |
| TBPS | T-butylbicyclophosphorothionate |
| TM | Transmembrane |
| TRN | Thalamic reticular nucleus |
| Tris | Tris(hydroxymethyl)aminomethane |
| VLPO | Ventrolateral preoptic nucleus |
| XN | Xanthohumol |

# List of Original Publications

This dissertation is based on the following original publications, which are referred to in the text by their Roman numerals:

I    Benkherouf, A. Y., Soini, S. L., Stompor, M., & Uusi-Oukari, M. (2019). Positive allosteric modulation of native and recombinant $GABA_A$ receptors by hops prenylflavonoids. *European journal of pharmacology*, 852, 34–41.

II    Benkherouf, A. Y., Logrén, N., Somborac, T., Kortesniemi, M., Soini, S. L., Yang, B., Salo-Ahen, O. M. H., Laaksonen, O., & Uusi-Oukari, M. (2020). Hops compounds modulatory effects and 6-prenylnaringenin dual mode of action on $GABA_A$ receptors. *European journal of pharmacology*, 873, 172962.

III    Benkherouf, A. Y., Eerola, K., Soini, S. L., & Uusi-Oukari, M. (2020). Humulone Modulation of $GABA_A$ Receptors and Its Role in Hops Sleep-Promoting Activity. *Frontiers in neuroscience*, 14, 594708.

The original publications have been reproduced with the permission of the copyright holders.

# 1    Introduction

In an increasingly sleepless world of digital distractions and shifting norms, insomnia affects millions globally, carrying far-reaching implications for individuals and society at large. This disorder is marked by difficulties in initiating and maintaining sleep, as well as non-restorative sleep patterns. Statistical estimates indicate that around 10–15% of adults experience chronic insomnia, while an additional 25–35% encounter intermittent episodes of sleeplessness (Cho and Shimizu, 2015). The consequences of insomnia are profound, leading to impairments in daily performance, cognitive deficits, and increased risks of physical and neurological issues (Javaheri and Redline, 2017; Li et al., 2016). These effects extend beyond individual experiences, contributing to increased healthcare costs, decreased productivity, and an elevated risk of accidents. Addressing insomnia necessitates a comprehensive approach, often involving pharmacological interventions that aim to reduce sleep onset latency, improve sleep maintenance, and enhance sleep quality. Ideal interventions should restore regular wakefulness, improve daytime functioning, and carry a minimal risk of dependence.

Achieving these goals requires recognizing the intricate interplay between gamma-aminobutyric acid (GABA) and its crucial role in sleep regulation. GABA, a vital neurotransmitter in the brain, maintains the balance between excitatory and inhibitory neuronal signals, which is essential for optimal brain performance and restorative sleep (Morgan et al., 2012). Specifically, GABA triggers rapid inhibitory actions by binding to $GABA_A$ receptors, which are ligand-gated ion channels that facilitate chloride ion influx, resulting in hyperpolarization and the consequent suppression of neuronal firing. These pentameric $GABA_A$ receptors are composed of five subunits and can be classified into distinct subtypes based on their composition. Expression of $GABA_A$ receptor subunits varies across brain regions and exhibits heterogeneity at the cellular level, resulting in unique receptor subtypes with varying functional and pharmacological features (Olsen and Sieghart, 2008). The α1β2γ2 combination stands as the prevailing receptor subtype, significant for its role in the actions of benzodiazepines and Z-drugs, which are commonly prescribed for treating insomnia. These medications function as positive allosteric modulators of $GABA_A$ receptors at a common binding site, enhancing the inhibitory effects of GABA by stabilizing the open conformation of the receptor channel. Such modulation induces substantial alterations in circuit activity, facilitating the transition from alertness to sleep. Nevertheless, the use of benzodiazepines and Z-drugs presents various limitations, encompassing the risk of dependence, withdrawal, and side effects such as altered cognition and amnesia (Capiau et al., 2023).

Given these challenges with conventional treatments, herbal remedies have gained popularity as alternatives to conventional hypnotic drugs for insomnia management. One notable option is *Humulus lupulus* L., commonly known as hops, which is also widely used in the brewing industry to impart flavor and aroma to beer. Belonging to the Cannabaceae family, hops have a longstanding tradition across cultures as a natural relaxant and sleep promoter. This traditional use is supported by meta-analyses of randomized controlled trials, which show that combinations of hops and valerian improve sleep latency, depth, duration, and quality (Shinjyo et al., 2020). Furthermore, the European Medicines Agency (2016) officially recognized hops for their plausible effectiveness as a sleep aid and mild stress reliever. This effectiveness can be attributed to the pharmacological action of hops, which modulates $GABA_A$ receptors, resulting in sedative and sleep-promoting effects. While the therapeutic value of hops is acknowledged, the specific constituents responsible and their mechanisms are yet to be fully understood. Phytochemical analysis has revealed that hop cones contain a complex mixture of bioactive substances, such as resins, flavonoids, and essential oils (Almaguer et al., 2014). These secondary metabolites, released from the female inflorescences' lupulin glands as a yellow substance, have piqued interest due to their roles in specialty beer production and their potential health benefits.

As a matter of ongoing research, it has been suggested that α-acids from hop soft resin may contain the most potent constituents determining hops' sedative and sleep-promoting properties (Karabín et al., 2016). However, the possible contributions of β-acids, essential oils, and other unexplored constituents are also taken into account. Flavonoids, which are commonly present in the human diet, belong to a widely distributed family of polyphenols with distinct anxiolytic, sedative, and hypnotic effects. These compounds exhibit remarkable structural diversity and interact with multiple targets in the brain, including $GABA_A$ receptors. One particularly promising flavonoid from hops is xanthohumol, which has demonstrated modulatory actions on $GABA_A$ receptors (Meissner and Häberlein, 2006), but its exact molecular mechanisms and subtype selectivity require further elucidation. Existing evidence indicates that xanthohumol and similar hop prenylflavonoids may influence these receptors differently from benzodiazepines, potentially contributing to the sedative and sleep-promoting effects of hops.

Expanding upon the therapeutic implications of hops, it's noteworthy that beer consumption is the primary source of hop intake in humans (Stevens and Page, 2004; Van Cleemput et al., 2009). Therefore, it's essential to consider the involvement of GABA$_A$ receptors in the intoxicating actions of alcohol. Positive modulators of GABA$_A$ receptors, such as benzodiazepines, exhibit similarities and synergistic interactions with alcohol in terms of their behavioral effects. As a result, specific hop compounds might interact with alcohol, possibly amplifying the GABA$_A$ receptor-mediated effects of beer. This aspect raises important considerations for the health and safety of beer consumers and also underscores the prospects for developing hop-based remedies.

To address these knowledge gaps, guided by the assumption that hop constituents modulate GABA$_A$ receptors, this thesis aims to identify specific hop-derived modulators, unravel their mechanisms of action, subtype selectivity, and combinational effects, thereby elucidating the influence of hops on sleep. A multidisciplinary approach is necessary to achieve this comprehensive understanding, as it can capture the complex and multifaceted nature of hop phytochemistry and pharmacology, as well as the diverse and dynamic responses of GABA$_A$ receptors to hop compounds. Additionally, this approach will enable the integration of results from different levels of analysis, from molecular to behavioral, and facilitate the translation of the findings into practical applications for insomnia treatment. Therefore, this research will employ a range of techniques, including radioligand binding assays, electrophysiology, computational simulations, and *in vivo* evaluations.

In line with this approach, Study **I** will explore hop flavonoid modulatory potential, focusing on selectivity and allosteric modulation via the classical benzodiazepine site. Building upon the initial findings, Study **II** will identify putative binding sites and poses of potent hop flavonoids within the α1β2γ2 GABA$_A$ receptor subtype. Additionally, the study will examine the modulatory effects of isolated hop α/β-acid fractions, degradation products, and volatile compounds. Study **III** will select a neuroactive phytochemical from hops based on its content and potency, validate its functional activity, assess its selectivity, and explore its interactions with hop flavonoids and alcohol. The study will establish the phytochemical's role in hops-mediated sedation and sleep promotion. Ultimately, by uncovering the complex interactions between hops and GABA$_A$ receptors, this research aims to contribute to the development of novel, effective, and safe treatments for insomnia. The findings offer potential benefits for improving interventions and enhancing the quality of life for those affected by this sleep disorder.

# 2    Review of the Literature

## 2.1   GABA$_A$ receptors: overview

### 2.1.1    GABA: The central inhibitory neurotransmitter

Gamma-aminobutyric acid (GABA) is a crucial neurotransmitter in the brain responsible for maintaining the balance between neuronal excitation and inhibition, which is necessary for optimal brain function (Sieghart and Sperk, 2002; Sieghart, 2006; Pallotto and Deprez, 2014). It mediates 20–50% of all inhibitory synaptic neurotransmission in the central nervous system (CNS), with varying levels depending on the brain region (Bloom and Iversen, 1971; Sieghart, 1995; Barnard et al., 1998; Cooper et al., 1999; Bateson, 2004). GABA is synthesized from glutamate, which is produced from glucose metabolism via the Krebs cycle. This process involves the transamination of α-ketoglutarate by GABA transaminase (GABA-T) to form glutamate, which undergoes decarboxylation to form GABA (Figure 1). The conversion of GABA to glutamate in the presynaptic neuron is catalyzed by glutamic acid decarboxylase (GAD), which uses pyridoxal 5′-phosphate (PLP), the bioactive form of vitamin B6, as a coenzyme. GAD is expressed only in cells that use GABA as a neurotransmitter (Bown and Shelp, 1997; Ghit et al., 2021). In the mammalian nervous system, two forms of GAD (GAD67 and GAD65) catalyze the decarboxylation of glutamate to GABA at the synaptic terminals (Erlander et al., 1991; Kaufman et al., 1991; Esclapez et al., 1994; Awad et al., 2007). GABA accumulates in synaptic vesicles via a vesicular neurotransmitter transporter (VGAT) and is released by calcium ion ($Ca^{2+}$)-dependent exocytosis (Liu and Edwards, 1997).

The inhibitory effect of GABA is achieved by the activation of two distinct receptor classes: ionotropic GABA$_A$ receptors and metabotropic GABA$_B$ receptors. The GABA$_A$ receptors are responsible for mediating fast inhibition by opening chloride ion channels, whereas GABA$_B$ receptors mediate slow and prolonged inhibition by activating downstream signaling cascades through Gi/o proteins (Simeone et al., 2003; Shaye et al., 2021). The action of GABA in the synapse is terminated by reuptake at nerve terminals for reuse or reuptake to glial cells. Within these cells, GABA is metabolized by GABA-T to succinic semialdehyde (SSA), which then undergoes oxidation to succinate via succinic semialdehyde dehydrogenase (SSADH) (Ghit et al., 2021). Through tightly regulated synthesis, receptor-mediated activation, and subsequent termination, GABA exerts precise control over various inhibitory mechanisms within neural circuits, forming the basis for its diverse functional roles and pharmacological targeting.

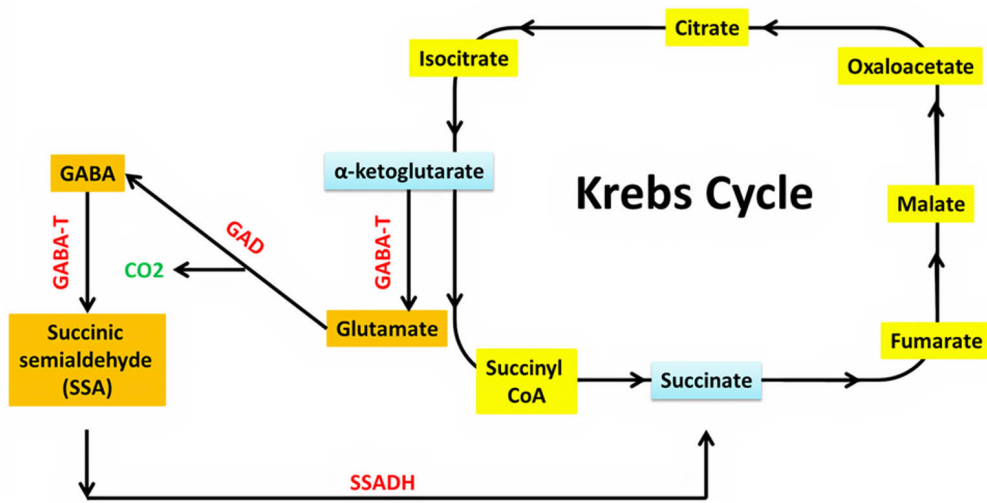

**Figure 1**. GABA biosynthesis and metabolism pathway in relation to Krebs cycle. GABA, γ-aminobutyric acid; GABA-T, GABA transaminase; GAD, glutamate decarboxylase; SSADH, succinic semialdehyde dehydrogenase. Reprinted from Ghit et al. (2021). CC BY 4.0.

## 2.1.2    Modes of neuronal inhibition

GABA$_A$ receptors are primarily responsible for mediating the fast inhibitory action of GABA in the CNS (Olsen and Sieghart, 2009). These receptors contain an intrinsic chloride channel that opens upon activation, leading to the influx of chloride ions that induce membrane potential hyperpolarization, ultimately inhibiting the neuron (Olsen and Sieghart, 2008). GABA$_A$ receptors mediate two modes of neuronal inhibition - phasic inhibition by synaptic receptors and tonic inhibition by extrasynaptic receptors (Figure 2). The synaptic GABA$_A$ receptors exhibit low affinity for GABA, responding to high levels of GABA (1-10 mM) that are released throughout the synaptic cleft, leading to a transient and fast desensitizing postsynaptic inhibition (Brickley and Mody 2012; Lee and Maguire 2014). Whole-cell patch clamp recordings can identify this form of phasic response as inhibitory postsynaptic currents (IPSC) (Brickley and Mody 2012; Hunt et al., 2013). Conversely, the extrasynaptic GABA$_A$ receptors display a high affinity for GABA and respond to ambient levels of the neurotransmitter that escape the synaptic cleft (known as spillover) or are released from astrocytes, resulting in persistent, slow-desensitizing tonic inhibition (Brickley et al., 1996; Mody, 2001; Walker and Semyanov, 2008; Lee and Maguire, 2014). Modulation of phasic and/or tonic inhibition forms the basis of GABA$_A$ receptor-targeting drugs, which offer therapeutic benefits for managing various conditions associated with neuronal excitability, including but not limited to anxiety, insomnia, epilepsy, amnesia, and alcoholism (Rudolph and Möhler, 2006; Olsen and Sieghart, 2008; Rudolph and Knoflach, 2011; Shen et al., 2012).

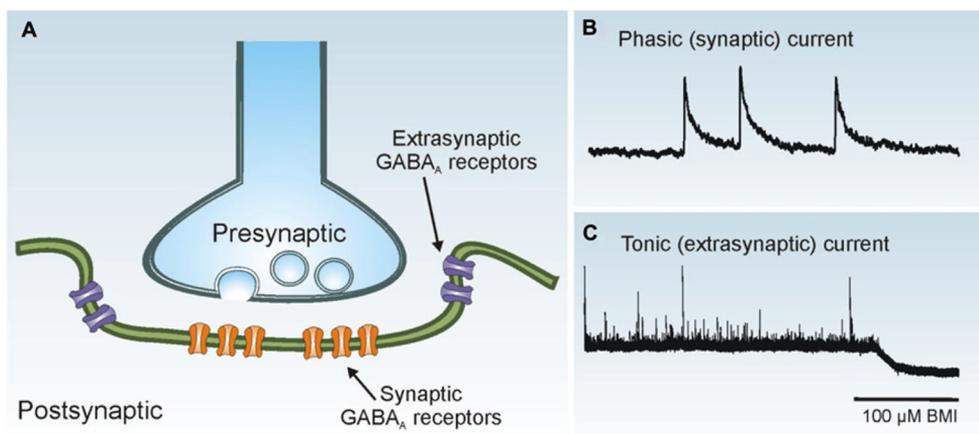

**Figure 2.** Modes of GABA$_A$ receptor activation. (A) Locations of synaptic (orange) and extrasynaptic (purple) receptors relative to the presynaptic GABAergic terminal. (B) Whole-cell voltage-clamp recording for the inhibitory postsynaptic currents (IPSCs) in a dentate granule cell. (C) Isolated tonic currents from a dentate granule cell, as revealed by the shift in the holding current in response to the presence of a high concentration of GABA$_A$ receptor antagonists bicuculline methiodide (BMI; 100 μM). Reprinted from Hunt et al. (2013). CC BY 3.0.

### 2.1.3    Receptor structure and subunit composition

GABA$_A$ receptors are predominantly heteropentameric membrane-spanning subunit complexes that belong to the Cys-loop ligand-gated ion channel (LGIC) superfamily, including nicotinic acetylcholine (nAChR), glycine (Gly), and serotonin type 3 (5-HT3) receptors (Sieghart, 1995; Johnston, 1996). GABA$_A$ receptors, like most members of the LGIC superfamily, typically consist of five protein subunits positioned surrounding a central pore that serves as the ion channel (Chebib and Johnston, 2000). Each of these subunits comprises various domains: a large extracellular amino N-terminal domain (ECD) that includes a component of the orthosteric agonist binding site, four transmembrane α-helices (TM1–4) that span the lipid bilayer, one extracellular loop between TM2 and TM3, two intracellular loops between TM1 and TM2 and between TM3 and TM4, and a small extracellular carboxy C-terminal domain (Johnston, 2005). Notably, the second membrane-spanning domain (TM2) lines the channel pore's wall and its net charge dictates whether the channel conducts anions or cations (Figure 3) (Jensen et al., 2002). Recent advancements in GABA$_A$ receptor structural understanding have been driven by Cryogenic transmission electron microscopy (Cryo-EM) techniques. This method has enabled researchers to capture detailed structures of the receptor in near-physiological states, notably revealing the human GABA$_A$ receptor with α1β2γ2 subunit composition at a 3.9 Å resolution (Zhu et al., 2018). Through this approach, significant insights into ligand-receptor interactions and their preferences for distinct subtypes have been uncovered (Laverty et al., 2019; Masiulis et al., 2019), potentially influencing drug design for optimized affinity and selectivity (de Oliveira et al., 2021).

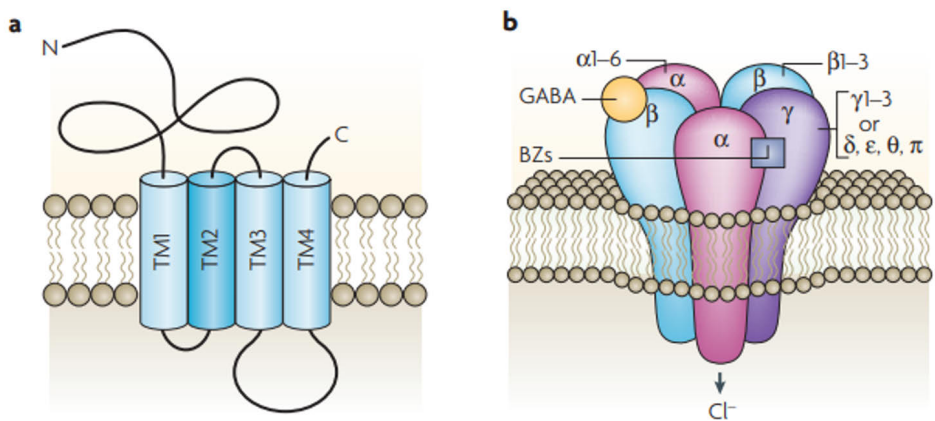

**Figure 3.** The structure and subunit composition of a typical GABA$_A$ receptor represented schematically. (A) The mature subunit is comprised of a large hydrophilic extracellular amino N-terminus, four hydrophobic transmembrane domains (TM1–4), and a small hydrophilic extracellular carboxy C-terminus. TM2 lines the channel pore. (B) The heteropentameric chloride permeable channel is formed by the assembly of five subunits from seven subfamilies (α, β, γ, δ, ε, θ and π). Modified with permission from Jacob et al. (2008) © Springer Nature.

## 2.1.4 Subunit diversity and assembly

Molecular cloning investigations have shown that GABA$_A$ receptors consist of five subunits, each displaying distinctive brain localization patterns (Olsen and Tobin, 1990; Lüddens and Wisden, 1991; Wisden et al., 1992). These receptor complexes are composed of 8 subunit classes that are encoded by a collective of 19 genes found in mammals: α1-α6, β1-β3, γ1-γ3, δ, ε, π, θ, and ρ1-ρ3 (Figure 4; Whiting et al., 1999; Korpi et al., 2002; Rudolph and Möhler, 2006; Olsen and Sieghart, 2008). In the endoplasmic reticulum, GABA$_A$ receptors are assembled from their individual subunits in a multistep process that regulates the heterogeneous composition of receptor subtypes on the surface of neuronal cells (Klausberger et al., 2000; Jacob et al., 2008). The predominant configuration of the majority of GABA$_A$ receptors involves α, β, and γ subunits in 2α:2β:1γ stoichiometry, with each receptor containing two GABA-binding sites and arranged γ-β-α-β-α anticlockwise around the channel pore (Sigel and Buhr, 1997; Tretter et al., 1997; Farrar et al., 1999). Additionally, the ρ1-ρ3 subunits often form homopentameric or heteropentameric complexes (Cutting et al., 1991; Enz and Cutting, 1998; Sallard et al., 2021), with five GABA-binding sites per receptor (Bormann, 2000). The γ subunit may be replaced by specific subunits such as δ, ε, θ, or π, thereby contributing to the receptor's molecular diversity (Jacob et al., 2008). Despite the proposed alteration from 2α:2β:1γ to 2α:2β:δ, the definitive stoichiometry and arrangement of δ-GABA$_A$Rs have yet to be conclusively determined (Botzolakis et al., 2016; Hartiadi et al., 2016).

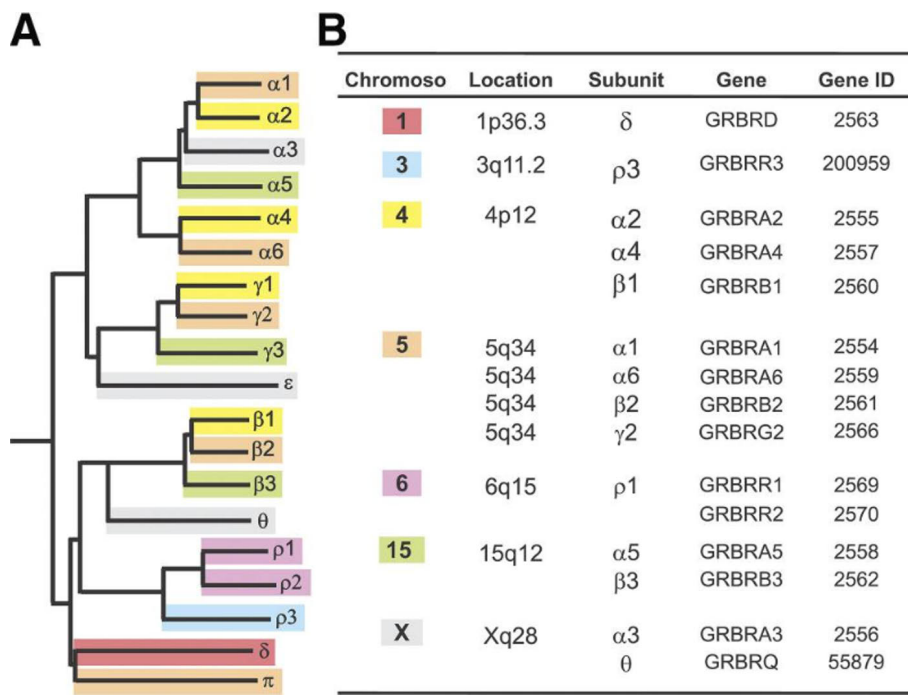

**Figure 4.** GABA$_A$ receptors subunit classes and genes. (A) Phylogenetic tree depicting amino acid sequence interconnections in 19 mammalian GABA$_A$ receptor subunits and the ancestral gene. (B) Mapping positions of GABA$_A$ receptor subunit-encoding genes in the human genetic code. Reprinted with permission from Chuang and Reddy (2018) © American Society for Pharmacology and Experimental Therapeutics.

Recent advances in single-cell RNA sequencing (scRNA-seq) have revealed new insights into the diversity and co-expression of GABA$_A$ receptor subunits in the brain. Analysis of scRNA-seq data from the human cortex (Tasic et al., 2018; Hodge et al., 2019) unveiled the simultaneous presence of mRNAs for up to 14 different GABA$_A$ receptor subunits within individual cell types. Although the abundance of mRNA may not precisely reflect protein levels, the distinct co-expression pattern of these 14 subunits could hypothetically generate as many as 62,847 unique receptor subtypes (Sente et al., 2022). These findings suggest that the diversity of GABA$_A$ receptor subtypes and responses to GABA may be much greater than previously appreciated. Therefore, further scRNA-seq studies will be instrumental in uncovering cell-type-specific and region-specific patterns of GABA$_A$ receptor subunit expression, reflecting the functional specialization and adaptation of GABAergic signaling in different neuronal circuits.

## 2.1.5    Subtype regional distribution

The GABA$_A$ receptor mRNA expression varies across brain regions and cell types (Laurie et al., 1992; Wisden et al., 1992), resulting in the receptor isoforms with distinct functional and pharmacological characteristics. GABA$_A$ receptors consisting of α1 and γ2 subtypes are found in almost all regions of the brain, with the α1β2γ2 combination representing the most prevalent subtype (Rabow et al., 1995; Hevers and Lüddens, 1998). Notably, the γ2 subunit is incorporated into over 90% of GABA$_A$ receptor subtypes, and its presence is essential for benzodiazepine binding (Pritchett et al., 1989; McKernan and Whiting, 1996; Tretter et al., 1997). GABA$_A$ receptor assemblies of α1/2/3/5 subunits in conjunction with β and γ subunits are mostly localized at post-synaptic regions. In these locations, they mediate rapid phasic inhibition and primarily demonstrate high sensitivity to benzodiazepines (Olsen and Sieghart, 2008). Benzodiazepine-sensitive receptors carrying the α1 subunit are abundantly expressed in the cortex, thalamus, and cerebellum (Sieghart, 1994; Rudolph et al., 1999), whereas those containing the α2 subunit are prevalent in the limbic system, spinal dorsal horn, and motor neurons (Fritschy and Möhler, 1995; Bohlhalteret al., 1996).

Naturally, the δ subunit is selectively localized to extrasynaptic sites and preferentially associates with α6/α4 and β subunits. The α6β2/3δ isoforms predominate in the cerebellar granule cells (Nusser et al., 1998), while α4β2/3δ isoforms are expressed in thalamic relay neurons, neocortical pyramidal cells, hippocampal dentate gyrus granule cells, and the striatum (Pirker et al., 2000; Belelli et al., 2005; Brickley and Mody, 2012). These δ-GABA$_A$Rs are insensitive to benzodiazepines but exhibit high GABA affinity (Saxena and Macdonald, 1996; Brown et al., 2002), mediating tonic inhibition in various neuronal populations across the brain (Semyanov et al., 2004; Glykys and Mody, 2007; Belelli et al., 2009). The ρ-GABA$_A$Rs are predominantly expressed in the retina, specifically on the axon terminals and dendrites of bipolar cells (Enz et al., 1995; Qian and Dowling, 1995), and can co-assemble with α1 and/or γ2 subunits in the cerebellar cortex and brain stem neurons (Milligan, 2004; Harvey et al., 2006).

## 2.1.6 Molecular interactions and mechanisms

### 2.1.6.1 Orthosteric site-mediated activation

Based on site-specific mutagenesis and radioligand binding studies, it is accepted that the GABA binding site, referred to as the orthosteric site, is situated at the two extracellular β+/α− interfaces of the GABA$_A$ receptor (Figure 5B) (Smith and Olsen, 1995; Ernst et al., 2003; Sieghart, 2015). When GABA binds to one of the sites, the intrinsic chloride channel opens, but the likelihood increases considerably when both sites are occupied (Macdonald et al., 1989; Twyman et al., 1990; Baumann et al., 2003; Chua and Chebib, 2017). The binding pockets at the β+/α− interfaces are made up of compact aromatic residues from discontinuous subunit components, including the amino acids α1F65, β2Y97, β2Y157, and β2Y205 (Figure 5C) (Padgett et al., 2007; Masiulis et al., 2019; Kim and Hibbs, 2021). Orthosteric agonists, such as the GABA analogs tetrahydroisoxazolopyridinol (THIP/gaboxadol) and muscimol (Figure 5A), trigger direct activation of GABA$_A$ receptors by binding to the GABA binding site, with high affinity for δ-GABA$_A$Rs (Mihalek et al., 1999; Meera et al., 2011; Benkherouf et al., 2019). The action of GABA and other orthosteric agonists can be inhibited by orthosteric antagonists, including bicuculline and SR 95531 (gabazine) by competing for binding at the same site (Johnston, 2013).

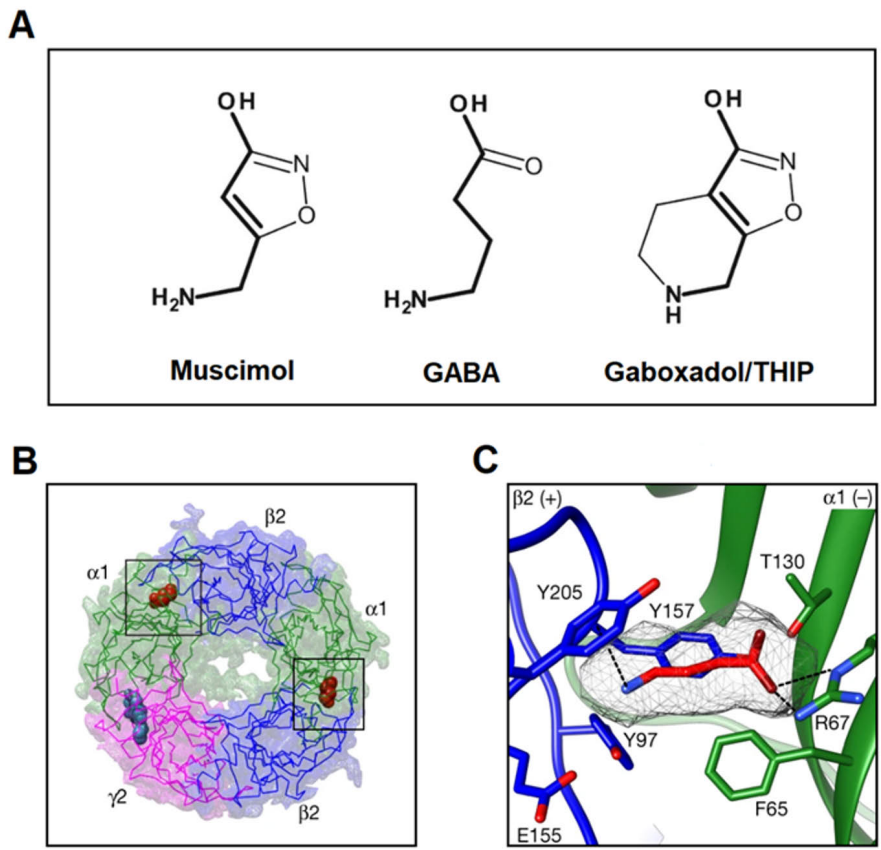

**Figure 5.** The orthosteric binding site. (A) Chemical structures of GABA and its analogs muscimol and gaboxadol/THIP. (B) Top view of the heteropantemeric assembly of the human α1β2γ2 GABA_A receptor where the boxes indicate two equivalent GABA binding sites at the β2+/α1− subunit interface. (C) Magnified view of the orthosteric site with bound GABA and its interacting residues. Modified with permission from Zhu et al. (2018) © Springer Nature.

### 2.1.6.2    Allosteric sites and multifaceted modulation

For a wide range of ligands, GABA_A receptors exhibit allosteric interacting sites, which are remote from the orthosteric binding site (Olsen, 2015; Sieghart, 2015). These allosteric sites mediate various modes of action on GABA_A receptors and can be restricted to particular receptor subtypes, offering greater selectivity and diversity in the interacting residues compared to the orthosteric sites (Christopoulos, 2002; Johnston, 2005; Olsen, 2015). Numerous clinically used ligands exert their actions through specific allosteric sites, including anxiolytics, anticonvulsants, sedative-hypnotics, general anesthetics, and muscle relaxants (Johnston, 2005).

Positive allosteric modulators (PAMs) are ligands that enhance the function of the $GABA_A$ receptor by stabilizing the open conformation of the receptor channel in the presence of GABA (Figure 6) (Chua and Chebib, 2017). Benzodiazepines, alcohol (ethanol), loreclezole, barbiturates, anesthetics (e.g. propofol and etomidate), neurosteroids (e.g. allopregnanolone and tetrahydrodeoxycorticosterone; THDOC), flavonoids, and terpenes are examples of these modulators (Wingrove et al., 1994; Johnston, 2005; Skilbeck et al., 2010). Certain PAMs, such as barbiturates, neurosteroids, and anesthetics, exhibit GABA-mimetic properties at higher concentrations, thereby directly activating $GABA_A$ receptors (Stephens et al., 2017; Alanis et al., 2020). Negative allosteric modulators (NAM) or "inverse agonists" have the contrary effect of PAMs in that they decrease GABA-induced responses (Figure 6) (Henschel et al., 2008; Rudolph and Knoflach, 2011). Non-selective NAMs, such as picrotoxin, inhibit $GABA_A$ receptor activity by binding non-competitively within the chloride channel pore, resulting in anxiogenic and convulsive effects (Olsen, 2006; Kalueff, 2007; Olsen, 2018). Interestingly, selective NAMs, particularly for α5-$GABA_A$Rs (e.g., α5IA and basmisanil), have been found to exhibit cognitive-enhancing benefits without anxiogenic or convulsive/ proconvulsive effects (Braudeau et al., 2011; Hipp et al., 2021). Ligands that have limited or no influence on the GABA-mediated response, yet can block the actions of PAMs or NAMs by competing for binding to their corresponding allosteric sites, are termed null or silent allosteric modulators (SAM) (e.g. flumazenil) (Figure 6) (Alanis et al., 2020; Ghit et al., 2021; Zhu et al., 2022).

It is not ruled out that these various classes of allosteric modulators can bind to multiple overlapping and non-overlapping sites on $GABA_A$ receptors and influence receptor function in an additive or synergistic manner (Kim and Hibbs, 2021). The pyrazoloquinoline 2-p-methoxyphenylpyrazolo [4,3-c] quinolin-3(5H)-one (CGS 9895) was demonstrated to bind to $GABA_A$ receptors at two distinct allosteric sites. It acted as a PAM at the ECD α+/β− interface and as a SAM at the ECD α+/γ2− interface, the classical benzodiazepine binding site (Ramerstorfer et al., 2011). Furthermore, methyl-6,7-dimethoxy-4-ethyl-β-carboline-3-carboxylate (DMCM) was found to exhibit biphasic effects on GABA-evoked currents, acting as a NAM through the benzodiazepine site at low nanomolar and as a PAM through the loreclezole site at high micromolar concentrations (Im et al., 1995; Stevenson et al., 1995). These observations underscore the complexity of allosteric modulation on $GABA_A$ receptors, wherein certain ligands exhibit diverse modes of action dependent on both concentration and the specific binding site with which they interact.

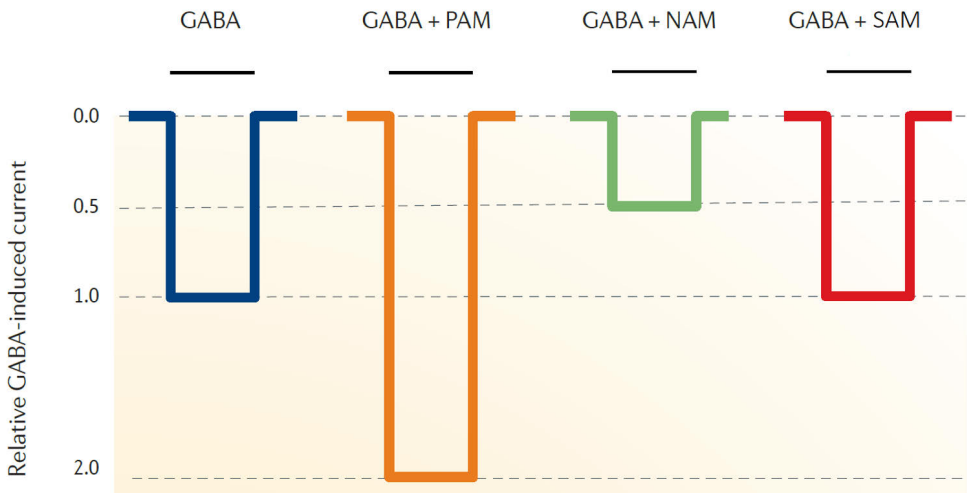

**Figure 6.** A graphical representation of the effects on GABA-evoked currents by GABA$_A$ receptor positive allosteric modulators (PAM) (enhancing), negative allosteric modulators (NAM) (diminishing), and silent allosteric modulators (SAM) (no effect). Modified with permission from Rudolph and Knoflach (2011) © Springer Nature.

## 2.2 GABA$_A$ receptors: Allosteric modulators

### 2.2.1 Benzodiazepines: Functions and binding dynamics

Benzodiazepines, including chlordiazepoxide and diazepam, were adopted clinically in the 1960s for their effectiveness as anxiolytics, anticonvulsants, sedatives, hypnotics, and muscle relaxants (Sieghart, 2015). Currently, benzodiazepines stand as some of the most commonly prescribed therapeutic drugs. Benzodiazepines modulate GABA$_A$ receptors allosterically and exert their inhibitory effects by increasing the flux of GABA-induced chloride ions (Rudolph and Möhler, 2005; Sigel and Lüscher, 2011; Ghit et al., 2021). Most GABA$_A$ receptors in the nervous system, spinal cord included, are benzodiazepine-sensitive GABA$_A$ receptors, which are made up of two α (α1, α2, α3 or α5) subunits, two β subunits, and a single γ2 subunit (Zeilhofer et al., 2009). The α subunit class primarily determines benzodiazepine pharmacology (Pritchett et al., 1989; Sieghart, 1995). Studies using GABA$_A$ receptor knock-in mouse lines identified behavioral correlations between a specific α subunit isoform and various pharmacological effects of benzodiazepines (Figure 7) (Rudolph et al., 2011; Tan et al., 2011). For example, sedation and amnesia induced by benzodiazepines are mediated by α1-GABA$_A$Rs while anxiolysis is mediated by α2-GABA$_A$Rs (Rudolph et al., 1999; Low et al., 2000; Möhler et al., 2002).

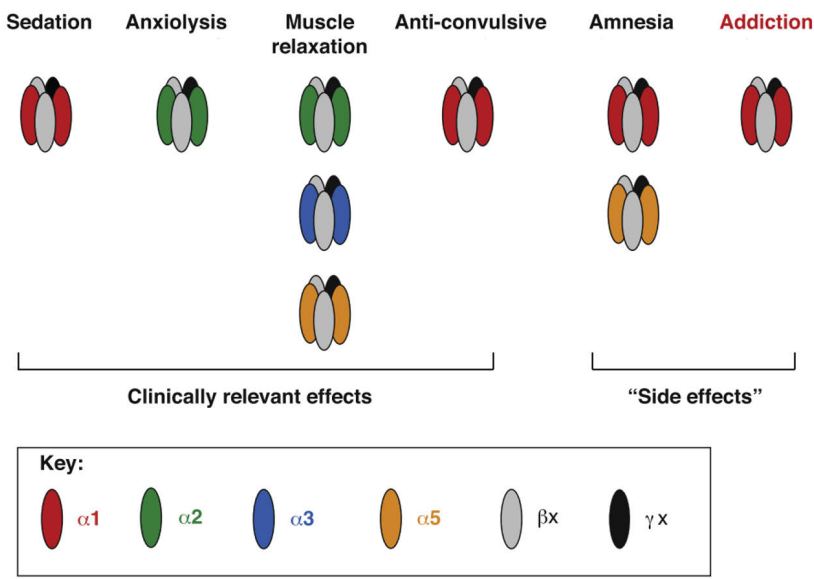

**Figure 7.** Pharmacological effects of benzodiazepines and the associated α subunits of GABA$_A$ receptors. Adapted with permission from Tan et al. (2011) © Elsevier.

### 2.2.1.1 The classical benzodiazepine site

At the single-channel level, analysis indicated that benzodiazepines exert their positive modulatory effects through an increase in the frequency of the chloride channel opening in bursts, without altering burst duration or channel conductance (Macdonald and Twyman, 1992; Macdonald and Olsen, 1994; Sieghart, 1995). This activity is mediated by a binding site known as the high-affinity or classical benzodiazepine site at the extracellular α+/γ− interface (site 1) (Figure 9), which is closely homologous to the two GABA-binding sites at the extracellular β+/α− interfaces (Ernst et al., 2003; Sigel and Ernst, 2018; Ghit et al., 2021). GABA$_A$ receptor knock-in mouse lines indicated that benzodiazepine-sensitive receptors have a histidine residue in a conserved position at the α+/γ− interface of their extracellular N-terminus (α1H101, α2H101, α3H126, and α5H105) (Wieland et al., 1992; Benson et al., 1998). Benzodiazepine-insensitive receptors with α4 or α6 subunits, on the other hand, possess an arginine residue at the corresponding position (α4R99 and α6R100) (Wieland et al., 1992; Dunn et al., 1999; Korpi et al., 2002).

Classical benzodiazepines, including diazepam and flunitrazepam, show comparable nanomolar affinity for the benzodiazepine-sensitive GABA$_A$R subtypes (α1, α2, α3, and α5), and this interaction can be inhibited by the antagonist flumazenil (Sieghart, 1995; Lüddens et al., 1995; Atack et al., 1999). Not only drugs with a benzodiazepine structure can interact with the high-affinity benzodiazepine site but also compounds with non-benzodiazepine structures, including β-carbolines (e.g. DMCM and abecarnil), imidazopyridines (e.g. zolpidem), triazolopyridazines (e.g. CL 218,872), cyclopyrrolones (e.g. zopiclone and suriclone ) and pyrazoloquinolinones (e.g. CGS 8216) (Figure 8) (Sieghart, 1995; Lüddens et al.,1995; Sigel, 2002; Sigel and Ernst, 2018). In addition, other ligands targeting the benzodiazepine site, such as Z-drugs (zolpidem, zaleplon, and zopiclone), quinolones, triazolopyridazines and β-carbolines, exhibit a greater affinity for α1-GABA$_A$Rs compared to α2- or α3-GABA$_A$R subtypes, while having almost no influence on α5-GABA$_A$Rs (Pritchett and Seeburg, 1990; Lüddens et al., 1995; Olsen and Sieghart, 2008; Olsen and Sieghart, 2009).

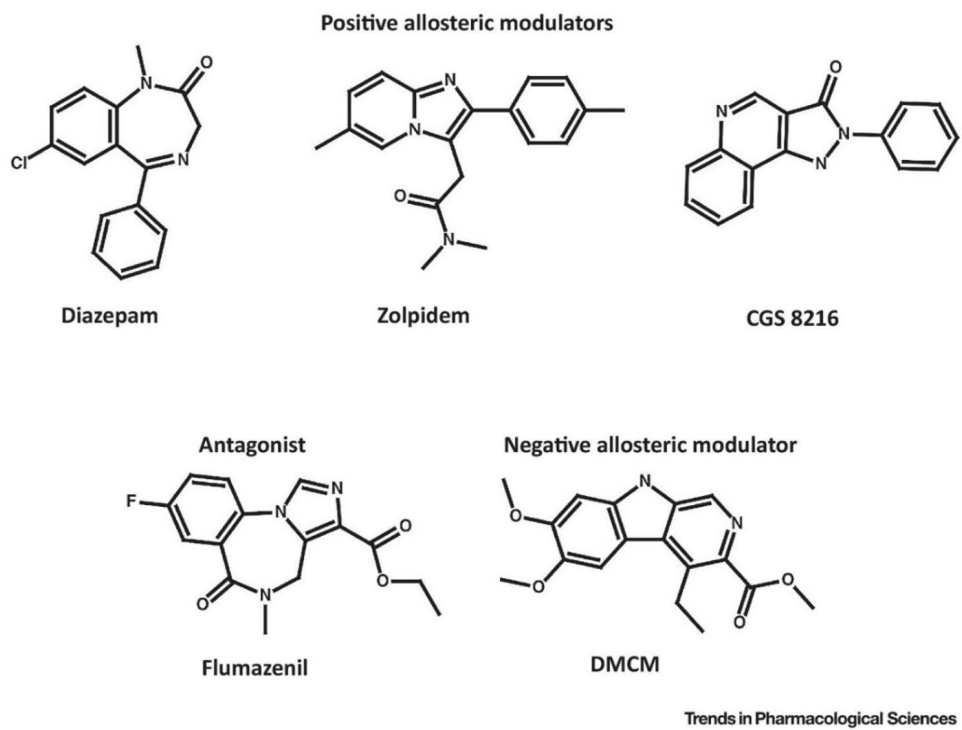

**Figure 8.** Chemical structures of prominent benzodiazepine and non-benzodiazepine compounds targeting the high-affinity benzodiazepine binding site. Reprinted with permission from Sigel and Ernst (2018) © Elsevier.

## 2.2.1.2    The non-classical benzodiazepine sites

Certain ligands for the high-affinity benzodiazepine binding site have been shown to interact at additional sites in the GABA$_A$ receptor complex with lower affinity (Olsen, 2008; Ghit et al., 2021). CGS 9895, identified as a silent allosteric modulator at the high-affinity benzodiazepine site (site 1), has been found to act as a positive allosteric modulator at a corresponding homologous site at the extracellular α+/β− interface (site 2) (Figure 9) involving α1Y209 as a key interacting residue (Ramerstorfer et al., 2011; Varagic et al., 2013; Maldifassi et al., 2016). Furthermore, Ro 15-4513 (ethyl-8-azido-5,6-dihydro-5-methyl-6-oxo-4H-imidazo-1,4-benzodiazepine-3-carboxylate) is an imidazobenzodiazepine that functions as a partial inverse agonist at the high-affinity benzodiazepine site (Bonetti et al., 1988; Hadingham et al., 1993; Korpi et al., 2002). Notably, Ro 15-4513 has also been shown to bind to α4/6βγ2 receptor subtypes, where it functions as an agonist (Bonetti et al., 1988; Knoflach et al., 1996). Homology modeling suggested that Ro 15–4513 is capable of binding to GABA$_A$ receptors at the extracellular α+/β3− interface, with α6R100 and β3Y66 residues contributing to the interaction at this additional benzodiazepine binding site (Wallner et al., 2014).

Classical benzodiazepines can also display low-affinity components on GABA$_A$ receptors and independently of the γ subunit. For example, diazepam was found to potentiate GABA-evoked currents in recombinant α1β2 and α1β2γ2 receptors at micromolar concentrations, in a manner insensitive to flumazenil antagonism. At nanomolar concentrations, diazepam enhanced GABA-induced response only in α1β2γ2 receptors and in a flumazenil-sensitive manner (Walters et al., 2000). This behavior suggested an additional 'low-affinity' site and suggested that the action of benzodiazepines on α1β2γ2 GABA$_A$ receptors is biphasic, consisting of nanomolar (high-affinity) and micromolar (low-affinity) components (Ramerstorfer et al., 2011; Sigel and Ernst, 2018; Lian et al., 2020).  Mutations in the amino acid residues α1S269, β2N265, and γ2S280 within the TM2 domain abolished the micromolar, rather than the nanomolar, action of diazepam (Walters et al., 2000). These residues contribute to allosteric modulation of GABA$_A$ receptors by therapeutically relevant anesthetics, including etomidate and propofol (Chiara et al., 2012; Maldifassi et al., 2016). In recent years, high-resolution cryo-EM structures of the human α1β3γ2L GABA$_A$ receptor uncovered this low-affinity benzodiazepine-binding site at the TM2 β3+/α1− interface (site 3), which is likely implicated in the high dose anesthetic actions of diazepam (Masiulis et al., 2019; Kim et al., 2020).

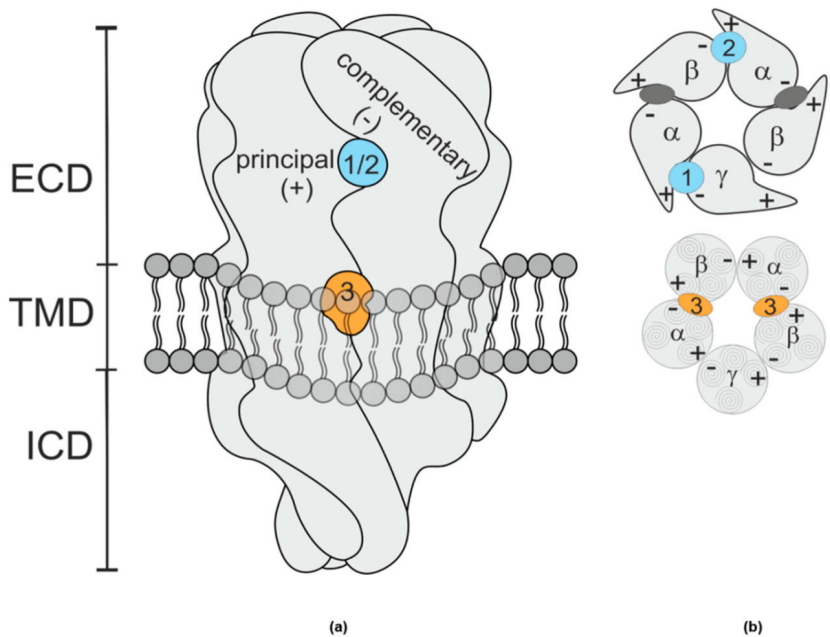

**Figure 9.** A graphical illustration of GABA<sub>A</sub> receptor benzodiazepine-binding sites. (a) A side view of the membrane-spanning receptor displaying the extracellular domain (ECD), the transmembrane domain (TMD), and the intracellular domain (ICD). A principal subunit face (+) and a complementary subunit face (−) constitute the binding sites at the interfaces. (b) Top view of the ECD of αβγ GABA<sub>A</sub> receptor, indicating GABA binding sites (grey). The classical high-affinity benzodiazepine binding site is shown at the ECD α+/γ− interface (site 1: blue) and a homologous site is indicated at the ECD α+/β− interface (site 2: blue). The low-affinity site is shown at the two TM β+/α− subunit interfaces (site 3: orange). Reprinted from Iorio et al. (2020). CC BY 4.0.

## 2.2.2    Ethanol: Behavioral and neurological impacts

Ethanol, as a CNS depressant, produces several behavioral effects such as anxiolysis, impaired motor coordination, sedation, and hypnosis; it also exhibits anticonvulsant properties (Grobin et al., 1998). It has been linked with cognitive deficits (Matthews et al., 1995; Givens and McMahon, 1997) and, at large doses, can induce anesthesia and respiratory depression (Koch-Weser et al., 1976). Chronic alcohol intake results in neuroplastic alterations in the brain, which underlie the emergence of tolerance and addiction, along with withdrawal effects (Enoch, 2008; Koob and Le Moal, 2008; Pietrzykowski and Treistman, 2008). Additionally, there is a correlation between alcohol intake and cerebellar ataxia, manifested by symptoms such as unsteady gait and impaired posture (Laureno, 2012; Dar, 2015). In chronic alcoholism, these deficits continue, primarily attributed to the pronounced atrophy observed in the cerebellar vermis (Torvik and Torp, 1986; Baker et al., 1999; Sullivan et al., 2000).

### 2.2.2.1    GABA$_A$ receptor-mediated responses

GABA$_A$ receptors have long been thought to be involved in the intoxicating effects of alcohol. Positive modulators of GABA$_A$ receptors, including barbiturates and benzodiazepines, showed obvious similarities and synergistic interactions with ethanol in terms of their behavioral responses (Isbell et al., 1950; Hu et al., 1987; Tauber et al., 2003). Moreover, muscimol, a GABA$_A$ receptor agonist, enhanced the sedative effects of ethanol, while picrotoxin and bicuculline, GABA$_A$ receptor blockers, diminished ethanol-induced sedation as well as ethanol-induced motor incoordination (Liljequist and Engel, 1982; Wallner, 2006). Long-sleep (LS) and short-sleep (SS) mice, that vary in their genetic response to ethanol, have been shown to exhibit distinct behavioral sensitivity to the GABA$_A$ receptor agonist gaboxadol (Martz et al., 1983; Lobo and Harris, 2008). In laboratory rats, gaboxadol augmented voluntary alcohol consumption (Smith et al., 1992), while Ro 15-4513 and picrotoxin showed the opposite effect on such intake (Samson et al., 1987; Boyle et al., 1993).

Ethanol was first found to potentiate GABA-mediated inhibition using cortical neurons in cats (Nestoros, 1980). Later studies showed that intoxicating levels (10–50 mM) of ethanol potentiated GABA-evoked chloride flux in various preparations, including cultured spinal cord neurons, rat cortical synaptoneurosomes, and microsacs (Allan and Harris, 1986; Suzdak et al., 1986; Ticku et al., 1986). Cerebellar neurons have been essential in investigating the influence of ethanol on GABA$_A$ receptors. Specifically, *in vivo* measurements showed that ethanol decreased the activity of purkinje neurons in rats, an inhibition that was counteracted by Ro 15–4513 and bicuculline (Palmer et al., 1988; Palmer and Hoffer, 1990; Förstera et al., 2016). However, the first evidence that ethanol enhanced GABA-evoked chloride current emerged from patch clamp recording in hippocampal and cortical neurons (Aguayo, 1990).

Ethanol modulates the activity of GABA$_A$ receptors in an allosteric manner, resulting in enhanced GABA-evoked currents (Wallner et al., 2003; Olsen et al., 2007). A study at the level of a single GABA$_A$ receptor channel reported that these enhancements are attributed to the increase in both frequency and duration of ion channel open state. There was also an extension in the number of opening bursts and the duration of these bursts (Tatebayashi et al., 1998). Furthermore, the duration of the ion channel's closed state was reduced. The combined effects of these actions, with GABA and ethanol present, result in an enhanced chloride ion flux across the open channel (Tatebayashi et al., 1998; Davies, 2003).

## 2.2.2.2    Sensitivity of extrasynaptic δ-GABA$_A$Rs

The major subset of GABA$_A$ receptors is synaptic and typically not sensitive to ethanol levels experienced while social alcohol drinking (≤30 mM) (Wallner et al., 2006). Conversely, these receptors can be modulated by high ethanol concentrations (50–200 mM) that correspond with anesthetic or lethal doses (Mihic et al., 1997; Wallner, 2006; Olsen, 2018). This modulation is mediated by a specific binding site for ethanol within the GABA$_A$ receptor TM2 and TM3 transmembrane domains, a site also essential for the anesthetic actions of etomidate and propofol (Wallner, 2006). The proposed ethanol-binding pocket is comprised of residues α2S270 and β1S265 in the TM2 and α2A291 and β1M286 in the TM3 domains (Mihic et al., 1997; Mascia et al., 2000; Ueno et al., 2000).

Only particular subtypes of the GABA$_A$ receptor appear to respond to ethanol at doses attained during social drinking (3–30 mM) (Wallner, 2006; Nie et al., 2011). Extrasynaptic δ-GABA$_A$Rs, which mediate the tonic inhibitory currents, have been identified as being responsive to low ethanol levels (≤30 mM) in numerous reports (Sundstrom-Poromaa et al., 2002; Wallner et al., 2003; Wei et al., 2004; Hanchar et al., 2005; Santhakumar et al., 2006; Glykys et al., 2007). The remarkable sensitivity of GABA-evoked tonic currents to ethanol has been shown in several investigations using dentate granule and cerebellar granule cells (Carta et al., 2004; Wei et al., 2004; Hanchar et al., 2005; Liang et al., 2006). However, only recombinant α4/6β3δ GABA$_A$ receptor subtypes expressed in Xenopus oocytes exhibited responsiveness to ethanol at levels as low as 3 mM (Wallner et al., 2003). This low ethanol concentration was demonstrated to compete with the binding of [$^3$H]Ro 15–4513 to α4/6βxδ subtypes, suggesting a novel site with high ethanol binding affinity (Hanchar et al., 2006; Wallner et al., 2006; Santhakumar et al., 2007). At the α+/β3− subunit interface, an extracellular binding pocket was proposed between the α4/6R100 residue and β3Y66, forming the ethanol/Ro 15-4513 binding site (Wallner et al., 2014). However, the increased behavioral sensitivity to ethanol was only observed in α6R100Q in alcohol non-tolerant (ANT) rats (Hanchar et al., 2005), a line that also showed remarkable sensitivity to diazepam (Uusi-Oukari and Korpi, 1990; Korpi et al., 1993).

The claim that δ-GABA$_A$Rs are sensitive to ethanol at low millimolar concentrations was not universally observed and remains a topic of debate (reviewed by Förstera et al., 2016). Subsequent studies were unable to detect the binding of [$^3$H]Ro 15-4513 or ethanol-induced displacement in recombinant α4/6β3δ subtypes (Borghese and Harris, 2007; Korpi et al., 2007). Indirect modulation of native GABA$_A$ receptors by ethanol, possibly through protein kinase phosphorylation activation and enhanced presynaptic GABA release, complicates the precise modeling of ethanol sensitivity (Harris et al., 1995; Weiner et al., 1997; Aguayo et al., 2002; Carta et al., 2004). Several studies have used different cRNA or cDNA ratios of α, β, and δ subunits to express recombinant receptors in Xenopus oocytes and heterologous cell lines such as Human embryonic kidney cells (HEK293), giving rise to variations in receptor expression levels or assembly patterns (Wagoner and Czajkowski, 2010; Botzolakis et al., 2016; Hartiadi et al., 2016). These variations could impact the functional properties of the receptors, influencing their GABA sensitivity and modulation by ethanol and other PAMs. Thus, methodological discrepancies in sample preparation and receptor expression systems could contribute to the diverging findings concerning ethanol.

### 2.2.3    Picrotoxin: From nature to research significance

Picrotoxin (PTX), a naturally occurring plant toxin with an equimolar combination of active picrotoxinin and the considerably less active picrotin, is a universal "reference" blocker of GABA$_A$ receptor chloride channels (Das et al., 2003; Olsen, 2006; Kalueff, 2007). PTX is not structurally similar to GABA but inhibits chloride flux by binding non-competitively to the ionophore site at the N-terminus of TM2 that lines the chloride channel pore (Figure 10A) (Olsen, 2006; Kalueff, 2007). Furthermore, it has been shown to allosterically modulate GABA$_A$ receptors (Xu et al., 1995; Sedelnikova et al., 2006; Erkkila et al., 2008), with a hypothetical allosteric site of PTX proposed at the C-terminus of TM2 (Olsen, 2006; Kalueff, 2007). PTX can interact with GABA-bound receptors as well as resting receptors, with a roughly tenfold higher affinity for the latter (Dillon et al., 1995). Recent high-resolution cryo-EM structures of the human α1β3γ2L GABA$_A$ receptor have elucidated the mode of action of PTX (Masiulis et al., 2019). The TMs were found to assume the same conformation in both PTX-bound and PTX/GABA-bound complexes, with the five TM2 helices pointing their 9′ Leu hydrophobic side chains towards the center of the pore, decreasing its radius to 1.5 Å (Figure 10B, C). This indicated that the inhibition of PTX is not the result of obstructing an open pore but rather the induction and stabilization of a closed pore state. PTX needs to first bind to an open channel pore and then maintain a closed/resting receptor conformation, which explains its channel-blocking and allosteric antagonistic properties.

Targeting the PTX-binding site has historically led to many early discoveries about the actions, behavior, and ligand interactions of GABA receptors (Ticku and Olsen, 1978; Mehta and Ticku, 1986; Carpenter et al., 2013). [$^{35}$S]t-butylbicyclophosphorothionate (TBPS) is a radioligand that acts as a non-competitive blocker of the PTX site (Squires et al., 1983). Due to its high specific activity, [$^{35}$S]TBPS has been employed to label GABA$_A$ receptors in the CNS using autoradiography, proving to be a reliable *in vitro* marker for predicting the efficacy and potency of ligands targeting GABA$_A$ receptors (Squires et al., 1983; Korpi and Lüddens, 1993; Sinkkonen et al., 2001). Ethynylbicycloorthobenzoate (EBOB) is structurally comparable to TBPS but it is radiolabeled with $^3$H rather than $^{35}$S. This characteristic provides [$^3$H]EBOB with greater stability and a longer radioisotopic half-life than [$^{35}$S]TBPS (12 years vs. 87 days) (Squires et al., 1983; Lawrence et al., 1985; van Rijn et al., 1990; Im and Blakeman, 1991; Cole and Casida, 1992). Notably, [$^3$H]EBOB has at least 15 times higher affinity for the picrotoxin site of the vertebrate GABA$_A$ receptor than [$^{35}$S]TBPS, as indicated by its equilibrium dissociation constant ($K_D$ ~2 nM) (Cole and Casida, 1992; Huang and Casida, 1996; Peričić et al., 1998; Yagle et al., 2003). Such radioligand developments would deepen our understanding of GABA$_A$ receptors and enhance precision in detecting ligand interactions across brain regions and receptor systems.

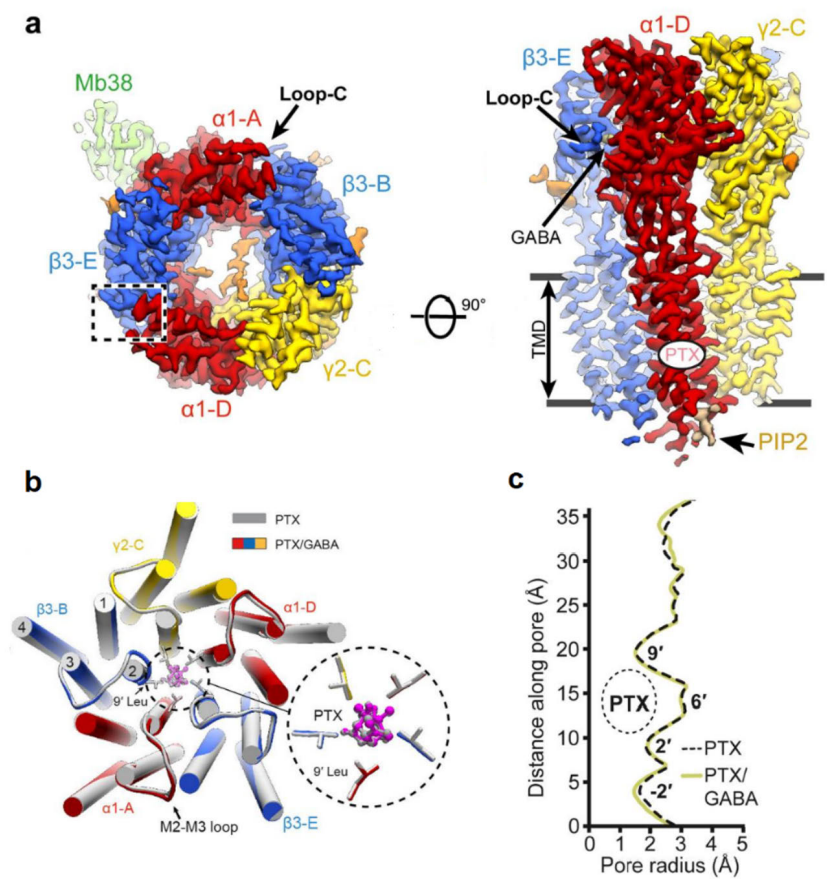

**Figure 10.** Cryo-electron microscopy analysis of the human α1β3γ2L GABA$_A$ receptor structure in the presence of Picrotoxin (PTX) and GABA. (a) Extracellular view (left) of PTX/GABA-bound GABA$_A$ receptor in parallel to the membrane view (right). (b) Magnified extracellular view of PTX- and PTX/GABA-bound GABA$_A$ receptor transmembrane domains superposed globally. A stick represents the side chain of 9' Leu residues, whereas a ball and a stick depict PTX. (c) Pore radius plots for PTX and PTX/GABA bound receptor. Modified with permission from Masiulis et al. (2019) © Springer Nature.

## 2.2.4    Flavonoids: Dietary neuroactive polyphenols

Flavonoids constitute a diverse and widely distributed family of secondary plant metabolites. These polyphenolic compounds are ubiquitous in the human diet and are found in sources such as red wine, beer, green tea, and various herbal remedies (Johnston et al., 2015). Due to their potential therapeutic advantages for human health, flavonoids have attracted significant interest. These include anti-inflammatory (Pan et al., 2010), anticancer (Batra and Sharma, 2013), antioxidant (Heim et al., 2002), sedative-hypnotic (Shanmugasundaram et al., 2018), anxiolytic (Ognibene et al., 2008), analgesic (Ferraz et al., 2020), and neuroprotective properties (Vauzour et al., 2018). Flavonoids exert their effects on the CNS by interacting with various receptors and signaling pathways, including the $GABA_A$ receptors, which are important for flavonoid-induced anxiolytic, sedative-hypnotic, and anti-alcohol intoxication effects (Hanrahan et al., 2011; Shen et al., 2012; Wasowski and Marder, 2012).

### 2.2.4.1    Early flavonoid-$GABA_A$ binding discoveries

The association of flavonoids with $GABA_A$ receptors was first established when isoflavone, isolated from bovine urine at Roche in 1983, was found to displace [³H]diazepam binding to rat brain membranes. S-(–)-Equol emerged as the most potent isoflavone in the study, with the concentration producing half-maximal inhibition ($IC_{50}$) of 80 µM (Luk et al., 1983). Subsequent research demonstrated that several plant-derived flavonoids can influence the binding of [³H]diazepam and other classical benzodiazepine-site ligands, such as [³H]flunitrazepam (Medina et al., 1997; Paladini et al., 1999; Marder and Paladini, 2002; Hanrahan et al., 2011). Presently, various subclasses of natural flavonoids (Figure 11) are known to target $GABA_A$ receptors, such as flavones (e.g., apigenin, luteolin, and hispidulin), isoflavones (e.g., genistein), flavonols (e.g., quercetin), flavanones (e.g., naringenin), flavan-3-ols (e.g., epigallocatechin gallate (EGCG) and catechin), flavanonols (e.g., taxifolin, dihydromyricetin), and chalcones (e.g., isoliquiritigenin) (Cho et al., 2011; Johnston, 2015; Hinton et al., 2017).

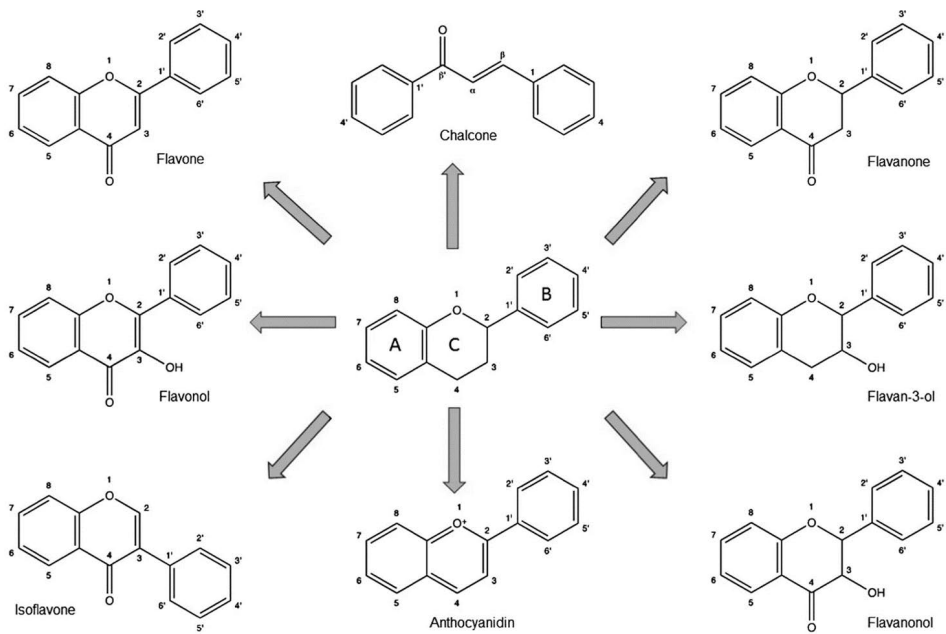

**Figure 11.** Chemical structures of flavonoid subclasses, depicting their basic structure and numerical arrangement Reprinted with permission from Muschietti et al. (2018) © Springer Nature.

### 2.2.4.2    Flumazenil-sensitive modulation

Flavanoids have been shown to exhibit a variety of actions on GABA$_A$ receptors, primarily acting as positive, negative, and silent modulators (reviewed in Hanrahan et al., 2015). Many of these interactions were sensitive to flumazenil antagonism, indicating their binding to the high-affinity classical benzodiazepine site (Campbell et al., 2004). Isoliquiritigenin (2′,4′,4′-trihydroxychalcone; ILG), a compound isolated from the roots of *Glycyrrhiza uralensis* (licorice), demonstrated notable affinity ($K_D$: 0.4 nM), which was 65-fold greater than that of diazepam in dorsal raphe neurons (Cho et al., 2011). In these neurons, ILG was found to enhance GABA-evoked currents in a flumazenil-sensitive manner. In a mouse model of pentobarbital-induced sleep, ILG (25, 50 mg/kg, i.p.) reduced sleep latency and prolonged sleep duration in a dose-dependent fashion. These effects were notably inhibited by flumazenil. Moreover, in the same test, the activity of ILG at a dose of 50 mg/kg proved to be comparable to that of diazepam at 2 mg/kg (Cho et al., 2011). In a subsequent study, ILG (50 mg/kg, i.p.) was observed to induce sleep and increase the amount of NREMS (non-rapid eye movement sleep) in the first 3 hours after treatment (Cho et al., 2012). Separately, a study with rats demonstrated that ILG (15, 25 mg/kg, i.p.) produces anxiolytic effects in the elevated plus-maze test (Jamal et al., 2008). Taken together, these findings suggest that the hypnotic effects of ILG are facilitated through a positive modulation of the GABA$_A$ receptors at the classical benzodiazepine-binding site. Functional studies employing stable or recombinant expression systems would be beneficial to elucidate its selectivity across various GABA$_A$ receptor subtypes and correlate these binding interactions with observed behavioral outcomes.

Dihydromyricetin ((2R,3R)-3,5,7,3',4',5'-hexahydroxyflavanone; DHM) is a flavonoid isolated from *Hovenia dulcis*, an herb used in traditional East Asian medicine to alleviate alcohol intoxication and hangover symptoms (Shen et al., 2012; Kim et al., 2017). DHM has been demonstrated as an effective positive modulator of extrasynaptic and synaptic GABA$_A$ receptors, enhancing both GABA-induced inhibitory tonic and postsynaptic currents in hippocampal neurons. Additionally, DHM inhibited the ethanol-induced enhancement of GABA$_A$ receptor activity and counteracted the ethanol-induced increase in the expression of the α4 subunit in the rat hippocampus. The modulatory effects of DHM, both in the presence and absence of ethanol, were found to be sensitive to flumazenil antagonism. DHM was also found to competitively displace the binding of [$^3$H]flunitrazepam, implying the involvement of benzodiazepine sites in its interactions with GABA$_A$ receptors (Shen et al., 2012; Silva et al., 2020). However, the behavioral effects of DHM appeared distinct from those of classical benzodiazepines. Anxiolysis and sedation were not observed at the dose (1 mg/kg, i.p.) shown to block the effects of ethanol. Instead, DHM alleviated acute alcohol intoxication and withdrawal symptoms, and it also reduced excessive alcohol consumption *in vivo* (Shen et al., 2012; Silva et al., 2020; Carry et al., 2021). This distinction underscores the unique pharmacological profile of DHM, suggesting that it might offer therapeutic advantages without the side effects commonly associated with classical benzodiazepines.

### 2.2.4.3    Flumazenil-insensitive modulation

Flumazenil-insensitive responses have also been observed in electrophysiological recordings and radioligand binding assays, indicating that certain flavonoids can act at sites other than the high-affinity classical benzodiazepine site (reviewed in Hinton et al., 2017). Both 6-methoxyflavanone and 6-methoxyflavone have been shown to positively modulate α1/2β2γ2 GABA$_A$ receptors expressed in Xenopus oocytes in a flumazenil-insensitive manner (Hall et al., 2014). Unlike 6-methoxyflavanone, the modulatory activity of 6-methoxyflavone was not dependent on the γ2 subunit. This is because it had a significant influence on GABA-induced responses in the α1β2 subtype, which contains low-affinity, flumazenil-insensitive benzodiazepine binding sites (Walters et al., 2000; Ramerstorfer et al., 2011; Hall et al., 2014; Maldifassi et al., 2016; Masiulis et al., 2019). Unexpectedly, despite being unresponsive to flumazenil antagonism, 6-methoxyflavanone displaced [$^3$H]flunitrazepam binding in rat brain membranes (Hall et al., 2014). These findings suggest a novel allosteric modulatory site for 6-methoxyflavanone that is dependent on the γ2 subunit but distinct from the known high- and low-affinity benzodiazepine binding sites. Behavioral studies with mice revealed that 6-methoxyflavanone possesses anxiolytic properties, whereas 6-methoxyflavone alleviates cognitive impairment associated with alcohol withdrawal (Akbar et al., 2017; Arif et al., 2022). Given the flumazenil-insensitive actions observed, further screening of flavonoids may uncover a broader spectrum of modulatory effects on GABA$_A$ receptors with notable therapeutic benefits.

#### 2.2.4.4    Flavonoids as allosteric agonists

Flavonoids have been proposed to activate GABA$_A$Rs by interacting with an allosteric agonist site, independent of GABA presence. Electrophysiological recordings conducted using recombinant GABA$_A$Rs have shown that 3-hydroxy-2′-methoxy-6-methylflavone (3-OH-2MeO6MF) can directly and selectively activate α4β1-3δ subtypes as an allosteric agonist (Karim et al., 2011; Falk-Petersen et al., 2017). The effects of 3-OH-2MeO6MF on α4β3δ were found to be non-competitively antagonized by bicuculline, indicating that the binding of the flavonoid occurs at an allosteric site. Interestingly, these extrasynaptic δ-GABA$_A$R subtypes did not show GABA-mediated enhancement when exposed to 3-OH-2MeO6MF. In contrast, the α1/2/4/6βγ2L subtypes exhibited distinct positive modulatory effects in response to 3-OH-2MeO6MF. Notably, the potency of 3-OH-2MeO6MF was observed to be significantly higher (10-fold) for δ-GABA$_A$Rs compared to γ2L-GABA$_A$Rs, which could be attributed to the distinct allosteric effects influenced by the δ- and γ-subunits. In animal studies, the administration of 3-OH-2MeO6MF (1–100 mg/kg, i.p.) to mice demonstrated anxiolytic effects in the elevated plus maze and light dark box tests, while a high dose (100 mg/kg, i.p.) prolonged the duration of sleep induced by barbiturates (Karim et al., 2011).

Subsequent research revealed that 2′-methoxy-6-methylflavone (2MeO6MF), a derivative of 3-OH-2MeO6MF, exhibits direct activation of α2β2/3γ2L and α2β2/3 subtypes. Additionally, it displays positive modulatory effects on α2β1γ2L, α1β1–3γ2L, and α1β2 subtypes (Karim et al., 2012; Chua et al., 2015). In mice, 2MeO6MF induced anxiolysis at low doses (1, 10 mg/kg, i.p.), sedation at intermediate doses (30, 100 mg/kg, i.p.), and sleep promotion at high doses (100 mg/kg, i.p.). These observations suggested a higher efficacy of 2MeO6MF in activating α2-GABA$_A$Rs compared to modulating α1-GABA$_A$Rs. The activation induced by 2MeO6MF was non-competitively antagonized by bicuculline and gabazine. Although 2MeO6MF weakly displaced the binding of [$^3$H]flunitrazepam, it exhibited insensitivity to the antagonism of flumazenil. These findings indicated the presence of novel flavonoid site(s) involved in the subtype-dependent activation and modulation of GABA$_A$R function (Karim et al., 2012; Hinton et al., 2017). Notably, when serine (S) was substituted for asparagine (N) at the 265$^{th}$ position of the β2 subunit, 2MeO6MF exhibited a shift in its effects, positively modulating α2β2N265Sγ2L GABA$_A$Rs at concentrations comparable to those required for activation. This finding suggests the presence of a common binding site mediating the dual actions of 2MeO6MF. Further investigations using site-directed mutagenesis and homology modeling with high-resolution GABA$_A$R structures (Zhu et al., 2018; Masiulis et al., 2019) would facilitate the characterization of the flavonoid binding pockets and their poses at these yet to be revealed flumazenil-insensitive and allosteric agonist sites.

### 2.2.4.5    Prospects for flavonoid research in CNS

As widespread components of the human diet, flavonoids represent a complex array of natural secondary metabolites with the potential to alter brain function and behavior. While individual flavonoids have been shown to interact with GABA$_A$ receptors through modulation, direct activation, or antagonism, significant knowledge gaps persist concerning their effects upon consumption via foods, beverages, and herbal remedies. Moreover, the actions of specific flavonoids exhibit a multifaceted nature that extends beyond the mere modulation of the benzodiazepine binding site. Further research is warranted to characterize the binding sites, allosteric mechanisms, and potential synergies underlying the modulation of specific GABA$_A$ receptor subtypes by different flavonoids and their combinations. Advancing our understanding of how these common dietary compounds impact neuronal excitability may yield important nutritional and pharmacological implications for sleep disorders and other neurological conditions.

## 2.3    GABA$_A$ receptors in insomnia pharmacotherapy

### 2.3.1    Insomnia characteristics and impacts

Insomnia is a prevalent sleep disorder that affects a significant segment of the population. Chronic insomnia affects around 10–15% of the adult population, while 25–35% experience occasional episodes of sleep disturbance (Doghramji, 2006; Cho and Shimizu, 2015). It is characterized by difficulties in initiating and maintaining sleep as well as non-restorative sleep (Mendelson et al., 2004; Roth, 2007). Insomnia is often experienced by individuals as acute episodes without any apparent cause. This is commonly attributed to poor sleep hygiene or a disturbance in the body's natural sleep-wake cycle. However, it is important to note that insomnia can also be indicative of an underlying health condition, a psychiatric/behavioral disorder, or medication use (Buysse, 2006; Leach and Page, 2015). This can have a profound impact on an individual's daily functioning, resulting in irritability, fatigue, low motivation, and cognitive deficits (Buysse et al., 2007; Fortier-Brochu et al., 2012; Morin and Jarrin, 2013). Additionally, chronic insomnia can trigger severe physical and neurological problems, including cardiovascular disease (Javaheri and Redline, 2017) and depression (Li et al., 2016). It can significantly raise mortality rates in otherwise healthy elderly individuals (Foley et al., 2004). Sleep disorders also contribute to increased healthcare costs, decreased productivity and increased accident risk (Kuppermann et al., 1995; Bhat et al., 2008). Treating insomnia requires a multifaceted approach including pharmacological treatments to reduce sleep latency, increase sleep maintenance and improve sleep quality. Optimally, these treatments should also result in normal wakefulness, improved daytime function and minimal risk of dependence (Wafford and Ebert, 2008; Nutt and Stahl, 2010; Perrault et al., 2022).

## 2.3.2    Sleep architecture essentials

Sleep is a complex process that consists of two phases: non-rapid eye movement (NREM) and rapid eye movement (REM). Each phase exhibits specific electroencephalogram (EEG) signatures and varying levels of arousal (Brown et al., 2012). NREM sleep comprises four distinct stages, with stage 1 representing the transitional state between wakefulness and sleep. This stage is distinguished by slow rolling eye movements and a low-amplitude, mixed-frequency EEG (2-7 Hz). In Stage 2, spindle activity at a frequency range of 12-14 Hz becomes prominent. Stages 3 and 4 are defined by high amplitudes of slow wave activity (SWA), also known as delta waves, occurring at lower frequencies (0.5-4 Hz) (Winsky-Sommerer, 2009; Voss, 2010; Fiorillo et al., 2019). The amplitude of delta waves during NREM sleep is a crucial metric for assessing sleep depth, intensity, and restfulness (Borbély et al., 1981). On the other hand, REM sleep is characterized by theta waves (4-8 Hz) and is associated with dreaming and processing emotional memory (Nishida et al., 2009; Fiorillo et al., 2019). While a state of restfulness is achieved during this phase, the threshold for awakening is lower compared to slow-wave sleep (Bourne and Mills, 2004). Healthy people alternate between NREM and REM sleep cycles, collectively referred to as sleep architecture, with each cycle typically lasting 90 - 120 minutes (Sinton and McCarley, 2004; Markov and Goldman, 2014). In the pursuit of an optimal sleep aid, it becomes imperative to focus on methods that can extend NREM sleep and enhance delta wave activity without causing disruptions to the overall sleep architecture (Winsky-Sommerer, 2009). Striking this balance is crucial to promoting restorative and healthy sleep patterns.

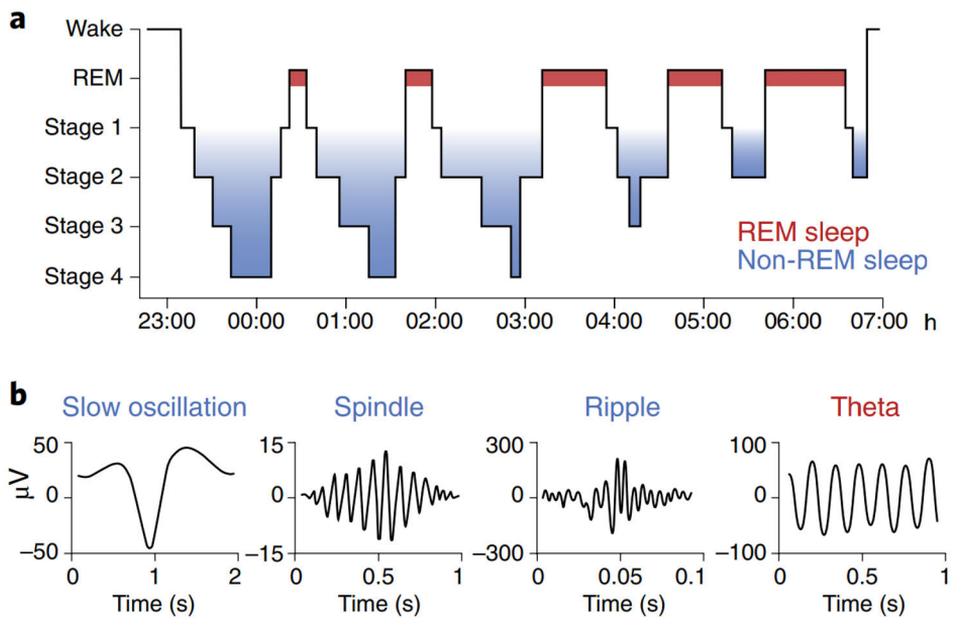

**Figure 12** Sleep architecture and electroencephalogram (EEG) rhythms. (a) Nocturnal sleep pattern in humans with different sleep stages. (b) Electrophysiological markers of Non-Rapid Eye Movement (NREM) sleep, including neocortical slow waves, thalamocortical spindle activity and hippocampal sharp wave-ripples. Additionally, hippocampal theta band oscillations are notably observed in rodents during the Rapid Eye Movement (REM) phase of sleep. Reprinted with permission from Klinzing et al. (2019) © Springer Nature.

## 2.3.3    GABA$_A$ receptors in sleep regulation

GABAergic neurons encompassing the thalamus make up the thalamic reticular nucleus (TRN) in the brain. This nucleus contributes to synchronizing the neurons within the thalamocortical circuit, producing cortical delta oscillations with low frequency and high amplitude, as well as cortical spindles in NREM sleep (Brown et al., 2012; Lewis et al., 2015). Experiments on mice lacking the β3 subunit of the GABA$_A$ receptor (Gabrb3−/−) revealed that deletion of the β3 subunit localized in the TRN suppresses spindle activity and increases delta EEG power during NREM sleep (Wisor et al., 2002). Remarkably, similar alterations in NREM sleep have been observed in both humans and rodents as a result of sleep deprivation (Dijk et al., 1993; Schwierin et al., 1999) and have been replicated using the GABA$_A$ receptor agonists gaboxadol and muscimol (Lancel et al., 1996; Faulhaber et al., 1997). Collectively, these findings highlight the important role of GABAergic transmission mediated by GABA$_A$ receptors in maintaining sleep homeostasis.

The ventrolateral preoptic nucleus (VLPO) in the anterior region of the hypothalamus, predominantly composed of inhibitory GABAergic neurons, is one of the key centers involved in sleep promotion (Sherin et al., 1996; Saper et al., 2005). As evidenced by extracellular electrophysiological recording in rats, VLPO neurons exhibit a twofold increase in firing rate during sleep in comparison to wakefulness (Koyama and Hayaishi, 1994). Moreover, cats and rats with VLPO lesions experienced more frequent transitions between sleep and alertness, along with a significant decline in sleep duration and severe insomnia (Nauta, 1946; McGinty and Sterman, 1968; Lu et al., 2000). In humans, loss of VLPO cells during aging causes disruptions in the sleep-wake cycle, which in turn leads to sleep fragmentation and excessive daytime napping, two common complaints among the elderly (Lim et al., 2014; Saper, 2021).

Researchers examined the influence of various $GABA_A$ receptor ligands on the VLPO by analyzing the expression of c-Fos protein as an indicator of neuronal activity (Nelson et al., 2002; Lu et al., 2008). The administration of muscimol, pentobarbital, and ethanol in rats induced sedation and increased delta wave activity in NREM sleep. Consistently, the ligands increased c-Fos expression in the VLPO neurons, but the proportion of c-Fos-positive neurons in the VLPO was around one-third of what was detected during natural sleep (Lu et al., 2008).

The VLPO neurons co-express both GABA and galanin, which inhibit the arousal nuclei such as the locus coeruleus (LC) and dorsal raphe nucleus (DRN) of the brainstem as well as the tuberomammillary nucleus (TMN) of the hypothalamus, leading to sleep induction and maintenance (Sherin et al., 1996; Saper et al., 2005). The release of GABA from sleep-on neurons in the VLPO was shown to inhibit wake-promoting histaminergic neurons in the TMN region by activating β1-$GABA_A$Rs (Sergeeva et al., 2010; Yanovsky et al., 2012). Conversely, the VLPO neurons are inhibited by norepinephrine, serotonin, histamine, orexin released from LC, DRN, TMN, and hypothalamic orexin neurons, respectively, leading to a state of wakefulness (Gallopin et al., 2000; Gallopin et al., 2004; De Luca et al., 2022). Whereas the inhibition of TMN neurons by the VLPO is direct and GABA-mediated, histamine release from the TMN triggers an indirect inhibitory effect on the VLPO through GABAergic interneurons (Liu et al., 2010; Williams et al., 2014). The rapid transitions between sleep and wakefulness are suggested to be triggered like a "flip-flop switch", which is stabilized by orexin neurons and mediated by the mutual inhibitory interactions between the arousal and sleep centers in the brain (Saper et al., 2010; Nutt and Sahl 2010).

### 2.3.4 The pursuit of safe and novel GABA$_A$ receptor modulators for insomnia treatment

The treatment of insomnia is typically tailored to the individual's specific case, with various approaches available. While some sufferers may turn to traditional methods like reading, consuming alcohol, or using over-the-counter remedies, prescription drugs may be a viable option for those seeking medical assistance (Ohayon, 2002; Leach and Page, 2015). Among the primary mechanisms of the Food and Drug Administration (FDA)-authorized drug interventions for insomnia is the positive modulation of GABA$_A$ receptors. Other mechanisms include the agonism of melatonin (MT1, MT2), antagonism of histamine (H1), and antagonism of orexin (OX1, OX2) receptors (Avidan and Neubauer, 2017; Bruni et al., 2021). Positive allosteric modulators (PAMs) that enhance the inhibitory GABAergic tone may elicit significant changes in the circuit activity, which can shift the brain from a state of alertness to sleep or sedation, as well as general anesthesia (Brickley et al., 2018).

In the early 1900s, barbiturates were introduced as the first class of sedatives and hypnotics targeting the GABA$_A$Rs. They gained widespread use initially but were later abandoned for treating insomnia as a result of their high toxicity, as well as the emergence of tolerance and dependence (López-Muñoz et al., 2005). Subsequently, benzodiazepines, including diazepam, flunitrazepam, flurazepam, triazolam, and midazolam, were launched in the 1960s. These medications quickly became popular for their remarkable anxiolytic and sedative/hypnotic effects, along with a reduced toxicity profile compared to barbiturates (Visser et al., 2003; Mendelson, 2005). Benzodiazepines consistently proved effective in reducing sleep initiation time and promoting sleep maintenance (Lancel, 1999). However, they tend to decrease the amount of REM sleep while prolonging the time it takes to enter the REM sleep phase (Winsky-Sommerer, 2009). Benzodiazepines also have additional effects, such as next-day drowsiness, muscular relaxation, cognitive impairment, and amnesia. Moreover, they can induce tolerance, dependency, and withdrawal symptoms with prolonged use, along with rebound effects upon cessation (Reviewed in Korpi et al., 1997).

The necessity for hypnotics with a more selective action and fewer safety drawbacks prompted the development of novel therapeutic agents in the 1980s, known as Z-drugs. These include zolpidem (an imidazopyridine), zopiclone, and eszopiclone (cyclopyrrolones), and zaleplon (a pyrazolopyrimidine). Z-drugs have distinct chemical structures from benzodiazepines and show higher selectivity for α1-GABA$_A$Rs. However, they share a common binding site and their selectivity is not region-specific, as α1-GABA$_A$Rs are widely expressed in the CNS, constituting 60% of the GABA$_A$R population in the brain (Pirker et al., 2000). Therefore, Z-drugs and benzodiazepines exhibit similar hypnotic effects, reducing sleep initiation, increasing NREM sleep, and decreasing REM sleep (Winsky-Sommerer, 2009).

While Z-drugs are often claimed to have advantages and fewer adverse reactions compared to benzodiazepines, there is a lack of substantial data supporting notable disparities in terms of clinical efficacy and safety (Dündar et al., 2004; Alanis et al., 2020). Similar to benzodiazepines, Z-drugs are associated with adverse effects, including dizziness, altered cognition, amnesia, and the potential development of dependence and withdrawal (Gunja et al., 2013; Schifano et al., 2014; Capiau et al., 2023). Recognizing the potential risks associated with Z-drugs, the FDA added boxed warnings to prescription labeling and patient medication guidelines in 2019, highlighting the likelihood of complex sleep behaviors such as sleepwalking and sleep driving (Rosenberg et al., 2021).

Despite the safety drawbacks, the number of sleep medicine prescriptions grew by 293% between 1999 and 2010. (Ford et al., 2014). The proportion of doctor's visits that resulted in a prescription for Z-drugs (zopiclone, zolpidem, or zaleplon) (350%), benzodiazepines (430%), or any sleep medication (200%) showed significant increases (Ford et al., 2014; Atkin et al., 2018). In light of this trend, it becomes crucial to develop novel medications that offer safer therapeutic alternatives for both acute and chronic insomnia. These medications should exhibit high effectiveness, have a low potential for addiction, and a reduced incidence of paradoxical effects (Alanis et al., 2020).

## 2.3.5 The role of synaptic GABA$_A$ receptors in hypnotic mechanisms

Classical and non-classical benzodiazepines continue to be widely prescribed for the treatment of insomnia by targeting synaptic αβγ-GABA$_A$Rs (Wisden et al., 2019). Numerous studies have been undertaken to identify the specific GABA$_A$ receptor subunits that mediate the hypnotic effects of this class of drugs. Since the sedative effects of diazepam are mediated by the α1-GABA$_A$Rs (Rudolph et al., 1999), it was initially believed that these subtypes would also be primarily responsible for the hypnotic properties of benzodiazepines. However, experiments with knock-in mice bearing mutated diazepam-insensitive α1 subunit (H101R) have challenged this assumption, revealing that the α1 subunit is not essential for benzodiazepine-induced sleep (Tobler et al., 2001).

In both α1(H101R) and wild-type mice, diazepam at 3 mg/kg reduced initial REM sleep and suppressed delta wave activity in NREM sleep. However, the sleep continuity enhancement effects of diazepam were observed exclusively in α1(H101R) mice. In a subsequent study using α1 knockout mice, the administration of diazepam at a higher dose (33 mg/kg) resulted in a loss of the righting reflex, an indicator of sleep. This effect lasted 57% longer in mutant mice than in wild-type mice (Kralic et al., 2002). These findings suggested the involvement of other

GABA$_A$ receptor subtypes in the effects of benzodiazepines on sleep. Subsequent studies using knock-in mice with diazepam-insensitive subtypes showed a significant decrease in diazepam-induced alterations to REM and NREM sleep in mice with α2(H101R) mutations. These results underscore the role of α2 GABA$_A$ receptor subtypes in mediating the benzodiazepine-induced hypnotic effects and their EEG patterns (Kopp et al., 2004; Rudolph and Möhler, 2004).

While classical benzodiazepines show comparable high affinity for all benzodiazepine-sensitive GABA$_A$ receptor subtypes (α1, α2, α3, and α5), zolpidem, a hypnotic Z-drug, has a greater affinity for α1- than for α2- or α3-GABA$_A$Rs and lacks affinity for α5-GABA$_A$Rs (Dämgen and Lüddens, 1999). The selectivity of zolpidem is also influenced by the γ subunits, with the drug showing a decreased affinity for γ1- and γ3- in comparison to γ2-GABA$_A$Rs (Richter et al., 2020; Zhu et al., 2022). The sedative effects of zolpidem align with those mediated by α1-GABA$_A$Rs, as similarly established for diazepam (Rudolph et al., 1999). This correlation was evidenced by the zolpidem-induced decrease in motor activity in knock-in α1(H101R) mice relative to their wild-type counterparts (Crestani et al., 2000). The hypnotic effects of zolpidem are mediated by α1-GABA$_A$Rs, as indicated by a shorter duration of the loss of righting reflex in α1-knockout mice compared to wild-type mice (Kralic et al., 2002; Blednov et al., 2003). This observation is consistent with the absence of [$^3$H]zolpidem high-affinity binding in cortical membranes and the loss of zolpidem's ability to displace [$^3$H]flumazenil high-affinity binding in α1-deficient mice (Kralic et al., 2002; Sur et al., 2001). Interestingly, a compensatory upregulation of α2 and α3 subunit peptides in the cerebral cortex of α1-deficient mice was noted, corresponding to the increase in the low-affinity [$^3$H]flunitrazepam binding sites for zolpidem (Kralic et al., 2002; Blednov et al., 2003). Further research into the hypnotic effects of zolpidem showed that α1(H101R) knock-in mice did not decrease delta wave activity in NREM sleep, a trend prominently observed in wild-type mice (Kopp et al., 2004). However, both groups of mice exhibited enhanced NREM sleep and reduced REM sleep following drug administration. Since zolpidem's hypnotic effects were not completely abolished by the deletion and mutation of the α1 subunit, it is plausible that α2- and/or α3- GABA$_A$Rs might partly be involved in its action.

Benzodiazepines and Z-drugs effectively induce and maintain sleep, but they also reduce delta wave power in NREM sleep, limiting sleep depth and overall restfulness (Alexandre et al., 2008; Uygun et al., 2016). The reduction in delta wave power may stem from the enhanced activity of α3-GABA$_A$Rs, predominantly expressed in the sleep-regulating region of the TRN (Waldvogel et al., 2017; Sperk et al., 2020). A recent study employed CRISPR-Cas9 gene editing to knockdown α3-GABA$_A$Rs in the adult mouse brain, particularly in the parvalbumin neurons of the TRN. This targeted intervention promoted deep sleep by amplifying NREM thalamocortical delta oscillations (Uygun et al., 2022). Additionally, the inhibitory synaptic currents were significantly decreased in the affected parvalbumin neurons. Therefore, the negative modulation of α3-GABA$_A$Rs offers a potential mechanism for improving sleep quality by promoting natural, restorative delta waves, a feat not achievable with current hypnotic medications.

## 2.3.6 The potential of extrasynaptic δ-GABA$_A$ receptors in modulating sleep patterns

The modulation of tonic GABAergic inhibition, mediated by extrasynaptic δ-GABA$_A$Rs, holds significant potential in the targeted and sustained management of medical conditions associated with neuronal excitability (Ghit et al., 2021). These receptors bind GABA with high affinity, enabling their activation by low levels of GABA in the extracellular space, and they are insensitive to classical benzodiazepines (Saxena and Macdonald, 1996). While certain ligands, including barbiturates, neurosteroids, and etomidate, significantly potentiate the activity of δ-GABA$_A$Rs (Belelli et al., 2005), they lack substantial selectivity for receptors containing the δ subunit over those containing the γ subunit (Jensen et al., 2013). A notable exception to this is gaboxadol, a sedative-hypnotic compound acting as a selective agonist of the δ-GABA$_A$Rs. In studies involving recombinant GABA$_A$Rs expressed in Xenopus oocytes, gaboxadol exhibited low-potency agonism, partial agonism, and super-agonism at the receptor subtypes α1β3γ2, α4β3γ2, and α4/6β3δ, respectively (Brown et al., 2002; Stórustovu and Ebert, 2006).

It is worth noting that the expression of α4β3δ GABA$_A$R is prominent in brain regions associated with the regulation of sleep, particularly the neocortex and the thalamic ventrobasal complex (VB) (Pirker et al., 2000; Peng et al., 2002). This specific receptor subtype exhibits high sensitivity to gaboxadol, leading to enhanced tonic currents in thalamic relay neurons (Faulhaber et al., 1997; Cope et al., 2005; Winsky-Sommerer et al., 2007). These currents contribute to the production of cortical delta activity that is characteristic of NREM sleep (Brickley, 2018). Multiple studies involving human participants and gaboxadol administration have observed a reduction in alertness upon sleep initiation as well as a prolongation in sleep duration, with no significant impact on the REM stage. During NREM sleep, gaboxadol resulted in an augmentation of EEG power, particularly in the delta wave frequency range, while specifically reducing spindle oscillations (Faulhaber et al., 1997; Lancel et al., 2001; Mathias et al., 2001; Walsh et al., 2007).

Despite promising results, the clinical development of gaboxadol for insomnia management was discontinued in 2007 due to the emergence of unforeseen adverse events, such as hallucinations and confusion, observed during phase III trials (Rudolph and Knoflach, 2012). Nonetheless, gaboxadol continues to serve as a valuable lead for the development of insomnia-targeting drugs and the exploration of δ-GABA$_A$R functions (Winsky-Sommerer, 2009). The region-specific localization of the extrasynaptic receptors implies that δ-selective ligands, in contrast to benzodiazepines and Z-drugs, may exert a more precise influence on sleep (Harrison, 2007). However, there remains a scarcity of allosteric modulators that specifically interact with extrasynaptic δ-GABA$_A$Rs. Further screening and development of selective novel compounds are beneficial to fully exploit the potential of these receptors in the treatment of insomnia.

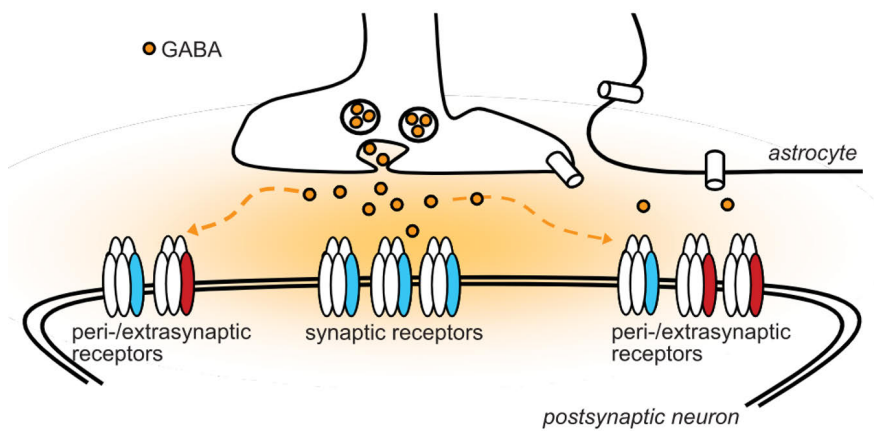

**Figure 13.** Summary of how sedative-hypnotic compounds target different GABA$_A$ receptors and their impact on sleep and related outcomes. The blue color represents the γ2 subtype, and the red color represents the δ subtype. The frequency is denoted by "Freq". Reprinted with permission from Winsky-Sommerer (2009) © John Wiley and Sons.

## 2.4 Herbal remedies targeting GABA$_A$ receptors for sleep

Insomnia patients may seek alternative remedies for several reasons, including concerns about the effectiveness and safety of the current medications and interest in testing novel treatments. Dissatisfaction with modern medicine may also lead some sufferers to explore complementary and alternative options, while others are motivated by a holistic approach to healthcare, active involvement in healthcare decisions, and a rise in health awareness (Leach and Page, 2015).

Over the past few decades, herbal remedies have gained popularity as an alternative to conventional hypnotic drugs for managing insomnia due to their effectiveness and low potential for adverse effects (Gyllenhaal et al., 2000; Meletis and Zabriskie, 2008; Chen et al., 2009). Various surveys indicate that approximately two-thirds of individuals with sleep disturbances search online for treatment recommendations, with medicinal herbs remaining a popular choice (Frass et al., 2012; Welz et al., 2018; Zhang, 2019; Bruni et al., 2021). This trend is unsurprising, given the longstanding tradition of employing medicinal herbs for their potential to promote relaxation and enhance sleep. Several herbs have received recognition from the German Commission E, a globally renowned scientific advisory board that has published numerous monographs on the safety and efficacy of herbal remedies (Blumenthal, 1998). Herbs recommended for sleep disturbances associated with anxiety or restlessness include *Valeriana officinalis* (valerian root), *Humulus lupulus* (hop cone), *Melissa officinalis* (lemon balm leaf), and *Lavandula angustifolia* (lavender flower). Other herbs, such as *Passiflora incarnata* (passion flower), *Piper methysticum* (kava root), and *Matricaria recutita* (chamomile flower), are known for their sedative properties, but the Commission E. has not specifically endorsed them for the indication of sleep (Taibi and Landis, 2009).

While clinical studies have examined the efficacy of herbal remedies in reducing sleep onset latency and improving sleep duration and quality, conclusive data supporting their use in treating insomnia remain limited, indicating the need for further research (Leach et al., 2015; Romero et al., 2017; Bruni et al., 2021). Despite this, hops and valerian are among the notable remedies that have shown promising results in sleep and related parameters. A recent meta-analysis of numerous randomized controlled trials (Shinjyo et al., 2020) found that this combination led to improvements in sleep latency, depth, duration, and quality (Lindahl and Lindwall, 1989; Morin et al., 2005; Koetter et al., 2007; Dimpfel and Suter, 2008). Other human studies reported increased EEG delta power (Vonderheid-Guth et al., 2000) and reduced caffeine-induced arousal (Schellenberg et al., 2004).

Traditional herbal remedies are complex mixtures of multiple active components. Consequently, it is likely that the pharmacological effects of these remedies result from the collective interactions among various compounds (Johnston et al., 2009). The composition of plant extracts used as herbal supplements is not always standardized, and the dosage depends on the whole quantity of the extract. The proportion of their active components, as determined, can be influenced by factors such as cultivation techniques, species differences, seasonal changes, and extraction methods (Wheatley, 2005). Nevertheless, plants and their constituents provide a vast and varied supply of therapeutic medicines for humans, presenting a compelling incentive to explore novel and effective molecules in herbal extracts (Shi et al., 2021a). Therefore, recent years have seen a surge in studies aiming to analyze the composition of herbal remedies to identify the active chemicals responsible for their sleep-promoting effects and elucidate their mechanism of action. Various mechanisms of action have been proposed, including modulation of GABAergic neurotransmission, which is typically achieved through interactions with GABA$_A$ receptors (Shi et al., 2014; Cicek, 2018; Bruni et al., 2021).

## 2.4.1    Chamomile and passionflower: Flavonoid apigenin in GABA$_A$ receptor modulation

Chamomile and passionflower are frequently included in herbal preparations aimed at promoting better sleep, and both of these plants are recognized as natural sources of apigenin, a flavonoid that has been extensively studied for its ability to interact with GABA$_A$ receptors, owing to its sedative and anxiolytic effects (Johnston et al., 2009; Gazola et al., 2015). An initial report indicated that apigenin competitively displaced [$^3$H]flunitrazepam binding ($K_i$: 4 μM) to bovine brain membranes without affecting [$^3$H]muscimol binding to GABA$_A$ receptors (Viola et al., 1995). Further research demonstrated that apigenin weakly displaced [$^3$H]flumazenil binding ($K_i$: 9 μM) to rat cerebellar membranes and inhibited GABA-evoked currents in rat cerebellar granule cells (Avallone et al., 2000). Apigenin also inhibited GABA-evoked currents in recombinant α1β1/2γ2 GABA$_A$ receptors, and this effect was insensitive to flumazenil (Goutman et al., 2003; Campbell et al., 2004; Kavvadias et al., 2004). In rodents, apigenin demonstrated a dose-dependent response, acting as an anxiolytic at low doses (mice: 3 mg/kg, i.p.; rats: 5 mg/kg, p.o.) and as a sedative at high doses (mice: 30, 100 mg/kg, i.p.; rats: 25, 50 mg/kg, i.p.) (Viola et al., 1995; Avallone et al., 2000). At a dose of 1 mg/kg (i.p.), apigenin did not exhibit any sleep-promoting effects in the thiopental-induced sleep model with mice. However, when combined with hesperidin (2 mg/kg, i.p.), a glycosylated flavonone derived from *Valeriana* species, it resulted in a significant increase in sleep time, indicating a synergistic interaction between the two compounds (Fernández et al., 2005).

Although the negative modulation of GABA$_A$ receptors by apigenin may seem counterintuitive in relation to its *in vivo* effects, other factors may influence its actions, such as dosage and potential interactions with endogenous ligands. For instance, apigenin was shown to enhance diazepam's positive modulation in $\alpha1\beta1\gamma2$ receptors, whereas this effect was absent in allopregnanolone and pentobarbitone modulation (Campbell et al., 2004). These results were obtained at concentrations lower than those required for apigenin to act solely as a modulator of GABA-induced responses. This second-order modulation of first-order modulation by benzodiazepines suggests that endogenous benzodiazepine ligands, such as endozepines, may be involved in the sedative and anxiolytic effects of apigenin (Johnston et al., 2009; Christian et al., 2013). However, apigenin's behavioral effects may not be solely attributed to its modulation of GABA$_A$ receptors, as other mechanisms have been proposed such as inhibition of N-methyl-D-aspartate (NMDA) receptors and calcium-dependent glutamate release (Losi et al., 2004; Chang et al., 2015). Therefore, the complex and multifaceted interactions of certain compounds in herbal remedies, including apigenin, highlight the need for further research to fully elucidate their mechanisms and therapeutic potential for mood and sleep disturbances.

## 2.4.2 Valerian root: Bioactive compounds and GABA$_A$ receptor interactions

Valerian root is primarily associated with the modulation of GABA$_A$ receptors and is a popular herbal remedy used to shorten sleep latency and improve sleep quality (Shi et al., 2014). Valerian root extract has been shown to inhibit the uptake and enhance the release of [$^3$H]GABA from brain synaptosomes and hippocampal slices in rodents (Santos et al., 1994; Ortiz et al., 1999). Further studies with radioligand binding assays have shown that valerian extract displaces [$^3$H]muscimol binding from rat cortical membranes (Cavadas et al., 1995). Initially, this effect was attributed to GABA naturally present in the extract. However, this could not account for valerian's *in vivo* effects on anxiety and sleep as the bioavailability of exogenous GABA is limited, and its capability to traverse the blood-brain barrier (BBB) remains questionable (Boonstra et al., 2015). Subsequently, GABA-free valerian extract was found to influence the binding of [$^3$H]flunitrazepam to rat cortical membranes, with a stimulating effect at low concentrations (EC$_{50}$: 4.13 x 10-10 mg/mL) and an inhibitory effect at high concentrations (IC$_{50}$ of 4.82 x 10-1 mg/mL) (Ortis et al., 1999). These dose-dependent results suggest the presence of two or more active constituents in valerian that differentially bind to GABA$_A$ receptors, prompting further research to identify the specific compounds responsible for these observed effects.

## Flavonoid: 6-Methylapigenin

Valerian extracts are composed of various constituents that were found to modulate GABA$_A$ receptor activity, including flavonoids and terpenoids (Manayi et al., 2016). One of these constituents is 6-methylapigenin (5,7,4'-trihydroxy-6-methylflavone), an apigenin derivative that is present in *Valeriana officinalis* and *Valeriana jatamansi* (previously classified as *Valeriana wallichii*). This flavonoid was reported to displace [$^3$H]flunitrazepam binding with a $K_i$-value of 495 nM, indicating its ability to modulate GABA$_A$ receptors at the benzodiazepine binding sites, and with a higher potency than apigenin (Wasowski et al., 2002). 6-methylapigenin showed anxiolytic properties with the elevated plus maze test in mice, where it increased the open arm exploration and duration at 1 mg/kg (Marder et al., 2003). Furthermore, in the thiopental-induced sleep model, a combination of 6-methylapigenin at 1 mg/kg and hesperidin at 2 mg/kg resulted in a synergistic prolongation of sleep duration, similar to the effect reported for apigenin (Marder et al., 2003; Fernández et al., 2005).

## Sesquiterpenoid: Valerenic acid

Valerian extract and one of its major constituents valerenic acid have been shown to inhibit the firing rate of rat brainstem neurons, with this effect being antagonized by bicuculline (Yuan et al., 2004). These findings suggest that valerenic acid may be a crucial component in the modulation of GABAergic function observed with valerian extracts. As a result, numerous studies have investigated the mechanism of action, subtype selectivity, and *in vivo* effects of this sesquiterpenoid compound. Valerenic acid has been reported to enhance GABA-evoked currents in cultured hippocampal neurons at 10 μM and recombinant GABA$_A$ receptors at 1-30 μM, indicating a positive modulatory action (Khom et al., 2007; Benke et al., 2009). Reported plasma levels of valerenic acid in humans were in a range that corresponds to its minimum effective concentration (1 μM) (Anderson et al., 2005). Valerenic acid's modulatory action was subunit-specific and affected β2/3-GABA$_A$Rs (β3 > β2) while having no significant influence on β1-GABA$_A$Rs (Khom et al., 2007; Khom et al., 2010). The enhancement of GABA-evoked responses by valerenic acid affected GABA$_A$Rs that are typically responsive to classical benzodiazepines, including α1,2,3,5-subunits. However, this modulation was distinct from classical benzodiazepines in that it was not dependent on the presence of the γ2-subunit, was insensitive to flumazenil antagonism, and influenced α4β2γ2 receptors. Additionally, at higher concentrations (≥30 μM), valerenic acid exhibited GABA-mimetic properties, directly activating α1β2γ2S GABA$_A$ receptors (Khom et al., 2007; Benke et al., 2009). These findings suggest that valerenic acid's observed effects on GABA$_A$ receptors do not occur through the classical benzodiazepine-binding site.

Radioligand binding assays confirmed the presence of two distinct binding sites for [$^3$H]valerenic acid in rat brain membranes, with one site exhibiting high affinity ($K_D$ = 25 ± 20 nM) and the other exhibiting low affinity ($K_D$ = 16 ± 10 µM). A published study utilized molecular docking and pharmacophore modeling to uncover the interacting amino acid residues with valerenic acid at GABA$_A$ receptors (Luger et al., 2015). This resulted in the identification of a binding pocket for valerenic acid located at the transmembrane (TM2) β+/α− interface, which included the β3N265 residue. Additional experiments involving mutational analysis have identified several key amino acid residues that contribute to the valerenic acid binding pocket. These residues, including β3N265, β3F289, β3M286, and β3R269, are located at or near the predicted binding site(s) for propofol and etomidate. However, it is worth noting that the study only examined specific homologous sites at the GABA$_A$ receptor interfaces, and as such, additional binding pockets for valerenic acid may exist.

Valerenic acid has been shown to possess potent anxiolytic activity in mice, as evidenced by elevated plus maze (10 mg/kg p.o.) and light/dark avoidance tests (1 - 6 mg/kg i.p. and 10 mg/kg p.o.) (Benke et al., 2009; Khom et al., 2010; Hintersteiner et al., 2014). Conversely, in mice with the β3 (N265M) point mutation, the anxiolytic effect of valerenic acid was not observed, indicating that β3-GABA$_A$Rs play a significant role in its anxiolytic properties (Benke et al., 2009). With regard to sedation and sleep-promoting effects, valerenic acid at a dose of 50 mg/kg (i.p.) was found to decrease spontaneous locomotor activity and prolong pentobarbital-induced sleep in mice (Hendriks et al., 1985). However, another study reported that the effect of valerenic acid on locomotion was not observed at doses up to 30 mg/kg (i.p.) (Khom et al., 2016). Furthermore, valerenic acid at doses up to 15 mg/kg (i.p.) did not show any effect in the thiopental-induced sleep model in mice. Nevertheless, when combined with linarin (4 mg/kg, i.p.), a glycosylated flavone isolated from *Valeriana* species, valerenic acid (5 mg/kg, i.p.) significantly increased sleep time (Fernández et al., 2004). These findings suggest that the *in vivo* effects of valerenic acid are dose-dependent and can be enhanced by other bioactive compounds present in valerian. Further studies are necessary to determine the doses and combinations that produce optimal and safe therapeutic outcomes in humans.

**Monoterpenoid: (+)-Borneol**

(+)-Borneol is a monoterpenoid alcohol that is abundant in *Valeriana officinalis* and is also found in the essential oils of chamomile and lavender (Granger et al., 2005; Johnston et al., 2009). This compound was shown to positively modulate α1β2γ2L GABA$_A$ receptors in a flumazenil-insensitive manner. Despite its low potency (EC$_{50}$: 250 µM), (+)-borneol demonstrated exceptionally high efficacy, as it potentiated

GABA (10 μM)-induced currents tenfold at 450 μM (Granger et al., 2005). Recently, a study demonstrated that borneol essential oil extracted from the camphor tree (*Cinnamomum camphora*) decreased the latency and increased the duration of sleep induced by pentobarbital in mice. Furthermore, the essential oil decreased locomotor activity in the open field test (Xiao et al., 2022). However, it should be noted that the sedative-hypnotic effects observed cannot be solely attributed to (+)-borneol, as the essential oil contained other compounds, such as limonene, β-caryophyllene, linalool, and α-pinene, at lower concentrations.

## 2.5    Hops: From brewing roots to neuropharmacological potential

### 2.5.1    Historical significance

*Humulus lupulus* L., a dioecious vine in the Cannabaceae family, is primarily cultivated for its female inflorescences, commonly known as hop cones or hops (Figure 14A). These cones have diverse applications in the beer brewing industry and are also valued for their medicinal properties (Jacquin et al., 2022). The presence of three species from the *Humulus* genus in China: *Humulus lupulus*, *Humulus yunnanensis*, and *Humulus japonicas*, suggests that ancestral hop species might have originated in Asia before spreading to Europe and North America (Murakami et al., 2006; Carbone and Gervasi 2022). Some theories suggest that the earliest Finnish settlers in Europe might have introduced hops from the East (Wilson, 1975; DeLyser and Kasper 1994). This perspective is supported, to an extent, by references to hopped beer in the Finnish national epic poem, Kalevala, and certain linguistic records, though current evidence is inconclusive (Korpelainen and Pietiläinen, 2021). One of the earliest records, by German monks from 736 AD, indicates the use of hops in brewing, as noted in a Bavarian monastic text (Hornsey, 2003). Yet, the widespread incorporation of hops in beer production wasn't prevalent until the 12th century (Burgess,1964; Turner et al., 2011). Over time, hops have been a vital component in defining the taste and scent of beer, offering a unique and enduring bitterness. Besides influencing the sensory qualities of beer, hops also serve as an effective antimicrobial agent, which extends the beverage's shelf life (Adamenko and Kawa-Rygielska 2022).

Historically, hops have been acknowledged for their medicinal benefits across various cultures. Native American tribes used hops to alleviate sleeplessness and pain (Hamel and Chiltoskey, 1975), while Indian Ayurveda and traditional Chinese medicine prescribed hops for ailments like stomach discomfort/cramps, loss of appetite, insomnia, and restlessness (Koetter and Biendl, 2010). Hops have also been associated with health advantages such as reducing fever, purifying blood, regulating bile secretion, and exhibiting anti-inflammatory effects. Such properties were noted in the 11th-century writings of the Arabic physician Mesue, believed to be among the earliest mentions of hops' medicinal application (Karabín et al., 2016). Avicenna's "The Canon of Medicine", published in 1025 AD, lists hops as an herbal remedy, and the 13th-century Arab botanist Ibn Al-Baytar highlighted its digestive and sedative properties (Koetter and Biendl, 2010). The consistent use of hops across diverse cultures, despite limited information exchange, provides compelling indirect evidence of their efficacy in medical applications (Carbone and Gervasi 2022). Today, the European Medicines Agency (2016) officially recognizes the longstanding use of hops as a sleep aid and mild reliever of mental stress, emphasizing their plausible effectiveness based on this tradition.

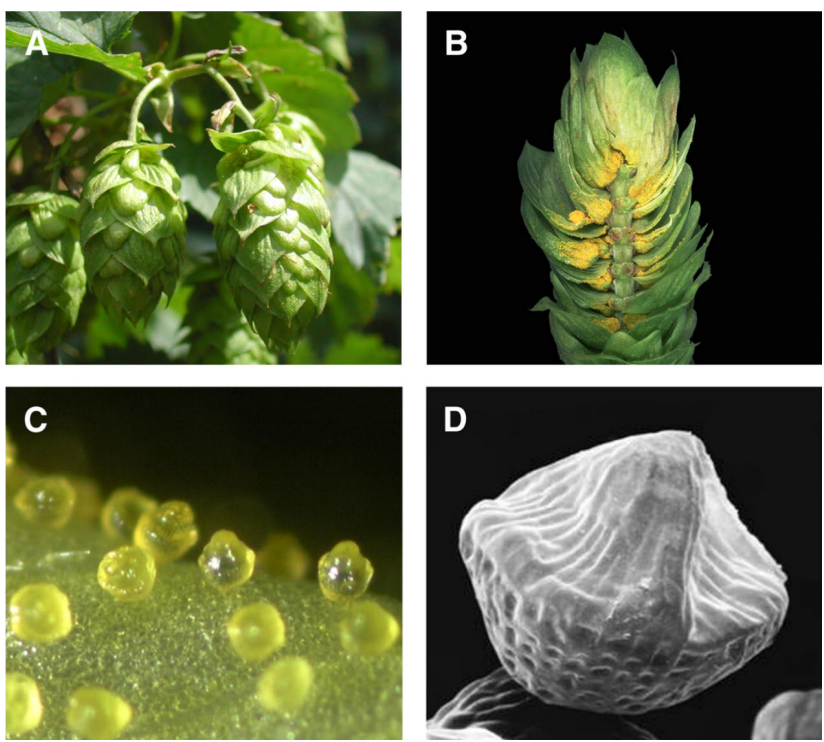

**Figure 14.** Structural features of hop cones and glandular trichomes  (A) Cones of the Taurus hop variety (length: ~5 cm). (B) Cross-sectional view highlighting basal glandular trichomes. (C) Visual obtained via brightfield microscopy showcasing mature glandular trichomes (Scale bar = 500 μm). (D) Electron microscopy scan revealing resin-filled subcuticular pockets (Scale bar = 100 μm). Reprinted from Nagel et al. (2008). CC BY 4.0.

## 2.5.2    Phytochemical composition

Phytochemical analysis has determined that whole hop cones comprise a diverse range of substances, including resins, polyphenols, volatile (essential) oils, waxes, monosaccharides, cellulose, soluble fiber, proteins, and amino acids (Almaguer et al., 2014). Among these constituents are bioactive secondary metabolites released in the form of yellow material from the lupulin glands (trichomes) of the female inflorescences (Figure 14B, C). These metabolites are non-biosynthetic in humans and possess significant value, particularly in the craft beer industry and for their therapeutic potential (Luzak et al., 2017; Keskin et al., 2019; Lin et al., 2019). The secondary metabolites found in hops can be classified into three main categories: polyphenols, resins, and essential oils (Figure 15), which offers a useful framework for understanding their diverse properties and potential applications (Steenackers et al., 2015; Astray et al., 2020).

### 2.5.2.1    Hop polyphenols

Hops are a rich source of polyphenolic compounds, particularly flavonoids, which account for up to 5% of the dried hop cone weight (Karabín et al., 2015). The primary prenylflavonoid found in hop cones is xanthohumol (XN). This compound, which is characteristic of *Humulus lupulus* L., comprises up to 1.1% of the dry weight of hops and represents a significant proportion of its flavonoids (Stevens and Page, 2004; Magalhães et al., 2012). Other noteworthy flavonoids present in hops include catechin and flavonol glycosides such as rutin, quercitrin, and quercetin (Gorissen et al., 1968; Sägesser and Deinzer, 1996).

Although XN is the primary flavonoid present in hops, about 50–75% of this compound is thermally isomerized to form isoxanthohumol (IXN) during the wort boiling process. However, the extent of this conversion is influenced by the boiling conditions, which involve the pH and ethanol concentration in the heating solvent (Moriya et al., 2018). While IXN accounts for only 0.01% of the dry weight of hops, its significant increase during beer brewing makes it the predominant hop-derived flavonoid, with concentrations of up to 3.44 mg/L, and thus a constituent of the human diet (Stevens and Page, 2004; Mudura and Coldea, 2015; Tronina et al., 2020).

In humans, IXN undergoes demethylation mediated by hepatic Cytochrome P450 enzymes and gut microbiota, resulting in the formation of 8-prenylnaringenin (8PN) (Guo et al., 2006; Possemiers et al., 2006; Bolca et al., 2007). Notably, hops have been found to contain both 8PN and its isomer, 6-prenylnaringenin (6PN), at concentrations of 0.002% and 0.01%, respectively. Both prenylflavonoids can be formed non-enzymatically from the XN precursor desmethylxanthohumol (DMX) during the beer brewing process (Stevens and Page, 2004; Prencipe et al., 2014).

The levels of prenylflavonoids in beer and thus in the human diet are influenced by several factors during production, including the variety and amount of hops used, as well as the brewing technique. Quantitative analysis using liquid chromatography tandem-mass spectrometry (LC-MS/MS) has revealed a wide range of concentrations for hop prenylflavonoids in beer. Initially, XN concentrations were found to vary between 0.002 and 0.69 mg/L (Stevens et al., 1999). However, higher levels of XN (up to 17.2 mg/L) were further achieved by adding extracts that are enriched in XN and dark malt to the wort during the final stages of boiling, followed by fast cooling (Wunderlich et al., 2005; Magalhães et al., 2012). Moreover, IXN, 8PN, and 6PN concentrations were reported to range from 0.04 to 3.44 mg/L, 0.001 to 0.24 mg/L, and 0.007 to 0.2 mg/L, respectively (Stevens et al., 1999; Rothwell et al., 2013; Tronina et al., 2020). Therefore, the intake of hop prenylflavonoids varies among individuals depending on the type and amount of beer consumed, which may have implications for the potential health benefits associated with these compounds.

### 2.5.2.2 Hop resins

Hop resins comprise a complex mixture of various compounds that contribute to the characteristic bitter flavor of beer. The resins are classified into two primary subcategories: soft resin, which is soluble in paraffin hydrocarbons like n-hexane, and hard resin, which cannot dissolve in n-hexane but is soluble in cold methanol and diethyl ether (Taniguchi et al., 2014; Bertelli et al., 2018). The soft resin in hops contains two distinct groups of prenylated acylphloroglucinols (bitter acids), specifically α-acids (5–13% of hop dry matter) and β-acids (3–8% of hop dry matter) (Van Cleemput et al., 2009; Almaguer et al., 2014). On the other hand, the hard resin is made up mostly of fast-oxidized byproducts of α- and β-acids over the course of hop storage (Taniguchi et al., 2014).

The major α-acids in hops are humulone, cohumulone, and adhumulone, which account for 35–70%, 20–65%, and 10–15% of the total α-acid content, respectively (Shah et al., 2010; Ntourtoglou et al., 2020). During wort boiling, the addition of hops results in the partial isomerization of α-acids to iso-α-acids, contributing to the distinctive bitterness of beer (Sanz et al., 2019). In addition to isomerization, α-acids undergo oxidation to form humulinones, which exhibit higher polarity and are therefore more soluble in beer than iso-α-acids. The β-acids, which mainly consist of lupulone, colupulone, and adlupulone, are present in varying proportions, with lupulone accounting for 30–55% of the total β-acid content. These compounds are known for their resistance to isomerization during beer brewing, and they contribute a bitterness characteristic that is nine times less potent than that of α-acids (Rutnik et al., 2023). Beyond the prevalent co- and ad-analogs, researchers have isolated several minor bitter acids, such as post-humulone/post-lupulone, pre-humulone/pre-lupulone, and adpre-humulone (Shah et al., 2010; Sanz et al., 2019).

### 2.5.2.3    Hop essential oils

The glandular trichomes of hops produce a rich variety of volatile compounds, collectively known as essential oils, which impart a distinct aroma and flavor to beer (Steenackers et al., 2015; Astray et al., 2020). These oils make up approximately 0.5–3% of hop dry matter, consisting of more than 400 organic compounds (Almaguer et al., 2014). Hydrocarbons account for a significant proportion (50–80%) of hop essential oils and fall into three classes, namely monoterpenes, sesquiterpenes, and aliphatic compounds. The monoterpene myrcene and the sesquiterpenes α-humulene, β-caryophyllene, and β-farnesene, are considered to be the primary constituents of hop essential oils. In addition to hydrocarbons, the oxygenated fraction, which makes up around 30% of hop essential oils, contains terpene alcohols like linalool, myrcenol, geraniol, and 2-methyl-3-buten-2-ol (2M3B), as well as oxidized sesquiterpenes, among other compounds (Sharpe and Laws, 1981; Astray et al., 2020). The minor sulfur component, although only present in small amounts (up to 1% of the total oil), also contributes to the overall complex composition of hop essential oils (Karabín et al., 2016).

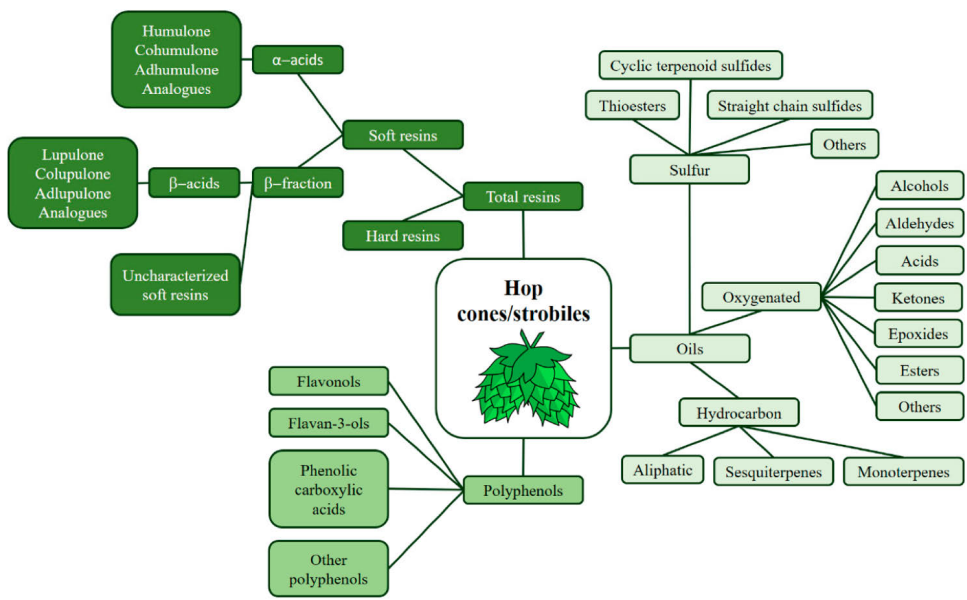

**Figure 15.** A classification scheme of different types of compounds present in hops, namely polyphenols, resins, and essential oils. Reprinted from Astray et al. (2020). CC BY 4.0.

## 2.5.3 Exploring hop neuroactive constituents and sleep-promoting mechanisms

In recent years, notable attention has been directed towards hops and their constituents, driven not only by the growing popularity of specialty beers but also by their promising bioactivities (Carbone and Gervasi, 2022). These include digestive, antimicrobial, anti-inflammatory, anticarcinogenic, antioxidant, cardioprotective, neuroprotective, and hypnotic activities, among others (Lin et al., 2019; Wang et al., 2020).

Research into the sedative and sleep-promoting effects of hops was motivated by observations of exhaustion and sleepiness among laborers involved in the collection and handling of hop cones (Tyler, 1987; Karabín et al., 2016). Since then, several studies have been carried out to explore the potential of this plant as a therapeutic intervention for sleep disorders (Zanoli and Zavatti, 2008; Franco et al., 2012). The sedative action of hop extract was initially demonstrated in mice by a reduction in locomotor activity upon administration of a high dose (200 mg/20 g, i.p.), without indications of muscle relaxation (Bravo et al., 1974). In a subsequent behavioral assessment, mice were administered lower intraperitoneal doses of hop extract, ranging from 100 to 500 mg/kg (Lee et al., 1993). The study found that the extract produced a dose-dependent suppression of spontaneous locomotion, as well as an increase in the duration of sleep induced by pentobarbital. Moreover, the extract exhibited hypothermic, antinociceptive, and anticonvulsant effects in the tested mice.

In an effort to uncover hop-derived neuroactive components, researchers have proposed that the sedative-hypnotic effects of the plant are derived from 2M3B. This volatile alcohol is a byproduct of bitter acids that undergo oxidation during the storage period of hops and is also identified in the oxidative fraction of hop essential oils (Wohlfart et al., 1983a; Rutnik et al., 2021). In mice, the administration of a high dose of 2M3B (800 mg/kg, i.p.) resulted in a state of narcosis that persisted for 8 hours (Hänsel et al., 1980). Moreover, a dose of 206.5 mg/kg i.p. of the same compound was found to reduce spontaneous locomotor activity in rats by 50% with no observed myorelaxation (Wohlfart et al., 1983b). The level of 2M3B in newly harvested hop cones is very low, and increases gradually over a two-year period when the hops are stored at room temperature, reaching a maximum concentration of 0.15% (Hänsel et al., 1982; Chadwick et al., 2006). Considering the high doses tested in rodents and the trace levels of 2M3B in hops, the sedative-hypnotic activity of hop extracts cannot be attributed solely to this compound. Therefore, other neuroactive constituents of hops are likely involved in the sedative-hypnotic activity of hop extracts, and their identification and characterization are a matter of ongoing research.

Expanding the scope to the primary constituents of hops, studies have assessed their neuropharmacological effects in various animal models. In terms of sleep promotion, rats were administered hop $CO_2$ extract and its abundant α-acid fraction at doses of 10 and 20 mg/kg (p.o.), which resulted in a dose-dependent prolongation of pentobarbital-induced sleeping time (Zanoli et al., 2005). Neither extract had an effect on the spontaneous locomotion of the animals tested in the open field test or on their anxiety responses in the elevated plus maze test. However, a further study conducted in mice showed that hop extract containing 36% α-acids at doses of 100 and 200 mg/kg decreased locomotor activity in the open field test, and the higher dose extended ketamine-induced sleeping time (Schiller et al., 2006). Additionally, in the same study, the behavioral effects of various fractions containing 84% α-acids, 60% β-acids, and pure essential oils were assessed in mice. α-acids at a dose of 21 mg/kg (p.o.) significantly prolonged the duration of sleep induced by ketamine, with no indication of anxiolytic activity. On the other hand, a similar sleep-promoting effect was only observed at relatively high doses of β-acids (119 mg/kg, p.o.) and essential oils (100 and 200 mg/kg, p.o.). Overall, it has been concluded that α-acids contain the most potent constituent(s) that determine the sedative and sleep-promoting properties of hops, without excluding the potential contribution of β-acids, essential oils, and other unexplored constituents (Schiller et al., 2006; Karabín et al., 2016). Therefore, further in-depth investigation is essential to isolate and characterize the specific compounds in hops that mediate their potential therapeutic effects and to elucidate the underlying mechanism of action.

Modulation of $GABA_A$ receptor function is the primary mechanism attributed to the sedative and sleep-promoting properties of hops. Although melatonin ML1 and serotonin 5-HT6 receptors have been proposed as potential targets, their involvement lacks functional confirmation. Current evidence for ML1 and 5-HT6 is based solely on binding affinities using a hydrophilic hop extract that lacks α- and β-acids (Abourashed et al., 2004). Beer extract and hop oil have been reported to enhance GABA-evoked currents in bovine α1β1-$GABA_A$Rs expressed in *Xenopus* oocytes (Aoshima et al., 2006). Consistently, hop extract (0.5–5 µg/mL) exhibited potent positive modulation of human α1β2-$GABA_A$Rs (Sahin et al., 2016). This was indicated by a dose-dependent increase in GABA-induced response, ranging from $82 \pm 27\%$ to $248 \pm 26\%$. Certain components of hop essential oils, specifically terpenoids, have demonstrated modulatory enhancement of recombinant $GABA_A$ receptors. However, these effects were detected at millimolar or high micromolar concentrations, which exceeded their natural presence in hops and may not be physiologically relevant. The constituents tested included myrcene (1 mM), β-caryophyllene (1 mM), linalool (300 µM), myrcenol (350 µM), geraniol (1 mM), and 2-methyl-3-buten-2-ol (2M3B) (600 µM) (Hossain et al., 2002; Aoshima et al., 2006; Kessler et al., 2014).

With the potential to offer novel mechanisms distinct from those of benzodiazepines, xanthohumol has emerged as a noteworthy hop flavonoid. A preliminary study using fluorescence correlation spectroscopy (FCS) reported a significant enhancement in the binding of muscimol to hippocampal neurons by xanthohumol at 75 nM (Meissner and Häberlein, 2006). This increase in GABA$_A$ receptor agonist binding is typically observed in PAMs and correlates with their ability to enhance agonist-induced inhibitory currents (Akk et al., 2020). Xanthohumol's effect on muscimol binding was found to be comparable to that of the benzodiazepine PAM midazolam at 7.5 nM. However, unlike midazolam, xanthohumol did not displace the binding of N-des-diethyl-fluorazepam (Ro 7-1986/602) (Hegener et al., 2002; Meissner and Häberlein, 2006). Moreover, xanthohumol was found to inhibit presynaptic excitatory glutamate release from isolated nerve terminals of the rat hippocampus, with an IC$_{50}$ of 15 μM (Chang et al., 2016). This action was blocked by the GABA$_A$ receptor antagonist, SR 95531 (gabazine), suggesting that xanthohumol might modulate both inhibitory and excitatory responses via GABA$_A$ receptors. However, no further studies have been conducted to confirm and elucidate the specific mechanisms and subtype selectivity underlying the interaction between xanthohumol and this target. Nevertheless, current evidence suggests that xanthohumol, along with its structurally analogous prenylflavonoids, may positively modulate GABA$_A$ receptors independently of the classical benzodiazepine-binding site. This modulation could potentially contribute to the sedative and sleep-promoting effects associated with hops.

In essence, hops harbor well-established sedative and somnolent properties, likely arising from neuroactive phytochemicals that modulate GABA$_A$ receptors. However, knowledge gaps persist regarding the precise identities, activities, subtype selectivity, and interactions of these hop bioactives. Various compounds in hops, including prenylated flavonoids, bitter acids, degradation products, and essential oils, may differentially modulate GABA$_A$ receptor function. Investigating these diverse constituents could provide vital insights into their utility in optimized and standardized plant-based formulations. Furthermore, considering that beer constitutes a primary dietary source of hops (Stevens and Page, 2004; Van Cleemput et al., 2009), interactions with alcohol warrant investigation, as certain hop bioactives may synergistically potentiate GABA-mediated intoxication in beer drinkers. The examination of these interactions is critical, given their significant implications for consumer health and safety. Alternatively, revealing the pharmacological basis underlying hops' sleep-promoting qualities represents a promising avenue for advancing safe and effective alternatives to existing sleep aids.

# 3    Aims

This thesis investigates the modulatory effects of hop constituents on $GABA_A$ receptors using a multidisciplinary approach to shed light on the intricate workings of hops' sedative and sleep-promoting properties. The objectives were to identify the major bioactive compounds in hops and unravel their modulatory actions, subtype selectivity, and combination effects within the $GABA_A$ receptor complex, as well as with other modulators such as alcohol. This advances our understanding of the molecular mechanisms underlying the modulation of $GABA_A$ receptors by hop-derived compounds, and facilitates drug development and optimization of natural formulations for insomnia and other sleep-related problems.

The studies focused on achieving the following aims:

- Investigate the modulation of GABA-mediated responses in native and recombinant $GABA_A$ receptors by hop flavonoids (**I**).

- Evaluate the sensitivity of potent hop flavonoids to flumazenil antagonism and assess their allosteric modulation through the classical benzodiazepine site (**I, II**).

- Utilize *in silico* methods to determine the putative binding sites and poses of potent hop flavonoids in the α1β2γ2 isoform of $GABA_A$ receptors (**II**).

- Examine the modulatory effects of isolated hop α/β-acid fractions, degradation products, and volatile compounds on $GABA_A$ receptors, and assess the contribution of the classical benzodiazepine site in their allosteric actions (**II**).

- Select a neuroactive phytochemical from hops based on its content and modulatory potency, validate its functional activity, and assess its selectivity as well as interactions with potent hop flavonoids and alcohol (**III**).

- Establish the role of the selected phytochemical in the sedative and sleep-promoting properties of hops *in vivo* (**III**).

# 4 Materials and Methods

## 4.1 Experimental reagents and compounds (I–III)

### 4.1.1 Radioligands

The tritium-labeled radioligands used were [propyl-2,3-$^3$H]EBOB (48 Ci/mmol), a non-competitive blocker of the GABA-gated chloride channel, in addition to the benzodiazepines [7,9-$^3$H]Ro 15–4513 (28 Ci/mmol), a partial inverse agonist, and [methyl-$^3$H]flunitrazepam (85.2 Ci/mmol), a positive allosteric modulator. All were procured from PerkinElmer Life and Analytical Sciences (Boston, MA, USA) (Figure 16).

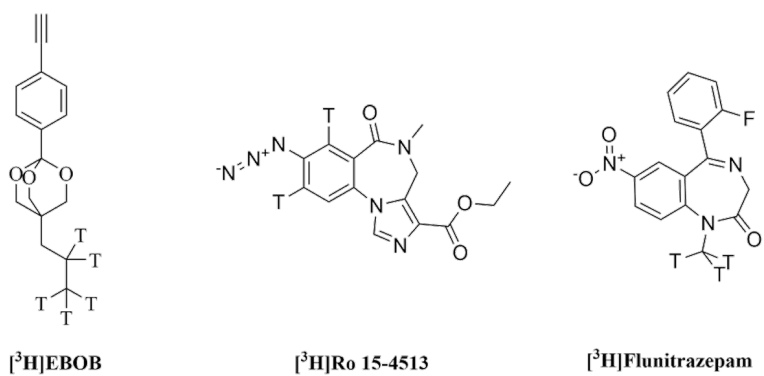

**[³H]EBOB**          **[³H]Ro 15-4513**          **[³H]Flunitrazepam**

**Figure 16.** Chemical structures of the radioligands used in Studies **I-III**. T indicates the tritium ($^3$H) radiolabel position (ChemDraw 20.1).

### 4.1.2 Drugs

The compounds diazepam, GABA, and picrotoxin were purchased from Sigma Chemical Co. (St. Louis, MO, USA). Flumazenil (Ro 15–1788) was supplied by Tocris Bioscience (Bristol, UK) and F. Hoffmann-La Roche Ltd (Basel, Switzerland). Orion Corporation (Espoo, Finland) provided the sodium pentobarbital injection solution (Mebunat Vet®, 60 mg/mL), and Altia (Rajamäki, Finland) provided the ethanol.

### 4.1.3 Hop constituents

#### 4.1.3.1 Source of hops

The primary source material for Study **I** consisted of by-products from the extraction of the Marynka hop variety using supercritical carbon dioxide ($CO_2$). These by-products were acquired from the New Chemical Syntheses Institute (INS, Puławy, Poland). Marynka is a dual-purpose Polish hop variety utilized in the brewing process to introduce bitterness during the boil and, subsequently, aroma and flavor during the dry hopping stages. Marynka contains a moderate amount of α-acids (6.61–11.1%) (Brudzynski and Baranowski, 2003; Leonardi et al., 2013). However, the high phenolic content of Marynka (Kowalczyk et al., 2013) makes it suitable for supercritical fluid extraction of xanthohumol, with yields up to 4.8% (Klimek et al., 2021). For Study **II**, hop pellets of the Citra variety cultivated in the USA were obtained from the Lappo craft brew outlet (Piikkiö, Finland). Citra, another dual-purpose hop variety, boasts a high α-acid content (11–13%) and is renowned for its distinctive tropical fruit flavor and aroma attributed to its elevated volatile oil content (2.2–2.8%) (Oliver, 2011).

#### 4.1.3.2 Prenylflavonoid isolation and conversion

Xanthohumol was isolated and then transformed into isoxanthohumol through an alkali-driven isomerization process, as explained by Anioł et al. (2008). Subsequently, isoxanthohumol was demethylated to yield 8-prenylnaringenin, according to the procedure detailed in Stompor et al. (2017). Spectral analysis of the hop prenylflavonoids was consistent with findings from previous studies (Stompor et al., 2013; Potaniec et al., 2014; Stompor et al., 2017), and their purity, as determined by high-performance liquid chromatography (HPLC), surpassed 98%.

#### 4.1.3.3 α/β-acid HPLC-guided isolation

From a 2-gram sample of hop pellets, α-acids, and β-acids were extracted using 10 mL of methanol/formic acid (99:1, v/v) over two separate extraction cycles. The mixture was thoroughly agitated for half an hour at $21 \pm 1°C$, followed by a 5-minute centrifugation at 1500 xg. The obtained extracts were pooled, brought to a 20-mL final volume using the extraction solvent, and passed through 0.2 μm polytetrafluoroethylene (PTFE) membrane filters. Ahead of subsequent semipreparative HPLC isolations, the refined extract was preserved at −20 °C.

The α/β-acid extracts underwent isolation via a semipreparative HPLC system from Shimadzu (Kyoto, Japan), according to the procedure described in Benkherouf et al., (2020). This system comprised a sample injector, solvent delivery unit, column oven, Ultraviolet-visible (UV-vis) detector, and degassing unit. Separation was performed on a Phenomenex Aeris 5 μm peptide XB-C18 column (250 × 10.0 mm; Torrance, CA, USA). Each sample was injected with an 80 μL volume, and chromatographic data were captured at a wavelength of 332 nm. Elution utilized binary solvent mixtures, with mobile phase A containing water with 0.1% formic acid and mobile phase B consisting of acetonitrile with 0.1% formic acid. These solvents were delivered at a 4 mL/min flow rate. Two distinct gradient programs were employed for sample collection: one for fractions 1-6 and another for fraction 7 to ensure optimal separation. Following separation, the fractions were dried at +30 °C using a rotary evaporator. After two consecutive rinses with methanol, the solvent was evaporated using nitrogen blowdown in a dry bath at +30 °C. The resulting dried samples were then quantified and solubilized in dimethyl sulfoxide (DMSO) to prepare 10 mM stock solutions.

**Table 1.** Summary of hop constituents, methods of extraction, and sources across studies

| Study | Compounds/Fractions | Source/Method | Basis of Identification |
|---|---|---|---|
| I | Xanthohumol (XN) | Isolation from supercritical $CO_2$ hop extract | HPLC |
| | Isoxanthohumol (IXN) | Base-catalyzed isomerization of XN | HPLC |
| | 8-prenylnaringenin (8PN) | Demethylation of IXN | HPLC |
| | Catechin, Quercetin, Quercitrin, Rutin | Purchase from Biopurify Phytochemicals Ltd (Chengdu, China) | HPLC |
| II | α-acids: Cohumulone (F2), Humulone (F3), Adhumulone (F4), Adprehumulone, Prehumulone, and Humulone (F5) | Extraction with acidified methanol and semipreparative HPLC from Citra hops pellets | UV–vis, ICE-3 standard, MS, NMR |
| | β-acids: Colupulone (F6), Lupulone (F7.1), Adlupulone (F7.2) | Same as above | UV–vis, ICE-3 standard, MS, NMR |
| | 2-methyl-3-buten-2-ol (2M3B) | Purchase from Sigma Chemical Co. (St. Louis, MO, USA) | GC |
| | 6-prenylnaringenin (6PN), Linalool | Purchase from Phytolab GmbH & Co. KG (Vestenbergsgreuth, Germany) | HPLC |
| III | Humulone (hum) | Purchase from Specs (Zoetermeer, Netherlands) | HPLC |
| | 6-prenylnaringenin (6PN), Isoxanthohumol (IXN) | Purchase from PhytoLab GmbH & Co. KG | HPLC |

### 4.1.3.4    Compound identification and analysis

Seven primary constituents were obtained from hop pellet extracts through a combination of chromatographic separation and spectral analysis, as detailed in Benkherouf et al., (2020). The characterized peak fractions encompassed the flavonoid xanthohumol (F1), the α-acids cohumulone (F2) and humulone (F3), adhumulone (F4), and the β-acids colupulone (F6), lupulone (F7.1), and adlupulone (F7.2). These constituents were differentiated using the ICE-3 calibration standard from Labor Veritas AG, Zürich, Switzerland, alongside UV-vis spectra and LC retention times. The ICE-3 standard contained α-acids (13.88% cohumulone, 30.76% N+ adhumulone) and β-acids (13.44% colupulone, 10.84% N+ adlupulone). Maximum absorbance wavelengths were set at 365 nm for xanthohumol, 320 nm for α-acids, and 330 nm for β-acids. Fraction F5, positioned between adhumulone and colupulone, exhibited a UV-vis absorption spectrum resembling that of the α-acids. Accurate masses were determined using an Ultra-High Resolution Qq-Time-Of-Flight (UHR-QqTOF) mass spectrometer equipped with an electrospray ion source (ESI). The separation was conducted on a Bruker Elute ultra-high performance liquid chromatography (UHPLC) system (Billerica, MA, USA), utilizing a Phenomenex Kinetix 2.6 μm column (100 × 4.6 mm; Torrance, CA, USA).

The composition and purity of the constituents obtained were further ascertained using 1H and 13C Nuclear Magnetic Resonance (NMR) spectroscopy on a Bruker Avance-III NMR instrument. Specifically, xanthohumol in F1 and humulone in F3 emerged as pure fractions. In contrast, F2 displayed a more intricate profile, comprising cohumulone and a combination of at least four other α-acids, underscoring the inherent challenges of isolating closely related compounds. F4 was predominantly composed of adhumulone (~85%), with humulone as a secondary component. F5 represented a mixture of α-acids in a 2:1:1 ratio, specifically adprehumulone, prehumulone, and humulone. F6 was characterized by the presence of colupulone, complemented by traces of linalool and lipids, whereas lupulone was verified in F7.1.

## 4.2    Study animals and biological samples (I–III)

### 4.2.1    Rodent sourcing and ethical considerations

The studies employed adult male Sprague-Dawley rats for binding assays (**I–III**) and adult male BALB/cAnNRj mice for behavioral tests (**III**). Both species were sourced from the Central Animal Laboratory of the University of Turku (UTUCAL), ensuring the ethical treatment and welfare of the animals. The rodents were maintained under standardized conditions, including a 12-hour light-dark cycle, a controlled temperature of 21 ± 1 °C, and a humidity level of 65%. They had unrestricted access to tap water and a regular chow diet. Commitment to ethical considerations and animal welfare was demonstrated by adherence to the 3Rs principles - Replacement, Reduction, and Refinement. These principles guided the selection and treatment of animals, minimizing their use and suffering while maximizing the generation of reliable and reproducible data.

In alignment with these ethical principles, our practices adhered to the Finnish Act on the Use of Animals for Experimental Purposes (62/2006), the EU Directive (2010/63/EU) on the Protection of Animals Used for Scientific Purposes, and the OECD's Principles of Good Laboratory Practice and Compliance Monitoring (ENV/MC/CHEM(98)17). The euthanasia of the rodents was carried out humanely using guillotine decapitation. For the binding assays, the fore/midbrain (loosely referred to as the forebrain) and the cerebellum were carefully extracted from the rats. These tissue samples were then rapidly frozen on dry ice and stably preserved at −70 °C before the membrane preparation procedure. For the behavioral tests (**III**), the mice were acclimated to the experimental environment within their home cages for 1 hour before the commencement of any treatments. All experimental protocols were authorized by the National Animal Experiment Board (project permit: ESAVI/25715/2018).

## 4.2.2    Brain membrane preparation

The protocol for preparing membranes from rat brain tissues was adapted with modifications from Squires and Saederup (2000) and Uusi-Oukari et al. (2014). The rat fore/midbrain regions were thawed and then homogenized for 20 seconds at a rotation speed of 9500 rpm in a solution composed of 10 mM tris(hydroxymethyl)aminomethane (Tris)-HCl at pH 8.5 and 2 mM ethylenediaminetetraacetic acid (EDTA). The resulting homogenates underwent centrifugation at 20,000 xg for 10 minutes while maintaining a temperature of +4 °C. The pellets obtained were subjected to three sequential washes using a solution composed of 10 mM Tris-HCl (pH 8.5), 0.2 M NaCl, and 5 mM EDTA. After the washes, the pellets were resuspended in chilled water while being kept in an ice bath and were subsequently centrifuged again. The pellets were subjected to an additional three washes using the Tris-HCl/NaCl/EDTA buffer (pH 8.5) before ultimately being suspended in 10 mM Tris-HCl at pH 7.4 and stored at -70 °C for further use.

## 4.2.3    Cell culture

Human embryonic kidney cells (HEK293 line), obtained commercially from Sigma-Aldrich (St. Louis, MO, USA), were cultured in Dulbecco's Modified Eagle Medium (DMEM) provided by Gibco (Gaithersburg, MD, USA). The growth medium was enriched with 10% fetal bovine serum (FBS), 50 units/mL of penicillin, and 50 µg/mL of streptomycin, all sourced from Sigma-Aldrich (St. Louis, MO, USA). The HEK293 cells were incubated under standardized conditions at a temperature of 37°C, 95% humidity, and 5% $CO_2$. In preparation for binding experiments and electrophysiology measurements, the cultured cells were split and seeded into 10/15 cm dishes and 12 mm coverslips in 24-well plates, respectively. Transfection procedures were initiated 24 hours after seeding.

## 4.3 Molecular Biology (I, III)

### 4.3.1 Plasmid DNA transformation and purification

A DH5-α competent strain of Escherichia coli (E. coli) was transformed with distinct pRK5 plasmids, controlled by the cytomegalovirus (CMV) promoter. Each plasmid carried a gene conferring ampicillin resistance and contained rat complementary DNA (cDNAs) encoding different $GABA_A$R subunits: α1 (L 08490), α6 (L 08495), β2 (X 15467), β3 (X 15468), γ2S (L 08497), γ2L (L 08497), and δ (L 08496) (Uusi-Oukari et al., 2000). To purify the plasmids, the NucleoBond Xtra Maxi kit for transfection-grade plasmid DNA (Macherey-Nagel GmbH & Co. KG, Düren, Germany) was used following the provided protocol. DNA concentration was quantified by measuring the capacity of a 2 µl sample to absorb UV light at a wavelength of 260 nm using a spectrophotometer.

### 4.3.2 Transient transfection of $GABA_A$Rs in HEK293 cells

The plasmids encoding distinct $GABA_A$R subunits (α1, α2, α6, β3, γ2S, δ) were employed in ratios of 1:1:1 and 1:1 for the transient expression of receptor complexes consisting of α1β3γ2S, α2β3γ2S, α6β3γ2S, α6β3δ, and α6β3 isoforms. For radioligand binding assays (**I, III**), HEK293 cells were transfected following the calcium phosphate co-precipitation protocol detailed in earlier works (Graham and Van, 1973; Lüddens and Korpi, 1997). In brief, a co-precipitation mixture was prepared for each 10/15 cm cultured dish, comprising 5 µg of plasmid cDNA for each subunit, 450 µL of sterile water, 50 µL of $CaCl_2$ (2.5 M), and 500 µL of 2X HEPES-buffered saline at pH 7.0. Integration of the γ2 subunit into the expressed subtypes was confirmed through competitive binding assessment using the benzodiazepine site-selective radioligand [$^3$H]Ro15-4513 (Figure 17), as previously outlined (Uusi-Oukari and Korpi 1990). In preparation for electrophysiology experiments (**III**), the K4® DNA Transfection Kit from Biontex (München, Germany) was employed following the provided instructions to optimize expression yield and the viability of HEK293 cells. Co-transfection of a plasmid encoding an enhanced variant of green fluorescent protein (pWPI-EGFP) (Addgene, 12254) facilitated the detection of GABA$_A$R-expressing cells for targeted recordings based on the emitted green signal.

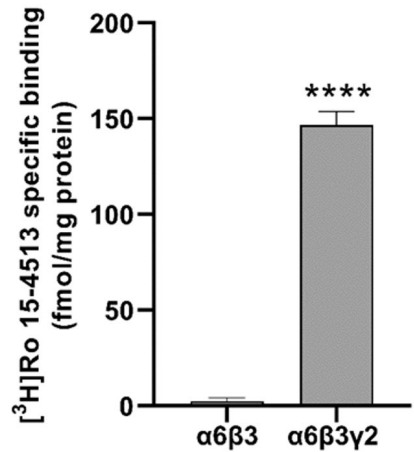

**Figure 17.** Evaluation of [$^3$H]Ro 15-4513 Binding to α6β3 and α6β3γ2 receptor subtypes in HEK293 Cells: Triplicate samples underwent a 60-minute incubation with gentle agitation at 4°C in an ice bath. The assay mixture contained 10 mM Tris-HCl buffer (adjusted to pH 7.4) and 2 nM of [$^3$H]Ro 15-4513, with a final volume of 300 µl. Flumazenil, at a concentration of 10 µM, was employed to measure nonspecific binding. A notable difference was observed in the specificity of [$^3$H]Ro 15-4513 binding to α6β3γ2 when compared to α6β ($****p < 0.0001$, student *t*-test).

## 4.4    Radioligand binding procedures (I–III)

### 4.4.1    [$^3$H]EBOB binding assay

The procedures for carrying out [$^3$H]EBOB radioligand binding assays followed Uusi-Oukari and Maksay's method (2006). Rat brain membranes were defrosted and then centrifuged at 20,000 xg (+4 °C) for 10 minutes in an initial washing step with assay buffer containing 50 mM Tris-HCl at pH 7.4, supplemented with 120 mM NaCl. The brain membrane pellets were subsequently resuspended in this buffer. For HEK293 cells, membrane harvesting was performed 48 hours post-transfection through the application of a detaching solution containing 10 mM Tris-HCl, 0.15 M NaCl, and 2 mM EDTA. This step was followed by centrifugation at 20,000 g (+4°C) for 10 minutes and resuspension of the obtained pellets in the assay buffer.

In Studies **I–III**, membranes from rat brains or HEK293 cells were incubated in triplicate or quadruplicate at room temperature for 2 hours with gentle agitation in a total volume of 400 μL assay buffer. This buffer was prepared with either 1 nM or 2 nM of [³H]EBOB, a range of hop compound concentrations (10 nM - 30 μM), and either 2 μM, 3 μM, or no GABA. For modulation assessment, the control was defined as the maximal [³H]EBOB binding observed with GABA alone, whereas for direct actions, it was the maximal [³H]EBOB binding without GABA.

In Study **III**, additional sets of triplicate samples of membranes from rat forebrain and cerebellum underwent a similar incubation process. These sets comprised varying GABA concentrations (ranging from 50 nM to 20 μM) and included either 1 μM humulone, 30 mM ethanol, or both. In experiments involving humulone and various recombinant receptor subtypes (α6β3γ2S, α6β3δ, α6β3), membranes from HEK293 cells were subjected to incubation with 1 nM [³H]EBOB and varying humulone concentrations (ranging from 30 nM to 30 μM), with or without the addition of 30 mM ethanol. Taking into consideration the potential presence of a maximum of 0.39 mM ethanol in 1 nM [³H]EBOB, the radioligand stock was evaporated, the residue was dissolved in 5 μL of DMSO, and the volume was subsequently adjusted with assay buffer to achieve the desired ethanol-free concentration.

The nonspecific binding in all [³H]EBOB binding experiments was ascertained using picrotoxin at 100 μM. Incubations were concluded by filtering the samples onto Whatman GF/B glass microfibers (Whatman International Ltd., Maidstone, UK) using a Brandel Cell Harvester (model M-24, Gaithersburg, MD, USA), to isolate the radioligand-receptor complexes from the free radioligand. The filters underwent a series of three washes using 5 mL of chilled Tris-HCl buffer (10 mM, pH 7.4), followed by a drying phase in ambient air. Subsequently, the filters were submerged in 3 mL of AquaLight Beta scintillation liquid (Hidex, Turku, Finland), and the resulting level of radioactivity was quantified with a Hidex 600 SL Automatic TDCR Liquid Scintillation Counter.

## 4.4.2    [³H]Ro 15–4513 and [³H]flunitrazepam binding assays

The binding assays for [³H]Ro 15–4513 and [³H]flunitrazepam were adapted from the methodology outlined by Uusi-Oukari and Korpi (1992). In brief, quadruplicate samples, each with a combined volume of 300 μL, were prepared. These samples were incubated at 4 °C in an ice bath with gentle agitation for a duration of 60 minutes. For the [³H]Ro 15–4513 binding, the assay buffer was 10 mM Tris-HCl at pH 7.4, whereas for [³H]flunitrazepam binding, it contained 50 mM Tris-HCl at pH 7.4, supplemented with 120 mM NaCl. Depending on the assay, the buffers included either 2 nM of [³H]Ro 15–4513 or 1 nM of [³H]flunitrazepam, both with and without the tested constituent. To assess nonspecific binding, flumazenil at a concentration of 10 μM was introduced into the assay. The incubation phase was concluded by filtering the samples through Whatman GF/B glass microfiber filters, followed by two subsequent washes using 5 mL of ice-cold buffer solution. Specifically, for [$^3$H]Ro 15–4513 binding, the washing buffer was composed of 10 mM Tris-HCl at pH 7.4, while for [$^3$H]flunitrazepam binding, the buffer comprised 10 mM Tris-HCl at pH 7.4 supplemented with 120 mM NaCl.

### 4.4.3    Protein measurements

The concentrations of proteins within brain membranes utilized in radioligand binding assays were quantified following the manufacturer's protocol, using a Bradford blue protein-dye assay kit (Bio-Rad Laboratories, CA, USA).

## 4.5    *In silico* docking (II)

A computational approach was employed to elucidate the potential binding interactions of hop prenylflavonoids within the α1β2γ2 subtype of the GABA$_A$ receptor. Molecular simulations were conducted using the Maestro software package (release 2019-1; Schrödinger, LLC) to model the docking of these phytochemicals within two proposed allosteric modulatory sites: the α1+/γ2− and α1+/β2− subunit interfaces. For comparison and validation, the benzodiazepine site antagonist flumazenil and the positive allosteric modulator CGS 9895 were also docked at these sites (protocol details are outlined in Benkherouf et al., 2020). The Glide XP algorithm and the Induced Fit protocol (IFP) were used to explore favorable ligand-receptor binding modes and conformational adjustments (Friesner et al., 2004, 2006; Sherman et al., 2006a, 2006b). Subsequent binding energy calculations were performed using Maestro's Prime/MM-GBSA tool to estimate the stability of the predicted complexes (Walters et al., 1998; Lyne et al., 2006; Sastry et al., 2013). Molecular inspection and analyses of the docking conformations were enabled through Maestro and PyMOL 3D visualization system (version 2.3).

## 4.6    Whole-cell patch clamp electrophysiology (III)

We employed whole-cell patch clamp electrophysiology to examine GABA-evoked responses in HEK293 cells expressing the fluorescent tag EGFP. Individual cells were voltage-clamped to -60 mV with an Axon Axopatch 200B amplifier (Molecular Devices, Sunnyvale, CA, USA). Their membrane currents were digitized by the NI USB-6341 acquisition system (National Instruments, Austin, TX, USA), and the data were recorded and analyzed using the WinWCP software (University of Strathclyde,

Glasgow, UK). Borosilicate glass capillaries (outer diameter 1.5 mm, inner diameter 1.12 mm, World Precision Instruments, Sarasota, FL, USA) were fabricated utilizing a P-87 Flaming/Brown-type micropipette puller (Sutter Instruments, Novato, CA). The micropipettes had a final resistance of 3-5 MΩ after filling with a cesium-based intracellular solution (in mM): 50 CsCl, 10 HEPES, 2 $MgCl_2$, 2 MgATP, and 1.1 EGTA, with a pH set at 7.2.

The extracellular HEPES-buffered Krebs solution consisted of (in mM): 140 NaCl, 2.52 $CaCl_2$, 4.7 KCl, 1.2 $MgCl_2$, 5 HEPES, and 11 D-glucose, with a pH set at 7.4. This solution was continuously perfused over the cells at a rate of 5 mL/min through a gravity-fed delivery system at room temperature ($21 \pm 1°C$). On experiment days, stock solutions of GABA and humulone were diluted in Krebs solution to 1 μM and 10 μM, respectively. The compounds were applied to the solution bath for 60 seconds, followed by a 120-second washout period between applications. The peak current of each cell was determined by measuring the increase from its baseline level to its maximum response and subsequently normalizing the obtained values to those evoked solely by GABA.

## 4.7   Behavioral studies (III)

### 4.7.1   Pentobarbital and ethanol sleep models

The behavioral models investigated the sleep-promoting effects of humulone in adult male BALB/cAnNRj mice. During the pre-treatment phase, mice received intraperitoneal (i.p.) injections of humulone at doses of either 10 mg/kg or 20 mg/kg. Humulone was dissolved in a solution containing 40% propylene glycol and 5% Tween 80. As a control, a vehicle solution comprising 40% propylene glycol and 5% Tween 80, devoid of humulone, was administered. Following a 45-minutes to allow for absorption and distribution, sleep was induced via i.p. dose of sodium pentobarbital (35 mg/kg) or ethanol (EtOH) (3.5 g/kg), both of which were diluted in 0.9% saline solution. Two key measurements were quantified using a chronometer to assess the CNS inhibitory effects of humulone. The onset of sleep latency was assigned as the interval between the administration of pentobarbital or EtOH and the point at which the mouse becomes unable to self-right when placed on its back, which is indicative of the loss of the righting reflex. Sleep duration was defined as the interval between the initial loss of the righting reflex and its subsequent recovery. Confirmation of the recovery of the righting reflex was established when the mouse successfully self-righted three consecutive times within one minute. The drug solutions were compounded immediately before administration, and a uniform total injection volume of 13 mL/kg was consistently maintained across all experimental conditions.

### 4.7.2 Open field test

The open field paradigm was employed to examine spontaneous locomotor behavior in humulone-exposed mice as a reflection of sedative effects. Adult male BALB/cAnNRj mice were randomly allocated to one of three groups and given i.p. injections of 10 or 20 mg/kg humulone or vehicle 45 minutes before the test (control). Subsequently, each mouse was positioned separately in the middle of a well-lit white Plexiglass arena (50x50x38 cm) and monitored non-stop for 15 minutes using an Ethovision XT 13 video tracking tool (Noldus, Wageningen, The Netherlands). For the entire arena and its sub-sections, the locomotor activity parameters of distance and velocity were measured and analyzed. Between subjects, the arena surface was disinfected with 70% ethanol and allowed to dry.

## 4.8 Statistical analysis (I-III)

The radioligand displacement measurements were subjected to statistical analysis and fitting using GraphPad Prism (version 7 for Studies **I** and **II**; version 8 for Study **III**) (GraphPad Software, San Diego, CA, USA). The dose-response data in Studies **I** through **III** were modeled using the below sigmoidal equation, with the Hill slope as a variable parameter. The concentration producing half-maximal inhibition ($IC_{50}$) on the non-competitive binding of [$^3$H]EBOB was determined using nonlinear least-squares regression.

$$Y = \text{Bottom} + \frac{\text{Top} - \text{Bottom}}{1 + \left(\frac{X}{IC50}\right)^{\text{HillSlope}}}$$

Here, Y represents the proportion of binding relative to the control. The term "Bottom" was assigned a value of 0 after subtracting nonspecific binding across all measurements of total binding. "Top" signifies the highest observed radioligand binding when the experimental sample is absent, while "X" indicates the concentration of the experimental sample.

In Study **II**, the $IC_{50}$ value derived from the competitive [$^3$H]flunitrazepam displacement curve underwent conversion into the inhibition constant ($K_i$) through the application of the Cheng-Prusoff equation (Cheng and Prusoff, 1973):

$$Ki = \frac{IC50}{1 + \left(\frac{[L]}{KD}\right)}$$

In this equation, "[L]" denotes the molar concentration of the radioligand employed in the assay ([$^3$H]flunitrazepam, 1 nM), while "$K_D$" represents the equilibrium dissociation constant of the radioligand, determined to be 1.1 nM (Bosmann et al., 1978).

To assess the statistical significance between two experimental groups, the Student's *t*-test was employed. In instances involving multiple comparisons, a one-way analysis of variance (ANOVA) was conducted, paired with an appropriate *post hoc* analysis (Dunnett or Tukey). All data were displayed as mean values ± standard error of the mean (SEM), and the significance of difference was established for p-values < 0.05.

# 5 Results and Discussion

## 5.1 Hop flavonoids (I)

Xanthohumol, the primary prenylflavonoid discovered in hops (0.2–1.1%, w/w) (Magalhães et al., 2011), along with its derivatives, isoxanthohumol and 8-prenylnaringenin, have attracted considerable scientific interest owing to their diverse array of biological effects against pathogenic microbes, oxidative stress, neuronal damage, inflammatory responses, and carcinogenesis (Botta et al., 2005; Yang et al., 2015; Wang et al., 2020). For example, the antioxidant properties of xanthohumol have been proposed to mitigate some of the negative consequences associated with binge drinking by protecting various tissues from ethanol-induced damage (Pinto et al., 2014; Elrod, 2018). Furthermore, isoliquiritigenin, a flavonoid sourced from licorice roots, is structurally similar to xanthohumol owing to its shared chalcone backbone. Experimental evidence has shown that isoliquiritigenin positively modulates GABA$_A$ receptors, displaying hypnotic and anxiolytic effects in mouse models (Jamal et al., 2008; Cho et al., 2011). Therefore, hop flavonoids deserve further investigation for their modulatory potential on GABA$_A$ receptors, as various naturally occurring flavonoids have been shown to mediate their neuroactivity through distinct GABA$_A$ receptor sites that are either sensitive or insensitive to flumazenil (Hanrahan et al., 2011; Johnston, 2015). In Study **I**, we examined the modulatory influence and selective targeting of hop flavonoids on native and recombinant GABA$_A$ receptors using radioligand binding assays. Subsequently, we probed whether the active flavonoids exhibit their modulation by interacting with GABA$_A$ receptor's benzodiazepine site.

### 5.1.1 Allosteric modulation of [$^3$H]EBOB binding to native GABA$_A$ receptors

We initially examined the displacement of [$^3$H]EBOB binding to rat forebrain membranes using $30\ \mu M$ of various prenylated and non-prenylated hop flavonoids with the co-application of $2\ \mu M$ GABA. The results revealed distinct binding characteristics among the tested compounds (Figure 18A). Notably, 8-prenylnaringenin, isoxanthohumol, and xanthohumol inhibited specific [$^3$H]EBOB binding by $97.6 \pm 0.3\%$, $85.2 \pm 0.8\%$, and $55.6 \pm 2.3\%$, respectively (Figure 18A). Additionally, when excluding GABA from sufficiently washed membranes, only 8-prenylnaringenin (30 µM) showed significant displacement, accounting for $51.3 \pm 3.9\%$ of [$^3$H]EBOB specific binding ($p < 0.001$) (Figure 18B). This suggests that 8-prenylnaringenin might induce direct activation of GABA$_A$ receptors, comparable to barbiturates, particularly at high doses (Sieghart, 1995). However, we cannot exclude the possibility of 8-prenylnaringenin potentiating the effect of residual trace amounts of endogenous GABA remaining in the brain membranes after washing.

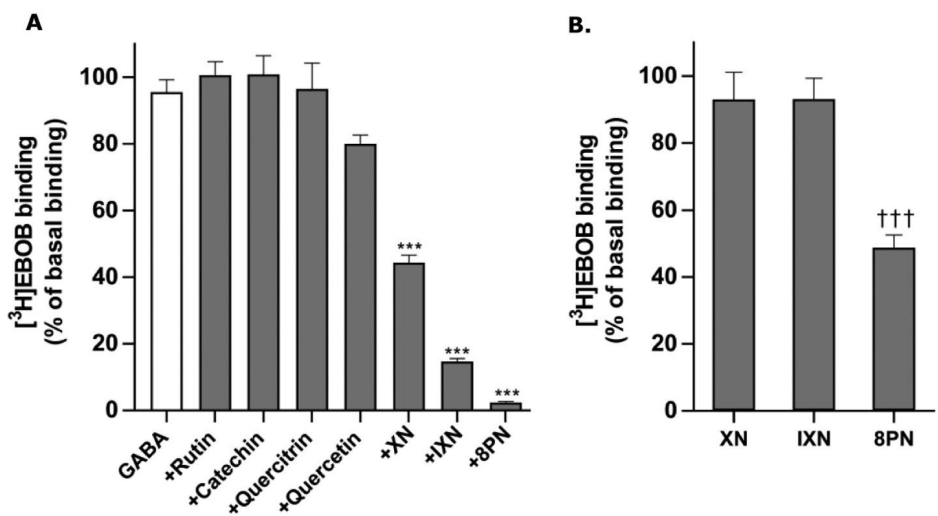

**Figure 18.** Modulation of various hop flavonoids (30 µM) on 1nM [$^3$H]EBOB binding to rat forebrain membranes with (A) or without (B) 2 µM GABA. Data points denote the mean ± SEM from three distinct experiments, each conducted in triplicate. Significance markers: ****p* < 0.001 indicates a pronounced inhibition of [$^3$H]EBOB binding compared to the basal binding with GABA only; †††*p* < 0.001 underscores the difference from GABA-free basal binding. Determined by one-way ANOVA paired with Dunnett's multiple comparison test.

Subsequently, we carried out a concentration series for the displacement of [$^3$H]EBOB binding with the co-application of 2 µM GABA for the three active prenylflavonoids. Xanthohumol and its structurally related compounds inhibited the binding of [$^3$H]EBOB binding to rat forebrain in a concentration-dependent fashion (Figure 19). This observation aligns with previous reports indicating that xanthohumol enhances muscimol-Alexa Fluor 532 (Mu-Alexa) binding to GABA$_A$ receptors in hippocampal neurons and inhibits triggered glutamate release through its activity on GABA$_A$ receptors (Meissner and Häberlein, 2006; Chang et al., 2016).

When examining the GABA$_A$ receptors' sensitivity to prenylflavonoid-mediated shifts in [$^3$H]EBOB binding, the sequence was as follows: 8-prenylnaringenin exerted the most influence, then isoxanthohumol, and lastly xanthohumol. With just 2 µM of GABA, a $10.2 \pm 1.6\%$ displacement in the basal binding of 1 nM [$^3$H]EBOB was observed ($p < 0.01$). However, co-application of 3 µM of either 8-prenylnaringenin or isoxanthohumol with GABA resulted in reductions of [$^3$H]EBOB binding by $26.6 \pm 2.8\%$ and $19.9 \pm 2.5\%$, respectively ($p < 0.001$ for both). In contrast, xanthohumol, at the same concentration, did not significantly alter the GABA-mediated displacement of [$^3$H]EBOB. The calculated IC$_{50}$ of these prenylflavonoids, when assessed in forebrain tissue samples, were at low micromolar values (Table 2). Compared to xanthohumol, isoxanthohumol exhibited 2.5-fold, and 8-prenylnaringenin 4.1-fold, greater inhibitory potency ($p < 0.001$). Similar low micromolar modulation was observed with natural prenylflavonoids from *Sophora flavescens* (Kushen), where kushenol I, sophoraflavanone G, (−)-kurarinone, and kuraridine increased GABA-evoked currents in recombinant α1β2γ2 GABA$_A$ receptors (Yang et al., 2011).

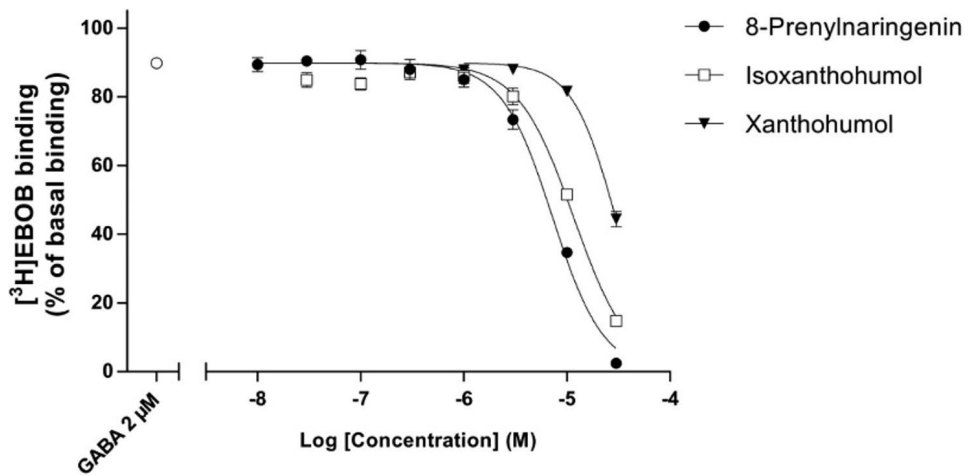

**Figure 19.** Dose-dependent effect of hop prenylflavonoids on 1 nM [$^3$H]EBOB binding to rat forebrain membranes, co-applied with 2 µM GABA. Data points denote the mean ± SEM from three distinct experiments, each conducted in triplicate.

**Table 2.** Potency assessment of prenylflavonoid enhancement of GABA-mediated shifts in the binding of [$^3$H]EBOB to native GABA$_A$ receptors

| Prenylflavonoid | IC$_{50}$ |
| --- | --- |
| Xanthohumol | $29.7 \pm 0.8$ µM |
| Isoxanthohumol | $11.6 \pm 0.7$ µM*** |
| 8-Prenylnaringenin | $7.3 \pm 0.4$ µM*** |

IC$_{50}$ are shown as mean values ± SEM (n = 3), derived via nonlinear regression analysis of the inhibition curves. ***$p < 0.001$, indicating a significant difference between the value of xanthohumol and the other prenylflavonoids, as determined by one-way ANOVA paired with Tukey's *post hoc* analysis.

## 5.1.2 Modulation of [$^3$H]EBOB binding to recombinant GABA$_A$ receptors

Consistent with the findings on native GABA$_A$Rs, we noted that hops prenylflavonoids enhanced the displacement of [$^3$H]EBOB binding to recombinant GABA$_A$Rs expressed in HEK293 cells in a concentration-dependent fashion. The concentration curves in Figure 20 highlight the impact of prenylflavonoids on the displacement of [$^3$H]EBOB binding in the various tested recombinant GABA$_A$R subtypes, and Table 3 presents the corresponding IC$_{50}$ values. The responsiveness of these subtypes was ranked as follows: α6β3δ > α2β3γ2 > α1β3γ2. Interestingly, our findings indicate that the modulatory activities of hop prenylflavonoids are not contingent on the presence of the γ or δ subunit. This suggests the existence of an allosteric binding site located at the α/β interfaces of the pentameric complex of the GABA$_A$R. Molecular docking experiments were conducted in the subsequent study to further validate this hypothesis and determine the probable binding sites and orientations of hop prenylflavonoids on the α1β2γ2 GABA$_A$R isoform (**II**).

Notably, hop prenylflavonoids displayed selectivity towards specific receptor subtypes. For instance, the inhibition potency of isoxanthohumol and 8-prenylnaringenin in the α6β3δ subtype was significantly higher (2.6-fold) than in the α1β3γ2 receptor subtype. In α6β3δ receptors, 8-prenylnaringenin displayed remarkable potency with an IC$_{50}$ of $3.6 \pm 0.5$ µM (n = 4). It was shown that the enhancement of the GABA-mediated response in αβδ subtypes was responsive to ethanol concentrations at low millimolar levels (3–30 mM), which typically elicit mild to moderate intoxication in humans (Sundstrom-Poromaa et al., 2002; Wallner et al., 2003; Hanchar et al., 2005). This implies that prenylflavonoids with potent positive modulatory effects on these particular subtypes may amplify the effects of ethanol on the CNS. In other words, the ingestion of heavily hopped beer could result in a more pronounced state of intoxication compared to unhopped beer containing equal ethanol concentration.

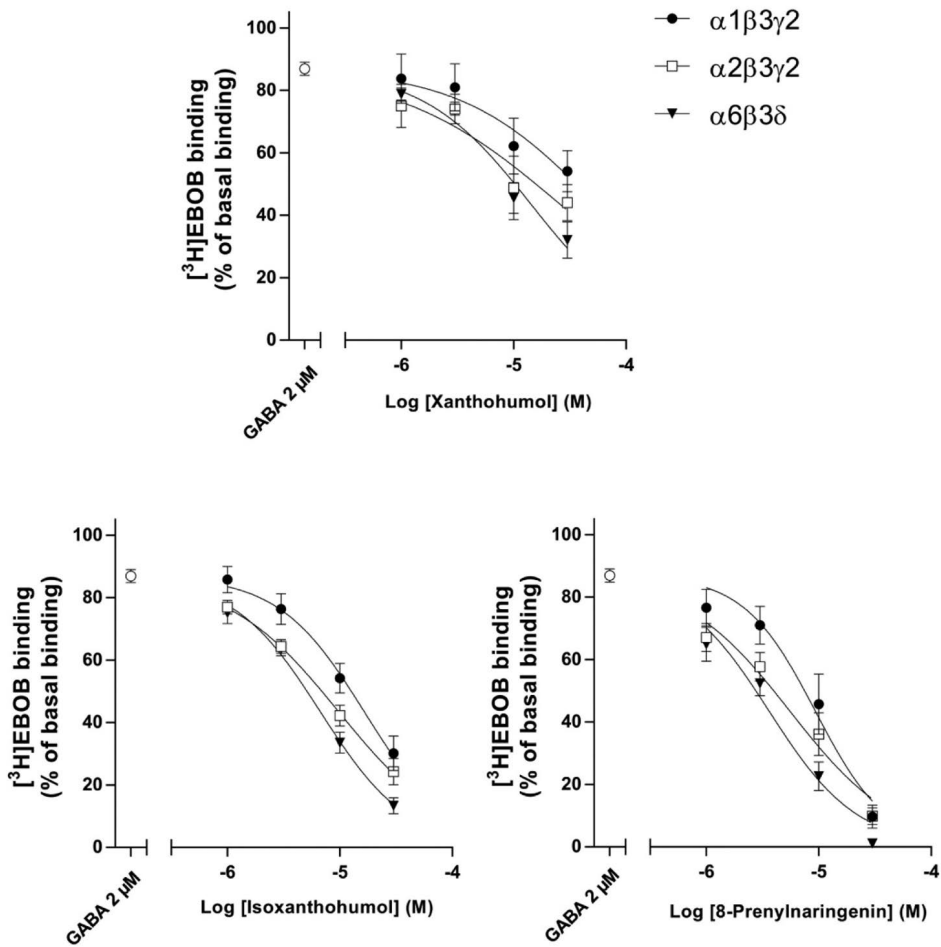

**Figure 20.** Dose-response relationship (1–30 µM) of xanthohumol, isoxanthohumol and 8-prenylnaringenin on the binding of 2 nM [$^3$H]EBOB to GABA$_A$Rs expressed in HEK293 cells when co-applied, with 2 µM GABA. Data points represent the mean ± SEM from four distinct experiments, each conducted in quadruplicate.

**Table 3.** Potency assessment of prenylflavonoid enhancement of GABA-mediated [$^3$H]EBOB displacement in HEK293 expressed recombinant GABA$_A$ receptors

|                        | α1β3γ2        | α2β3γ2      | α6β3δ        |
|------------------------|---------------|-------------|--------------|
| **Xanthohumol**        | 54 ± 31 µM    | 26 ± 11 µM  | 14 ± 2.1 µM  |
| **Isoxanthohumol**     | 16.5 ± 2.3 µM | 9.8 ± 1 µM  | 6.7 ± 0.6 µM |
| **8-Prenylnaringenin** | 9.3 ± 1.6 µM  | 5.5 ± 0.9 µM| 3.6 ± 0.5 µM |

The IC$_{50}$ values are shown as mean ± SEM (n = 4), determined through nonlinear regression fitting of the inhibition curves.

### 5.1.3  Flumazenil insensitivity to prenylflavonoid-induced [$^3$H]EBOB modulation

We assessed the potential of flumazenil to counteract the enhancement of GABA-mediated [$^3$H]EBOB displacement by prenylflavonoids at the classical benzodiazepine site (Figure 21). While hop prenylflavonoids increased the GABA-mediated displacement at benzodiazepine-sensitive subtypes ($\alpha2\beta3\gamma2$ and $\alpha1\beta3\gamma2$), this effect was unresponsive to flumazenil antagonism ($p > 0.05$). These observations are consistent with those for 6-methoxyflavanone, a synthesized flavonoid with an identical flavanone backbone as isoxanthohumol and 8-prenylnaringenin, which was found to positively modulate $\alpha1/2\beta2\gamma2$ GABA$_A$ receptor subtypes without sensitivity to flumazenil (Hall et al., 2014). Moreover, evidence from radioligand binding assays indicated that naringenin, a natural flavanone extracted from Mentha aquatic, exhibited a weak displacement of the binding of [$^3$H]flumazenil in rat cerebral cortical membranes, with an IC$_{50}$ value of 2.6 mM (Jäger et al., 2007).

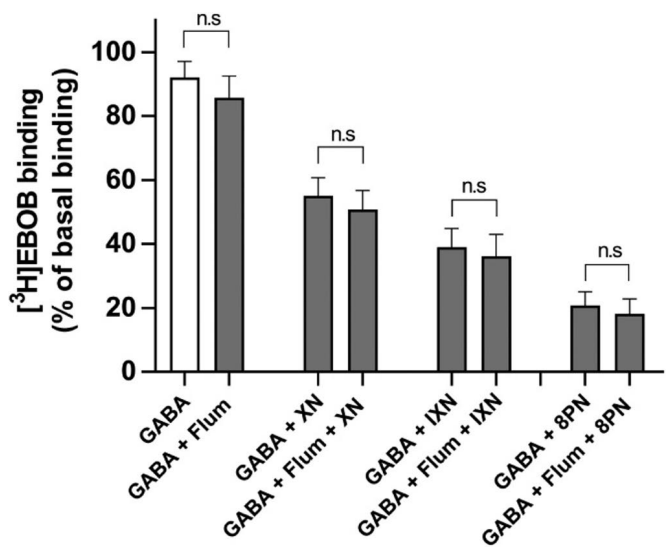

**Figure 21.** Effect of xanthohumol (XN) at 30 µM, isoxanthohumol (IXN) at 10 µM and 8-prenylnaringenin (8PN) at 10 µM on the binding of 1nM [$^3$H]EBOB to rat forebrain membranes, co-applied with 2 µM GABA and 2 µM flumazenil present. Data points denote the mean ± SEM from three distinct experiments, each conducted in triplicate. The label "n.s", indicates negligible difference from the respective binding without flumazenil, as assessed by unpaired *t*-test.

## 5.1.4 Competitive displacement of [³H]Ro 15–4513 binding at the benzodiazepine site

To ascertain whether the hops prenylflavonoids bind to the classical benzodiazepine site, we conducted [³H]Ro 15-4513 binding assays in rat forebrain membranes. The findings revealed that none of the tested prenylflavonoids displaced the radioligand at a concentration of 10 μM (Figure 22). Only at higher micromolar concentrations (100 μM) was there evidence of radioligand displacement. Specifically, 8-prenylnaringenin, isoxanthohumol, and xanthohumol, which belong to the flavanone subclass, inhibited 46.1%, 20.0%, and 12.9% of [³H]Ro 15-4513 specific binding, respectively (p < 0.001). This finding is noteworthy, considering that the GABA-enhancing activities of prenylflavonoids were evident at much lower micromolar values in the [³H]EBOB assay. Therefore, given the limited displacement of [³H]Ro 15-4513 binding and the lack of sensitivity to flumazenil antagonism, we propose that the positive modulatory actions of hop prenylflavonoids do not occur through the classical benzodiazepine binding site.

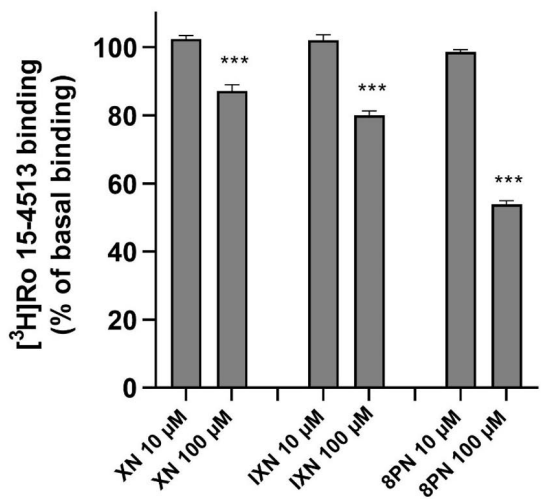

**Figure 22.** Influence of xanthohumol (XN), isoxanthohumol (IXN), and 8-prenylnaringenin (8PN) at 10 μM and 100 μM doses on the binding of 2 nM [³H]Ro 15-4513 2 nM to rat forebrain membranes. Data points represent the mean ± SEM from three distinct experiments, each conducted in quadruplicate. ***$p < 0.001$ indicates a significant inhibition of [³H]Ro 15-4513 binding compared to the basal binding, determined by one-way ANOVA paired with Dunnett's multiple comparison test.

## 5.2    Hop α/β-acid fractions, volatiles and derivatives (II)

The primary constituents of hop resins are α- and β-acids, comprising 5–13% and 3–8% of the dried weight of hops, respectively (Van Cleemput et al., 2009; Almaguer et al., 2014). Research has shown that α-acids possess sedative-hypnotic characteristics (Zanoli et al., 2005; Schiller et al., 2006), likely stemming from their capacity to modulate GABA$_A$ receptor function (Aoshima et al., 2006; Schiller et al., 2006; Sahin et al., 2016). However, the specific compounds responsible for this modulation remain unidentified. Through the oxidation process during storage, α- and β-acids transform into 2-methyl-3-buten-2-ol (2M3B), a volatile tertiary alcohol present in minute proportions (<0.01%) in commercial hop extracts (Hänsel et al., 1982). Although 2M3B has been found to induce sedation and narcosis in rodents (Hänsel et al., 1980; Wohlfart et al., 1983), the mechanism underpinning its brain-specific effects remains unexplored.

In the context of beer brewing, desmethylxanthohumol (DMX) undergoes non-enzymatic isomerization, resulting in the formation of 6-prenylnaringenin and, to a minor degree, 8-prenylnaringenin, which is attributed to its limited stability in aqueous conditions (Miles and Main, 1985; Hänsel and Schulz, 1988; Chadwick et al., 2006). Although it is present in hops at modest concentrations (0.01%) (Almaguer et al., 2014), 6-prenylnaringenin has the capability to traverse the BBB and confer analgesic effects through its role as a T-type calcium channel blocker (Du Nguyen et al., 2018; Sekiguchi et al., 2018). Additionally, linalool, a prominent component of hop essential oils, imparts a floral and fruity essence to beer and is detectable in quantities as high as 150 mg/kg within the hop oil composition (Štěrba et al., 2015). This monoterpene alcohol has been shown to induce hypnotic, sedative, and anxiolytic effects in rats, while preserving motor coordination (Linck et al., 2010; Guzman-Gutierrez et al., 2012).

To gain deeper insights into the mechanisms underlying the sedative and sleep-promoting attributes of hops, Study **II** investigated isolated α- and β-acid fractions of hops (F2–F7.2), in addition to 2-methyl-3-buten-2-ol, 6-prenylnaringenin, and linalool, to evaluate their potential for modulatory interactions with GABA$_A$ receptors.

### 5.2.1    Allosteric modulation of [³H]EBOB binding to native GABA$_A$ receptors

The modulation of [³H]EBOB binding to rat forebrain and cerebellum was initially studied using hop compounds at a concentration of 30 µM in the presence of 2 µM GABA (Figure 23A, B). With the exception of adhumulone (F4) and 2M3B in the forebrain, all the compounds tested showed a significant increase in GABA-mediated displacement of [$^3$H]EBOB. The measured displacement range of [$^3$H]EBOB was 19.7–80% in rat forebrain membranes and 6.2–92.4% in cerebellar membranes. Previous electrophysiological experiments with HEK293 cells and Xenopus oocytes have revealed that linalool, at concentrations of 0.3–2 mM, enhances GABA-evoked responses in recombinant GABA$_A$ receptors (Hossain et al., 2002; Aoshima et al., 2006; Kessler et al., 2014; Milanos et al., 2017). However, our radioligand binding data indicates that linalool exhibits high sensitivity to native GABA$_A$ receptors, with enhancement detected at concentrations as low as 30 μM. The observed dose-based discrepancy could potentially be explained by two factors. Firstly, the previous studies employed recombinant expression systems, which may not fully replicate the complexity and heterogeneity of native GABA$_A$ receptors in the brain. Secondly, linalool's volatile nature and its susceptibility to oxidation upon exposure to air could be another factor influencing the observed differences (Sköld et al., 2004). Nevertheless, our findings confirm the potential involvement of GABA$_A$ receptors in mediating the sedative and anxiolytic effects of linalool (Linck et al., 2010; Guzman-Gutierrez et al., 2012), suggesting further *ex vivo* studies to comprehensively elucidate its mechanism of action.

In the rat forebrain, the most potent compounds were humulone (F3), F5, and 6-prenylnaringenin (6PN), which displayed remarkable displacements of 80.0%, 80.3%, and 72.9% on the binding of [$^3$H]EBOB, respectively ($p < 0.001$). Interestingly, these compounds exhibited distinct activities in the cerebellum at the same 30 μM concentration ($p < 0.001$). Specifically, humulone (F3), F5, and 6-prenylnaringenin (6PN) inhibited 55.6%, 50.2%, and 92.4% of GABA-mediated displacement of [$^3$H]EBOB binding to the cerebellum, respectively ($p < 0.001$). Notably, 6-prenylnaringenin (6PN) exerted a substantial inhibitory effect on this displacement in the cerebellar compared to forebrain membranes. Previous research involving recombinant GABA$_A$ receptors expressed in HEK293 cells indicated that structurally related prenylflavonoids, such as 8-prenylnaringenin and isoxanthohumol, potently modulate α6β3δ subtype, particularly in comparison to α1β3γ2 (**I**). Given the predominant expression of the δ subunit in the form α6β2/3δ subtype within cerebellar granule cells, our findings suggest that 6-prenylnaringenin may exhibit a greater modulatory potency on the extrasynaptic δ-GABA$_A$Rs.

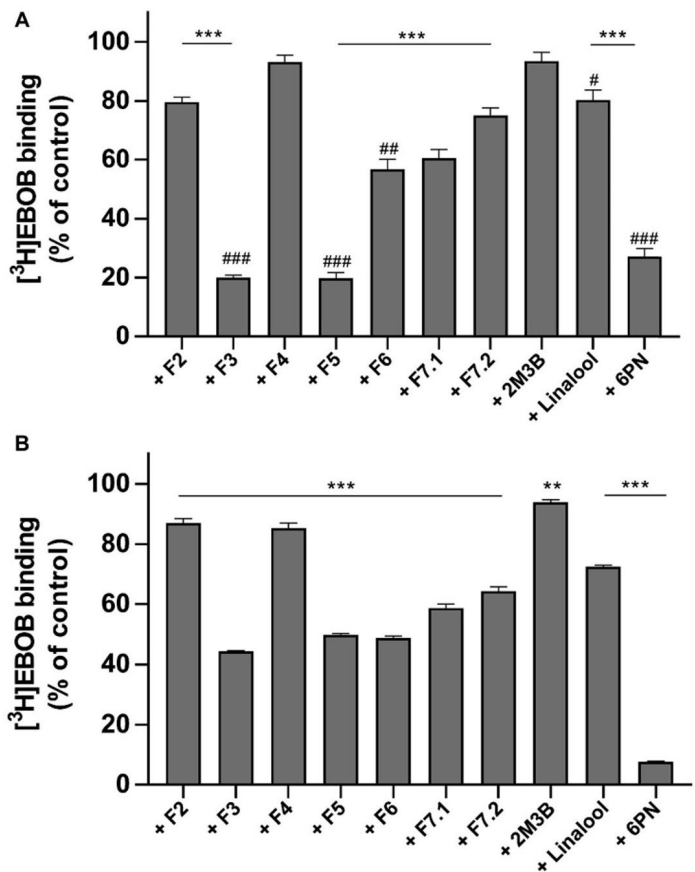

**Figure 23.** Effect of various hop constituents (30 µM) on the binding of [$^3$H]EBOB (1 nM) to rat forebrain (A) and cerebellar (B) membranes, co-applied with 2 µM GABA. The control is defined by the maximal [$^3$H]EBOB binding with GABA only. Data points denote the mean ± SEM from 3–5 distinct experiments, each conducted in triplicate. Significance markers: ***$p < 0.001$, **$p < 0.01$ indicate pronounced radioligand inhibition compared to the control, as determined by one-way ANOVA and Dunnett's multiple comparison test. Furthermore, ###$p < 0.001$, ##$p < 0.01$, #$p < 0.05$ signify differences between the forebrain and cerebellar [$^3$H]EBOB binding, determined by unpaired $t$-test.

Furthermore, we determined the direct effects of humulone (F3), F5, lupulone (7.1), methyl-3-buten-2-ol (2M3B), linalool, and 6-prenylnaringenin (6PN) on the binding of [$^3$H]EBOB to native GABA$_A$ receptors in the absence of GABA (Figure 24). Among these compounds, only 2M3B exhibited a notable effect, displacing 14.3% of specific [$^3$H]EBOB binding ($p < 0.01$) at a concentration of 30 µM. This modest yet measurable displacement by 2M3B suggests its potential to directly activate GABA$_A$ receptors. This mechanism may offer an explanation for the sedative and narcotic attributes associated with 2M3B, despite having been demonstrated only at high doses ranging from 200 to 800 mg/kg, which surpasses its natural concentration in hops (Hänsel et al., 1980; Wohlfart et al., 1983a, 1983b).

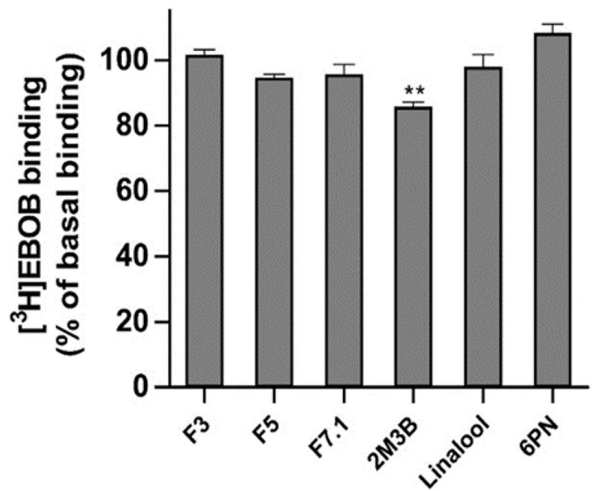

**Figure 24.** Effect of selected hop constituents (30 µM) on the binding of [$^3$H]EBOB (1 nM) to rat forebrain membranes without GABA. Data points represent the mean ± SEM from three distinct experiments, each conducted in triplicate. **p < 0.01 indicates a significant binding inhibition compared to the GABA-free baseline, as determined by one-way ANOVA paired with Dunnett's multiple comparison test.

We subsequently evaluated the compounds that exhibited substantial enhancement in GABA-mediated displacement of [$^3$H]EBOB across a concentration series. This experiment aimed to assess their potency to inhibit the binding of [$^3$H]EBOB during simultaneous GABA exposure. In rat forebrain membranes, 6-prenylnaringenin, humulone (F3), and F5 displayed pronounced modulatory effects, potentiating GABA-induced [$^3$H]EBOB displacement at low micromolar levels (Figure 25). For example, at a concentration of 1 µM, 6-prenylnaringenin and humulone (F3) led to displacements of $35.2 \pm 0.5\%$ (p < 0.001) and $39.6 \pm 1.6\%$ (p < 0.001), respectively, while the influence of F5 was lacking at 3 µM and lower. The estimated IC$_{50}$ for 6-prenylnaringenin, humulone (F3), and F5 were $3.7 \pm 0.4$ µM, $3.2 \pm 0.4$ µM, and $18.2 \pm 0.4$ µM, respectively. Based on these outcomes, it can be inferred that 6-prenylnaringenin emerges as the most potent among the hop prenylflavonoids, functioning as a positive allosteric modulator of GABA$_A$ receptors. Although 6-prenylnaringenin exists in hops in trace amounts (0.01%) (Almaguer et al., 2014), it is conceivable that it interacts synergistically or additively with abundant modulators like humulone, thereby amplifying the overall behavioral effects associated with hops.

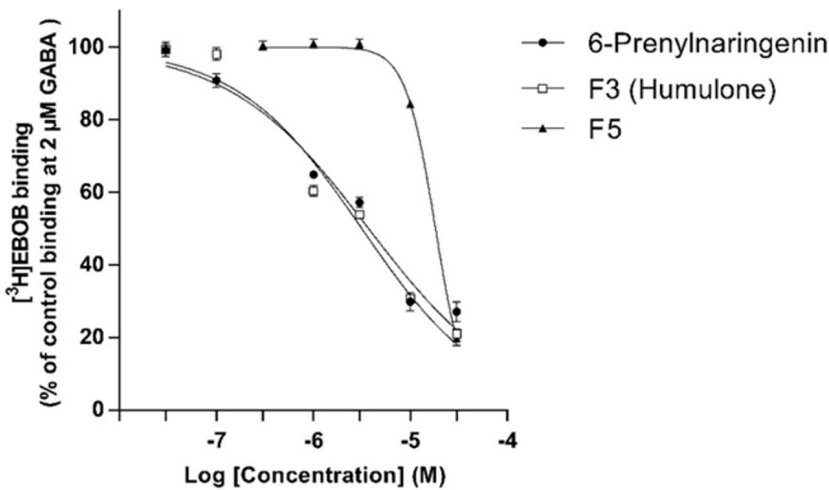

**Figure 25.** Influence of various doses of 6-prenylnaringenin, humulone (F3), and F5 on the binding of 1 nM [$^3$H]EBOB to rat forebrain membranes, co-applied with 2 μM GABA. The control is defined by the specific [$^3$H]EBOB binding with GABA only. Data points represent the mean ± SEM from three distinct experiments, each conducted in triplicate.

### 5.2.2    Resistance to flumazenil in [$^3$H]EBOB binding modulation

We probed if the benzodiazepine site sensitive to flumazenil is a determinant for the submicromolar modulatory effects of hop constituents. While flumazenil evidently antagonized diazepam's allosteric modulation (**$p < 0.01$), it did not block the potentiation of GABA-mediated displacement of [$^3$H]EBOB by 6-prenylnaringenin, humulone (F3), F5, and lupulone (7.1) in native GABA$_A$ receptors (Figure 26). Interestingly, the addition of flumazenil resulted in a marginal but statistically significant increase in the displacement of [$^3$H]EBOB, underscoring flumazenil's inherent partial agonism previously described in α4/6-GABA$_A$Rs (Hadingham et al., 1996; Knoflach et al., 1996; Hauser et al., 1997) and several animal studies (Dantzer and Pério, 1982; Kaijima et al., 1983; Vellucci and Webster, 1983; Belzung et al., 2000). However, the observed modulation of [$^3$H]EBOB binding by 6-prenylnaringenin (6PN), humulone (F3), and lupulone (F7.1) remained unaffected by flumazenil antagonistic effect. This suggests that these constituents exhibit this activity via a site distinct from the classical benzodiazepine binding site.

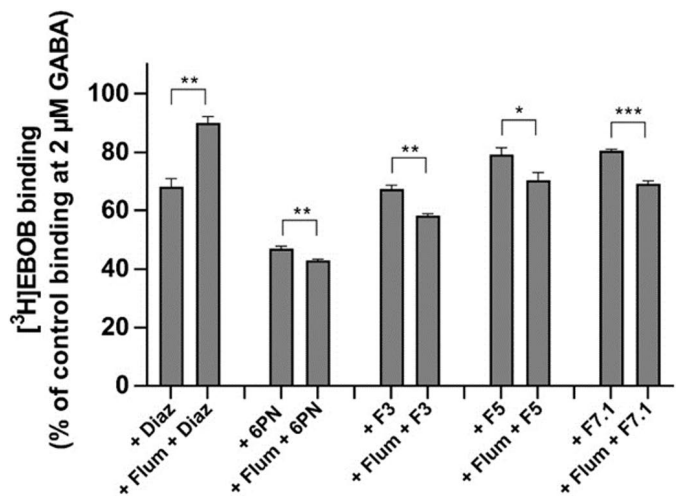

**Figure 26.** Effect of diazepam (Diaz, 0.1 µM), 6-prenylnaringenin (6PN, 5 µM), humulone (F3, 1 µM), lupulone (F7.1, 20 µM) and F5 (10 µM) on the binding of [³H]EBOB (1 nM) to rat forebrain membranes, co-applied with 2 µM GABA and flumazenil (Flum, 2 µM). The control represents the maximal [³H]EBOB binding with GABA only. Data points denotes the mean ± SEM from 3–5 distinct experiments, each conducted in triplicate. Significance markers: ***p < 0.001, **p < 0.01, *p < 0.05 indicate a significant binding inhibition relative to the control, assessed by unpaired *t*-test.

### 5.2.3 Competitive displacement of [³H]flunitrazepam and [³H]Ro 15–4513 binding

Furthermore, we evaluated the influence of hop compounds at a concentration of 30 µM on the binding of [³H]flunitrazepam (Figure 27A) and [³H]Ro 15–4513 (Figure 27B). Both radioligands target the benzodiazepine site of the GABA$_A$ receptor, with the former being a positive allosteric modulator and the latter a partial inverse agonist. Flumazenil (10 µM) was used to measure nonspecific binding for both radioligands. The results indicated a minor displacement of [³H]flunitrazepam by cohumulone (F2) ($15.6 \pm 2.0\%$, $p < 0.01$) and humulone (F3) ($17.0 \pm 4.8\%$, $p < 0.01$). When tested with [³H]Ro 15–4513, cohumulone (F2) and humulone (F3) resulted in displacements of $9.2 \pm 2.3\%$ ($p < 0.01$) and $7.5 \pm 2.0\%$ ($p < 0.01$), respectively. Despite its insensitivity to flumazenil's antagonizing action at the classical benzodiazepine site, 6-prenylnaringenin had a significant effect on the radioligands binding, displacing $47.1 \pm 4.1\%$ ($p < 0.001$) of [³H]flunitrazepam binding and $59.4 \pm 0.48\%$ ($p < 0.001$) of [³H]Ro 15–4513 binding. 6-prenylnaringenin displaced [³H]flunitrazepam in a concentration-dependent fashion (Figure 27C). The determined $K_i$ value for 6-prenylnaringenin was 16.5 µM, notably higher than the calculated $K_i$ value for diazepam in the same binding assay (14 nM) (Forbes et al., 1990). However, these findings revealed that 6-prenylnaringenin was more selective for the classical benzodiazepine binding site than its isomer 8-prenylnaringenin and other hop prenylflavonoids (**I**). This prompted further confirmation with molecular modelling to elucidate the binding selectivity of 6-prenylnaringenin's to $GABA_A$ receptors (Section 5.2.4).

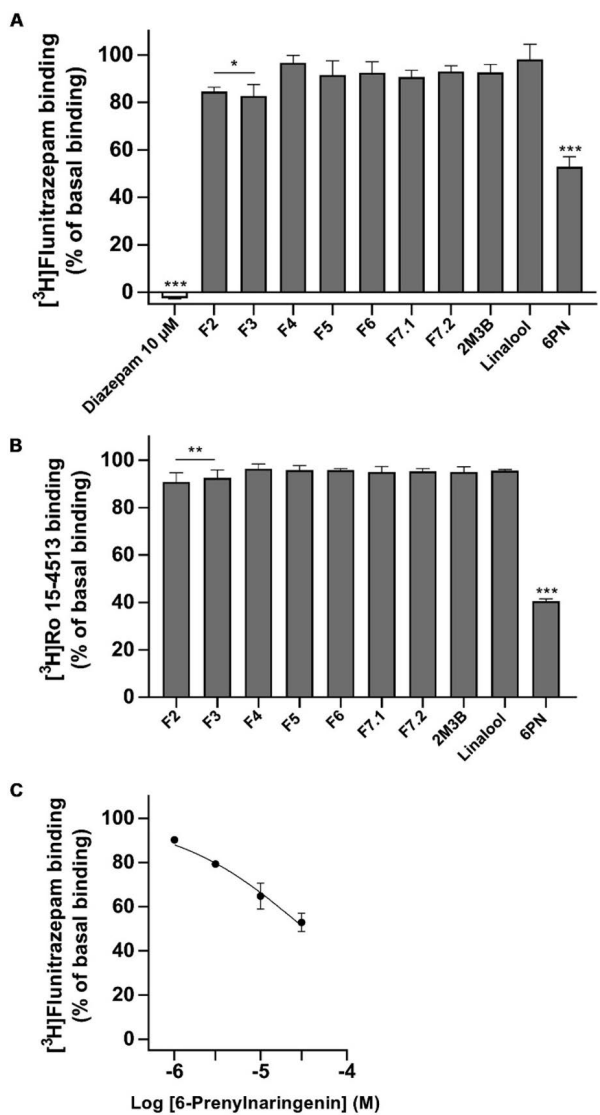

**Figure 27.** Influence of various hop constituents (30 µM) on the binding of (A) [³H]flunitrazepam (1 nM) and (B) [³H]Ro 15–4513 (2 nM) (B) to rat forebrain membranes. Data points represent the mean ± SEM from three distinct experiments, each conducted in quadruplicate. Significance markers: ***$p < 0.001$, **$p < 0.01$, *$p < 0.05$ indicate a significant binding inhibition relative to baseline, as determined by one-way ANOVA and Dunnett's multiple comparison test. (C) Displacement curve of [³H]flunitrazepam (1 nM) binding to rat forebrain membranes across different doses of 6-prenylnaringenin is also represented with data points denoting the mean ± SEM from six distinct experiments, each conducted in triplicate.

## 5.2.4    *In silico* simulations of hop prenylflavonoid binding at GABA_A receptor interfaces

Molecular docking experiments were conducted to investigate the potential binding sites and poses of hop prenylflavonoids at the human GABA_AR $\alpha1\beta2\gamma2$ isoform (PDB ID: 6D6U; Zhu et al., 2018). Flumazenil and CGS 9895 were selected as reference ligands for their binding energies and docking scores to their corresponding binding pockets (Figures 28A and B). The prenylflavonoids exhibited binding energies comparable to those of the reference ligands at the examined binding sites (Table 4, Supplementary Table S1). Particularly, 6-prenylnaringenin showed the most favorable Prime/MM-GBSA free energy of binding at both $\alpha1+/\beta2-$ and $\alpha1+/\gamma2-$ interfaces. The potential binding poses of 6-prenylnaringenin at the $\alpha1+/\gamma2-$ and $\alpha1+/\beta2-$ interfaces are presented in Figs. 23C and D, respectively. Further details regarding the conformation and orientation of 6-prenylnaringenin and other hop prenylflavonoids, along with their comprehensive interaction illustrations, are available in Supplementary Figures S1–S3.

The observed positive modulation in hop prenylflavonoids was not solely dependent on the $\gamma$ or $\delta$ subunit, implying that $\alpha/\beta$ interfaces may form potential sites for the prenylflavonoid binding within the GABA_A receptor pentameric complex (**I**). Previous studies involving point mutations in the mouse receptor's $\alpha1$ subunit revealed that the Tyr209 residue (equivalent to Tyr210 in the human receptor) at the extracellular $\alpha+/\beta-$ interface is a determinant for the positive allosteric modulation of GABA_A receptors by CGS 9895 (Ramerstorfer et al., 2011; Varagic et al., 2013; Maldifassi et al., 2016). In line with our docking results, Tyr210 interacted with CGS 9895 and all tested hop prenylflavonoids, thereby highlighting the $\alpha+/\beta-$ interface as a common allosteric modulatory site.

Docking results further revealed comparable residues and binding energies for 6-prenylnaringenin and flumazenil at the $\alpha1+/\gamma2-$ interface. Both compounds interacted with His102 on the $\alpha1$ subunit, an essential residue for the binding and positive modulation of classical benzodiazepines such as diazepam and alprazolam (Zhu et al., 2018; Masiulis et al., 2019; Kim et al., 2020). The optimal ligand poses of 6-prenylnaringenin displayed favorable aromatic stacking and hydrogen-bond interactions, aligning with a previously established pharmacophore model for flavonoids (Huang et al., 2001). This model comprises three features: a hydrophobic ring system, a hydrogen bond acceptor group, and a hydrogen bond donor group. These features correspond to the aromatic ring, the carbonyl oxygen, and the hydroxyl groups of 6-prenylnaringenin, respectively.

In comparison to its isomer 8-prenylnaringenin and other hop prenylflavonoids, 6-prenylnaringenin exhibited higher selectivity for the classical benzodiazepine binding site, as confirmed by [$^3$H]Ro 15–4513 and [$^3$H]flunitrazepam binding assays (**I, II**). This supports the notion that 6-prenylnaringenin and flumazenil bind to comparable sites at the α1+/γ2− subunit interface. However, it is important to note the resistance of 6-prenylnaringenin's positive modulatory action to flumazenil antagonism, specifically in relation to [³H]EBOB binding (Figure 26). Therefore, we propose that 6-prenylnaringenin also functions as a silent modulator, potentially blocking the action of benzodiazepines by competing for binding to their corresponding site at the α1+/γ2− interface. Notably, 6-prenylnaringenin significantly enhances GABA-induced responses via α+/β− interface, remaining unaffected allosterically by its silent modulation at the classical benzodiazepine site. This dual mode of action resembles CGS 9895, which interacts with GABAₐ receptors at two extracellular binding sites, acting as a positive allosteric modulator at the α1+/β− subunit interface and a silent modulator at the α1+/γ2− subunit interface (Ramerstorfer et al., 2011).

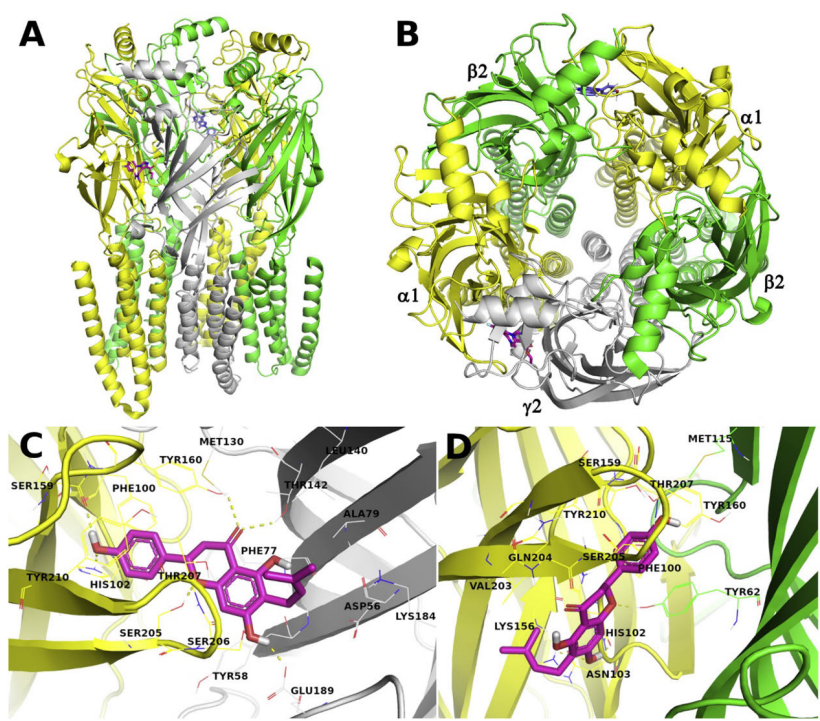

**Figure 28.** The α1β2γ2 GABAₐ receptor isoform (PDB ID: 6D6U) binding sites: α1+γ2- and α1+β2-. The subunits are colored differently in a simplified representation: yellow for α1, grey for γ2, green for β2. A) The receptor from the side with flumazenil's experimental pose (magenta sticks) bound at the α1+γ2- site and CGS 9895 (marine blue sticks) docked at the α1+β2- site. B) Overhead view of the receptor with the same reference ligands as in A). C) The α1+γ2- site with the predicted pose of 6-prenylnaringenin (magenta sticks). D) The α1+β2- site with the predicted pose of 6-prenylnaringenin (magenta sticks). Lines and labels show the amino acid residues that interact with the compounds. Nonpolar hydrogens are not shown for clarity. The atom color scheme is: carbon – magenta, nitrogen – blue, oxygen – red, fluorine – light blue, sulfur – dark yellow; hydrogen – white. Yellow dashed lines show polar interactions.

**Table 4.** Analysis of 6-prenylnaringenin docking at α1+/γ2- and α1+/β2- binding sites of the α1β2γ2 GABA$_A$ receptor experimental structure (PDB ID: 6D6U)

| Ligand | Interface α1+/γ2- | | | Interface α1+/β2- | | | |
| --- | --- | --- | --- | --- | --- | --- | --- |
| | Glide XP score (kcal/mol) | MM-GBSA ΔG bind (kcal/mol) | Key interacting residues from subunits α1/γ2[d] | Glide XP score (kcal/mol)[e] | IFD Score[f] | MM-GBSA ΔG bind (kcal/mol)[g] | Key interacting residues from subunits α1/β2[h] |
| 6-prenylnaringenin | -8.255 | -71.13 | His102, Ser159, Tyr160, Ser205, Ser206, Tyr210/ Thr142, Glu189 | -4.356 | -3002.55 | -29.95/-67.11 | Phe100, His102, Lys156, Gln204, Ser205, Tyr210/ Tyr62 |
| Flumazenil | -9.001[a] / -6.494[b] | -57.95[a]/ -64.71[b] | Phe100, His102, Ala161, Thr207, Tyr210/ Thr142 | N/A | N/A | N/A | N/A |
| CGS 9895 | N/A[c] | N/A | N/A | -2.246 | -2992.31 | -23.33/-69.64 | His102, Ser206, Tyr210/ Tyr62 |

[a] The selected docking pose with the top-scoring XP orientation which is rotated roughly 180° along the principal axis of the molecule relative to the experimental pose.

[b] The selected docking pose closely matches the experimental pose. In the original experimental pose, the MM-GBSA energy was -57.52 kcal/mol. During this initial pose, only the hydrogen atoms added to the receptor were minimized, while the heavy atoms remained fixed.

[c] Not applicable.

[d] HIS102 refers to the histidine residue at position 102 in the protein sequence with a proton at the epsilon nitrogen atom.

[e] Rigid binding site conformation (non-optimal).

[f] The composite score for induced fit docking was calculated as GlideScore + 0.05 x PrimeEnergy.

[g] Prime/MM-GBSA free energy of binding was computed for the corresponding Glide XP/induced fit docked poses.

[h] HIS102 refers to the histidine residue at position 102 in the protein sequence with a proton at the delta nitrogen atom.

## 5.3    Humulone as the principal α-acid in hops (III)

Humulone constitutes the predominant α-acid present in hops, accounting for 35–70% of total α-acids (Karabín et al., 2016). Its solubility in beer is approximately 14 mg/L, with concentrations reaching 28 mg/L (Fritsch and Shellhammer, 2007). In our previous investigation aimed at elucidating the specific constituents contributing to the sedative and sleep-promoting effects of hops, we identified distinct components that exhibited varying effects on the binding of [$^3$H]EBOB to native GABA$_A$ receptors (Study **II**). Among all the compounds tested, the humulone fraction demonstrated potent modulatory activity in rat forebrain membranes at low micromolar levels (IC$_{50}$ = 3.2 ± 0.4 μM). Notably, this effect exhibited resistance to flumazenil antagonism, paired with modest selectivity for the classical benzodiazepine binding site (Study **II**).

Given these findings, it became imperative to verify humulone's capacity to modulate the function of GABA$_A$ receptors through electrophysiological means, and to examine its potential role in hops' behavioral effects. Considering that the primary source of humulone intake in humans stems from hopped beer, it is plausible that humulone interacts with other positive modulators, including alcohol and hop flavonoids, thereby enhancing the effects mediated by GABA$_A$ receptors in beer. In Study **III**, we determined the modulatory action of pure humulone through electrophysiological recordings on GABA$_A$ receptors recombinantly expressed in HEK293 cells. Furthermore, we investigated the interactions between humulone and ethanol on both native and recombinant GABA$_A$ receptors, employing the [$^3$H]EBOB binding assay, along with the exploration of potential interactions with potentially neuroactive hop prenylflavonoids. Lastly, our research included an evaluation of the hypnotic and sedative actions of humulone using pentobarbital and ethanol-induced sleep models, in addition to open field tests conducted on BALB/cAnNRj mice.

### 5.3.1    Electrophysiology of humulone modulation on α1β3γ2 GABA$_A$ receptors

Previous studies have established that the α1 subunit is an essential determinant of sedation mediated by GABA$_A$ receptors (Rudolph et al., 1999; McKernan et al., 2000). Thus, we sought to examine the functional interaction between humulone and the highly prevalent α1β3γ2 isoform (Wisden et al., 1992; Fritschy and Möhler, 1995). Whole-cell patch-clamp electrophysiological recordings were conducted with HEK293 cells expressing recombinant α1β3γ2 GABA$_A$ receptors. A submaximal 1 μM concentration of GABA was applied to assess humulone's modulatory effects.

The represented current traces in Figure 29B demonstrate that 10 µM of humulone significantly potentiated nonsaturating GABA-evoked currents. By normalizing the peak current amplitudes to those stimulated by 1 µM GABA alone (Figure 29C), it was revealed that humulone potentiated the mean peak amplitude by $158 \pm 41\%$ (p < 0.01, n = 8 cells). This adds to the evidence supporting the mechanism of action of humulone as a positive allosteric modulator of $GABA_A$ receptors, consistent with prior research demonstrating the potentiation of GABA-evoked currents by hop and beer extracts (Aoshima et al., 2006; Sahin et al., 2016).

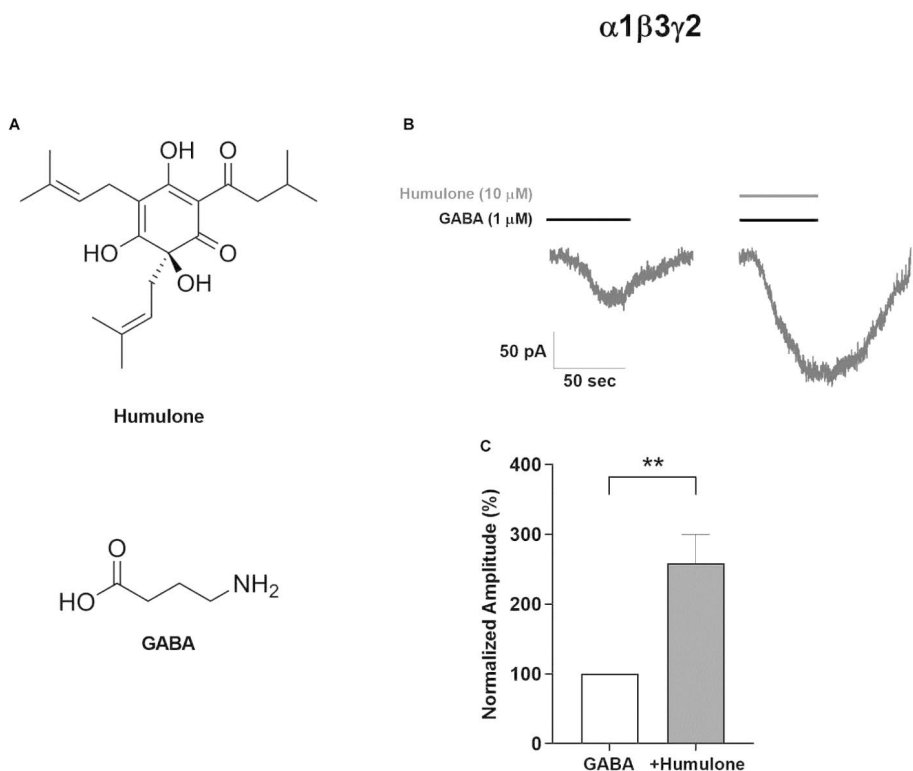

**Figure 29.** Humulone enhances GABA-evoked currents in α1β3γ2 $GABA_A$Rs recombinantly expressed in HEK293 cells. (A) Chemical structures of α-lupulic acid (humulone) and γ-aminobutyric acid (GABA). (B) Exemplary current traces depicting receptor activation/modulation during a 1-minute exposure to a submaximal GABA concentration (1 µM), with and without humulone (10 µM). (C) Vertical bar graph presenting normalized peak current amplitude relative to the control response evoked by GABA (1 µM) only. Each bar denotes the mean ± SEM of n = 8 recorded cells, voltage-clamped at -60 mV, and pH maintained at 7.4. **p < 0.01, indicating statistical significance between GABA-only exposure and combined application with humulone (paired *t*-test).

### 5.3.2 Humulone and ethanol: synergistic effects on [$^3$H]EBOB binding

We examined the influence of ethanol on the humulone modulation of [$^3$H]EBOB binding in membrane preparations from the forebrain and cerebellum. Prior research indicated that ethanol, at low millimolar levels ($\leq 30$ mM), does not directly displace the binding of [$^3$H]EBOB (Supplementary Figure S4; Höld et al., 2000; Zhao et al., 2014). However, our current findings demonstrate that ethanol enhances the modulatory effects of humulone in both distinct brain regions. Co-application of 1 µM humulone with 30 mM ethanol produced a marked leftward shift in the concentration-response curve for GABA-induced displacement of [$^3$H]EBOB (Figure 30). Quantitative analysis revealed a significant reduction in the IC$_{50}$ for GABA-mediated displacement of [$^3$H]EBOB in the presence of ethanol. Specifically, in the forebrain, the IC$_{50}$ decreased from $5.48 \pm 0.09$ µM to $4.60 \pm 0.08$ µM, while in the cerebellum, the IC$_{50}$ decreased from $2.40 \pm 0.06$ µM to $1.80 \pm 0.07$ µM (both $p < 0.01$) (Table 5). As a result, this non-competitive synergistic action proved to be more pronounced (11.2% increase) in the cerebellum relative to the forebrain.

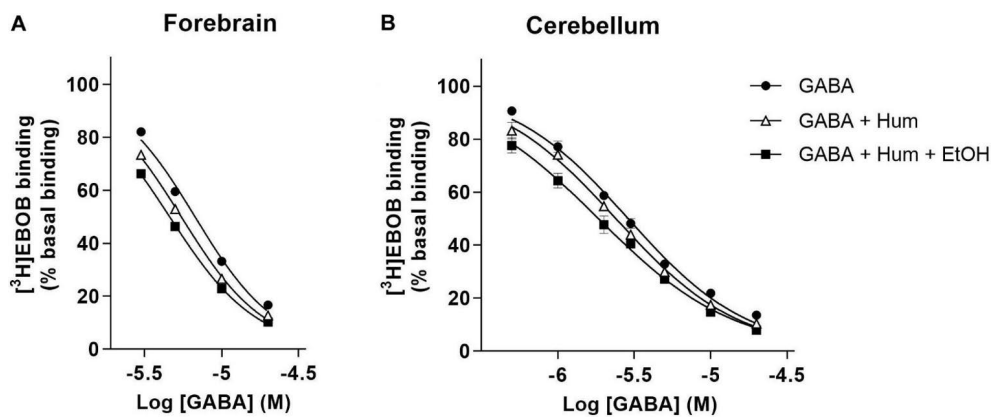

**Figure 30.** GABA dose-dependent shifts in [$^3$H]EBOB binding in membranes of the forebrain (A) and cerebellum (B), influenced by 1 µM humulone (Hum), either with or without 30 mM ethanol (EtOH). Data points denote the mean ± SEM from three distinct experiments, each conducted in triplicate.

**Table 5.** Potency assessment of GABA effects on [³H]EBOB binding with humulone in rat forebrain and cerebellar membranes with or without ethanol (EtOH)

| | IC$_{50}$ (µM) | | |
|---|---|---|---|
| **Membrane** | **GABA** | **GABA**<br>**+ 1 µM Humulone** | **GABA**<br>**+ 1 µM Humulone**<br>**+ 30 mM EtOH** |
| **Forebrain** | 6.73 ± 0.18 | 5.48 ± 0.09[**] | 4.60 ± 0.08 |
| **Cerebellum** | 2.89 ± 0.09 | 2.40 ± 0.06[**] | 1.80 ± 0.07 |

The IC$_{50}$ values are shown as mean ± SEM from three distinct experiments, each conducted in triplicate. **$p < 0.01$, indicating significant differences compared to the corresponding IC$_{50}$ values observed for GABA only and GABA+Humulone+EtOH (determined via one-way ANOVA paired with Tukey's *post hoc* analysis).

The significance of the cerebellum in controlling motor function and coordination is notable, with adverse implications stemming from both acute and chronic alcohol intake (Seiger et al., 1983; Engblom et al., 1991; Luo, 2015; Valenzuela and Jotty, 2015). Moreover, α6-GABA$_A$Rs are distinctly expressed in cerebellar granule cells, particularly in the form of extrasynaptic α6βδ combinations (Nusser et al., 1998). These subtypes exhibit a strong affinity for ambient GABA (Saxena and Macdonald, 1996) and high sensitivity to ethanol at low millimolar levels (Sundstrom-Poromaa et al., 2002; Wallner et al., 2003; Wei et al., 2004; Hanchar et al., 2005, 2006; Santhakumar et al., 2006; Wallner, 2006; Glykys et al., 2007). Yet, the precise extent of this sensitivity remains debatable, with varying studies reporting conflicting outcomes (as reviewed by Förstera et al., 2016). In light of these aspects, we conducted further investigation into the potential role of α6-GABA$_A$R subtypes in the potential modulatory interactions between humulone and ethanol (Section 5.3.3).

## 5.3.3 Differential potency of humulone across GABA$_A$ receptor subtypes

The modulatory effects of humulone were initially determined employing [³H]EBOB binding to recombinant α6β3γ2, α6β3δ, and α6β3 receptor subtypes expressed in HEK293 cells. As illustrated in Figure 31A, humulone dose-dependently enhanced the GABA-mediated displacement of [³H]EBOB across all the subtypes examined. Notably, such enhancement in α6β3δ subtype was detectable at humulone concentrations as low as 1 µM. This effect was only observed with humulone at concentrations of 10 µM and higher in α6β3, α6β3γ2 receptors. With 3 μM GABA present, the $IC_{50}$ values for humulone revealed superior inhibitory potency in the α6β3δ subtype ($8.45 \pm 0.92$ μM), markedly higher than in the α6β3γ2 ($49.03 \pm 10.11$ μM) and α6β3 ($56.85 \pm 12.79$ μM) subtypes ($p < 0.01$). However, the distinction between the α6β3γ2 and α6β3 subtypes lacked statistical significance. The modulatory influence of humulone remained largely consistent when the γ2 subunit was incorporated into the α6β3 receptor subtype. Nevertheless, with the presence of the δ subunit, there was a significant enhancement in humulone's potency, showing increases of 5.8 and 6.7 times relative to the α6β3γ2 and α6β3 subtypes, respectively. Such observations underscore the δ subunit's role in facilitating the pronounced effects of humulone at low micromolar levels on extrasynaptic $GABA_A$ receptors, as similarly noted with hop prenylflavonoids (**I**).

Subsequently, we examined if the modulation by humulone, previously noted in the forebrain and cerebellum, responds to low ethanol doses (30 mM) in recombinant α6β3, α6β3γ2, and α6β3δ subtypes (Figure 30A, B). In line with the reported low potency of ethanol in displacing [³H]EBOB binding ($IC_{50} = 370 \pm 4$ mM) (Höld et al., 2000; Zhao et al., 2014), our findings indicated that 30 mM ethanol did not influence GABA-mediated [³H]EBOB binding to α6β3, α6β3γ2, and α6β3δ receptor subtypes (Figure 31 B, C, and D). However, contrary to our findings with the forebrain and cerebellar membranes, the combined presence of 30 mM ethanol and 1 μM humulone showed no amplification of the humulone-potentiated [³H]EBOB displacement in the recombinant α6β3δ, α6β3γ2, and α6β3 receptor subtypes. Indirect modulation of native $GABA_A$ receptors by ethanol, potentially through protein kinase phosphorylation activation and increased presynaptic GABA release, complicates the precise modeling of ethanol sensitivity (Harris et al., 1995; Weiner et al., 1997; Aguayo et al., 2002; Carta et al., 2004). While humulone showed a pronounced potency towards the α6β3δ subtype, its specific contribution to the synergy between ethanol and humulone in the cerebellum remains elusive. Evidence suggests that humulone only marginally affects the binding of [³H]Ro 15-4513 to native αβγ2 $GABA_A$ receptors. Moreover, flumazenil's antagonistic actions don't appear to impact humulone's effect on the binding of [³H]EBOB (**II**). Taking these observations into account, and considering the distinct non-competitive synergy between ethanol and humulone in brain membranes as well as humulone's specific potency in extrasynaptic receptor subtypes, it is evident that humulone doesn't positively modulate via the classical benzodiazepine binding site at the α+γ2− subunit interface.

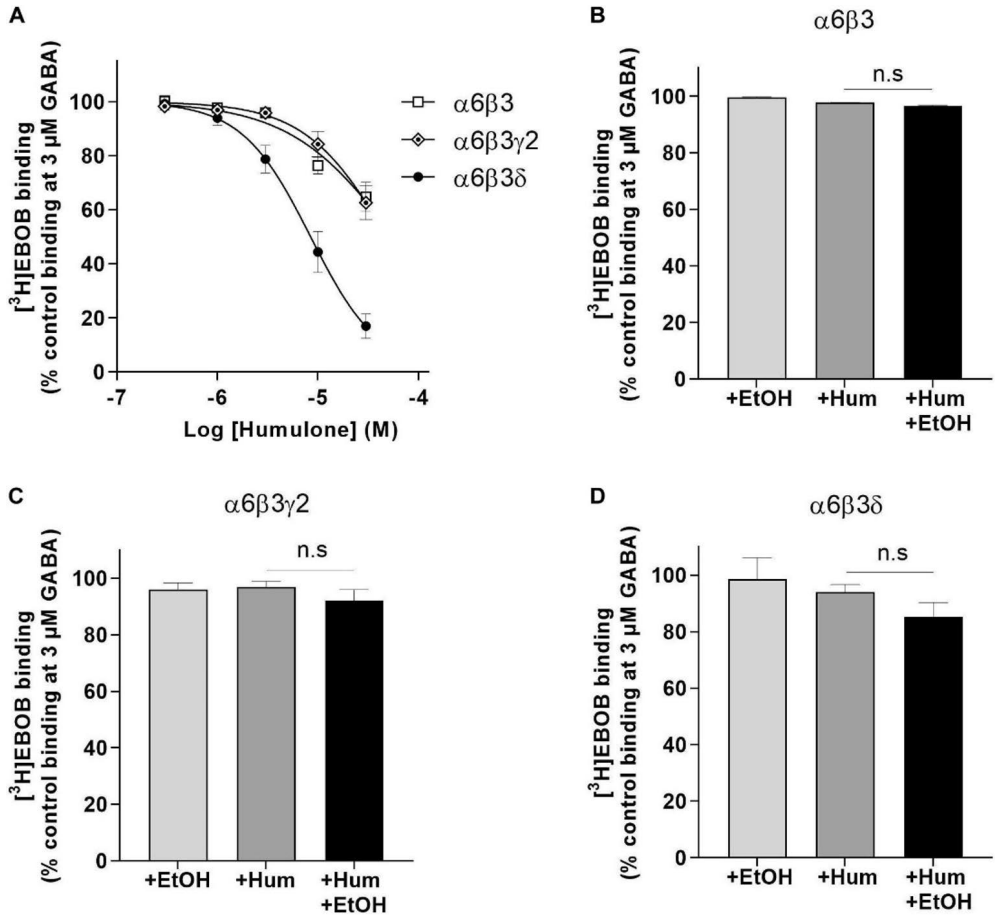

**Figure 31.** Influence of humulone (Hum) on GABA-mediated [³H]EBOB binding shifts in HEK293 cells expressing various GABA_A receptor subtypes: α6β3γ2, α6β3δ, and α6β3. (A) Binding displacement plots of 1 nM [³H]EBOB are depicted as a percentage of control across five humulone concentrations when co-applied with 3 μM GABA. (B-D) The effect of 1 μM humulone on the binding of 1 nM [³H]EBOB in combination with 3 μM GABA, expressed as a percentage of control, either with or without 30 mM ethanol (EtOH). The control is defined by the binding of [³H]EBOB with only GABA present. Data points denote the mean ± SEM from 3–6 distinct experiments, each conducted in triplicate. The label "n.s." indicates instances where the observed mean showed no significant difference from the control, as determined by a one-way ANOVA paired with Tukey's *post hoc* analysis.

### 5.3.4 Interaction of humulone with hop prenylflavonoids: additive modulation of [³H]EBOB binding

We examined the interactions between humulone and hop prenylflavonoids, which have demonstrated the ability to modulate the binding of [³H]EBOB to GABA$_A$ receptors at low micromolar levels (**I, II**) (Figure 32). The simultaneous application of 6-prenylnaringenin (6PN) and isoxanthohumol (IXN) at a concentration of 1 μM resulted in an additive enhancement of the GABA-mediated [³H]EBOB displacement in the rat forebrain (p < 0.01). This phenomenon appears to be mediated through the α1+/β2− interface. Molecular docking analysis predicted a more favorable binding free energy for 6PN at this interface. Notably, three specific residues—Lys156, Gln204, and Ser205—exhibited a binding affinity for 6PN but not for IXN (**II**; Supplementary Table S1). The introduction of humulone significantly amplified the effects observed with the 6PN+IXN pairing (p < 0.01). As detailed in Table 5, this combined influence is equivalent to the summative displacement of [³H]EBOB attributed to each compound individually.

The additive interactions between hop prenylflavonoids and humulone resemble those observed with specific flavonoids and other GABA$_A$ receptor modulators. For example, apigenin (from chamomile) and (–)-epigallocatechin gallate (from green tea) have been shown to amplify the modulatory effects of diazepam on α1β1γ2 receptor subtypes (Campbell et al., 2004). Moreover, select flavonoid modulators from *Valeriana* species, namely 6-methylapigenin and linarin, synergize with hesperidin and valerenic acid, respectively, to enhance sleep induction in animal models (Marder et al., 2003; Fernández et al., 2004). Considering that the binding of PAMs to distinct binding sites can elicit additive responses (Visser et al., 2003; McMahon and France, 2005; Paronis, 2006), it is plausible that hop neuroactive properties are driven by multiple compounds, collectively enhancing GABA$_A$ receptor function.

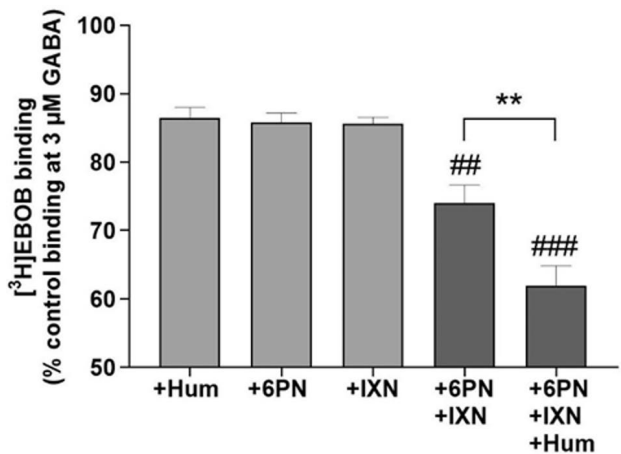

**Figure 32.** Enhancement of GABA-triggered [$^3$H]EBOB binding shifts in forebrain membranes becomes additive when humulone (Hum) is co-applied with 6PN (6-prenylnaringenin) and IXN (isoxanthohumol), all at a concentration of 1 µM. The control condition is represented by the maximum binding of [$^3$H]EBOB with only 3 µM GABA present. Every bar in the graph denotes the mean ± SEM from three distinct experiments, each conducted in triplicate. Significance markers: ###p < 0.001 and ##p < 0.01 indicate significant differences compared to their respective single constituents. Moreover, **p < 0.01 signifies the variance between the combined effects of 6PN, IXN and Hum versus the effects of just 6PN and IXN. This is determined by a one-way ANOVA paired with Tukey's *post hoc* analysis.

**Table 6**     Individual and combined modulation of [$^3$H]EBOB binding relative to 3 µM GABA control

| Compound (1 µM) | [$^3$H]**EBOB displacement (%)** |
|---|---|
| Hum | 13.5 ± 1.5 |
| 6PN | 14.2 ± 1.4 |
| IXN | 14.4 ± 0.9 |
| 6PN+IXN | 25.9 ± 2.6 |
| 6PN+IXN+Hum | 38.1 ± 2.9 |

Values are shown as mean ± SEM from three distinct experiments, each conducted in triplicate.

## 5.3.5     Influence of humulone on sleep parameters: pentobarbital and ethanol-induced sleep

We investigated the impact of humulone pretreatment on the latency and duration of sodium pentobarbital-induced sleep (35 mg/kg, i.p.) in mice. Although there are no single-dose toxicity assessments for humulone in mice, the doses examined were conservative, remaining below the reported median lethal dose (LD$_{50}$) for rats: 1,500 mg/kg, p.o.; 600 mg/kg, i.m. (Bejeuhr, 1993). In our study, administration of humulone at 20 mg/kg markedly extended sleep duration (p < 0.001) and reduced sleep onset latency (p < 0.01) induced by pentobarbital compared to control mice (Figure 33A, B). However, this influence was not noted at a dose of 10 mg/kg. In a prior study, an α-acid extract prolonged sleep duration induced by pentobarbital in rats at doses of 10 and 20 mg/kg without affecting sleep latency (Zanoli et al., 2005). Subsequent research has shown that a dose of 21 mg/kg of humulone-containing α-acid fraction increased ketamine-induced sleep duration in mice, with no observable enhancement at 42 mg/kg (Schiller et al., 2006). Interestingly, while humulone had no impact on the onset of ethanol-induced sleep (3.5 g/kg, i.p.), it significantly extended sleep duration in a dose-dependent manner at both 20 mg/kg (p < 0.001) and the lesser dose of 10 mg/kg (p < 0.05) relative to controls (Figure 33C, D). The Human Equivalent Dose based on body surface area for 10 mg/kg in mice is 0.81 mg/kg (Nair and Jacob, 2016). Thus, a dose of 10 mg/kg is equivalent to a 60 kg individual consuming 2 L of non-alcoholic hop-infused beer with a humulone concentration of 24.2 mg/L (Hahn et al., 2018). Based on our [$^3$H]EBOB binding assays (Figure 30), the enhancement in sleep duration at a low humulone dose may be ascribed to its synergistic modulation with ethanol in native GABA$_A$ receptors.

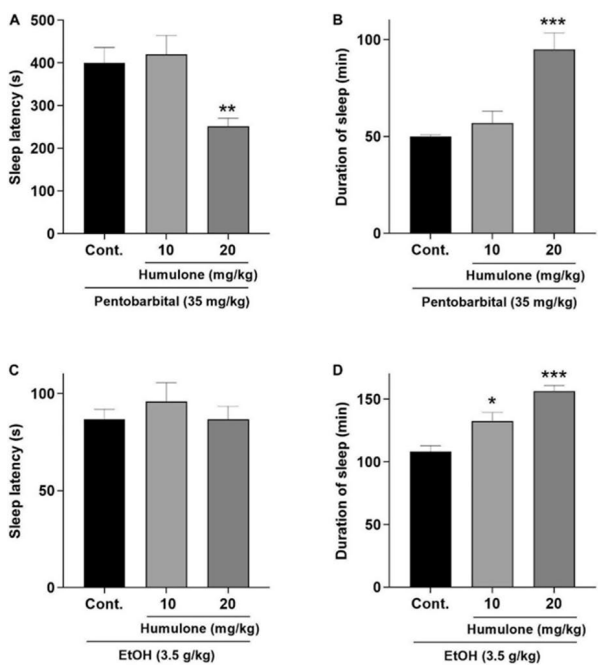

**Figure 33.** Humulone's influence on sleep onset latency and duration in mice following a 45-minute pre-treatment and sequential induction by sodium pentobarbital (35 mg/kg, i.p.) (A,B) or ethanol (3.5 g/kg, i.p.) (C,D). Every bar denotes the mean ± SEM from 6 to 7 mice per group. Significance markers: ***p < 0.001, **p < 0.01, *p < 0.05, indicating differences from the group receiving the vehicle only, as determined by a one-way ANOVA paired with Dunnett's multiple comparison test.

### 5.3.6 Locomotor impact of humulone: open field test

An initial investigation involving mice documented that hop extract doses of 100 and 200 mg/kg, comprising 36% humulone, reduced locomotor activity in the open field test, indicating potential sedation. However, the extract did not manifest anxiolytic effects in the elevated plus maze test (Schiller et al., 2006). In alignment with these findings, but at a lower dose of humulone in pure form, our results revealed a significant influence on the locomotor parameters tested. Specifically, mice treated with 20 mg/kg of humulone traversed reduced total distance ($p < 0.01$) and displayed decelerated average velocity ($p < 0.01$) in comparison to their vehicle-treated counterparts (Figure 34). This alteration does not necessarily indicate a modulation in anxiety-related behavior, as the durations spent in both the periphery and center regions exhibited no discernible differences across all mouse groups during the open field evaluation (Supplementary Figure S5). The occurrence of multiple weak modulators, which might compete with humulone for the same binding site, could diminish the potency of α-acids. In support of this, no decline in locomotion was identified in rats subjected to α-acid extract at doses of 10 and 20 mg/kg (Zanoli et al., 2005). This can be rationalized considering that α-acids also contain both cohumulone and adhumulone, molecules bearing structural resemblance to humulone but exhibiting only marginal modulatory influences on GABA$_A$ receptors (**II**).

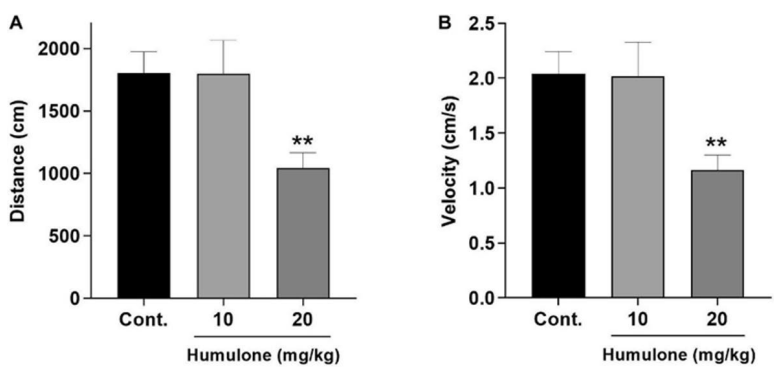

**Figure 34.** Impact of acute treatment with humulone (10 and 20 mg/kg) on mice's locomotion in the open field paradigm. Both the distance covered (A) and velocity (B) were measured over a 15-minute interval, initiated 45 minutes post-intraperitoneal injection of either humulone or the vehicle. Every bar denotes the mean ± SEM from 7 to 11 mice per group. Significance marker: **$p < 0.01$, indicating differences from the group receiving the vehicle only, as ascertained by a one-way ANOVA paired with Dunnett's multiple comparison test.

# 6    General Discussion and Future Directions

## 6.1    Structure-activity insights of hop prenylflavonoids in GABA$_A$ receptor modulation

Hop prenylflavonoids have promising characteristics as scaffolds for the development of GABA$_A$ receptor modulators. The prenylation of flavonoids is believed to enhance their lipophilic nature, facilitating permeability across cell membranes and improving interaction with biological targets (Hatano et al., 2000; Sasaki et al., 2011). This modification has also been shown to enhance modulatory effects on GABA$_A$ receptors, particularly in flavanones (e.g., 8-prenylnaringenin and isoxanthohumol; **I**) and other flavonoid classes, including chalcones (e.g., xanthohumol; **I**) and isoflavones (Çiçek et al., 2018). Recent advancements in the development of prenylflavonoids—through isolation and partial or complete synthesis—demonstrate their broad-spectrum therapeutic potential (Shi et al., 2021b).

The planar molecular structure of flavonoids is recognized to play an essential role in their interaction with benzodiazepine binding sites (Dekermendjian et al., 1999). Various natural and synthesized flavones with such structures have shown high affinity for these binding sites (Medina et al., 1998). Flavanones, on the other hand, do not contain a C2=C3 double bond, making them less planar (Figure 11). This distinction justifies that the positive modulatory actions of hop prenylflavonoids, belonging to the flavanone subclass, do not occur through the classical benzodiazepine binding site (**I, II**). These phytochemicals from hops offer an alternative avenue for drug design, potentially bypassing the dependence and withdrawal issues associated with benzodiazepines by enhancing GABAergic inhibition through allosteric sites different from those targeted by this class of drugs.

Although our studies identified prenylflavonoids with micromolar potency (**I, II**), the sleep-promoting effects of hops cannot be solely attributed to this class of compounds, given their relatively low abundance (up to 1.1%) compared to α-acids (up to 13%). Nonetheless, the polyphenolic structures of flavonoids found in hops offer versatility for medicinal chemistry, enabling the generation of analogs that leverage their modulatory properties. Structure-activity relationship studies with flavone derivatives highlighted the role of 6-substitution in determining their efficacies and binding affinities as GABA$_A$ receptor ligands (Ren et al., 2011). Given that 6-prenylnaringenin exhibited the highest modulatory activity among all tested flavonoids, this observation might also apply to the flavanone subclass, such as hop prenylflavonoids. Thus, a deeper understanding of the structure-activity relationship of hop prenylflavonoids is essential to facilitate the synthesis of analogs and derivatives optimized for insomnia treatment. Looking ahead, derivatizing multiple hydroxyl groups in hop constituents might enhance drug-like properties and pharmacokinetics. Moreover, understanding the impact of chiral centers and aliphatic moieties could reveal pharmacophoric determinants, guiding the synthesis of more potent flavonoid analogs that target specific subtypes involved in sleep regulation, notably the extrasynaptic GABA$_A$ receptors.

## 6.2 Hop prenylflavonoids and humulone as novel modulators of extrasynaptic GABA$_A$ receptors

Given the scarcity of allosteric modulators that selectively interact with extrasynaptic δ-GABA$_A$Rs, exploring selective novel compounds is crucial to fully exploit the therapeutic potential of these receptors in the treatment of insomnia. Humulone and hop prenylflavonoids have been shown to enhance the response to GABA-induced activity, demonstrating greater selectivity for δ-GABA$_A$Rs (**I, III**). In particular, the δ subunit is localized to extrasynaptic sites and preferentially co-assembles with α6/α4 and β subunits. This forms δ-GABA$_A$Rs that are insensitive to benzodiazepines but exhibit high GABA affinity (Saxena and Macdonald, 1996; Brown et al., 2002) and mediate tonic inhibition in various neuronal populations throughout the brain (Semyanov et al., 2004; Glykys and Mody, 2007; Belelli et al., 2009). Additionally, these receptors are major targets for various compounds, including anesthetics (e.g., propofol), hypnotic agents (e.g., gaboxadol), and neurosteroids (e.g., allopregnanolone). Ethanol, in particular, has been shown to enhance the effect of these receptors, resulting in moderate levels of intoxication in humans at millimolar concentrations (Sundstrom-Poromaa et al., 2002; Wallner et al., 2003; Hanchar et al., 2005). Thus, hop compounds that can positively modulate δ-GABA$_A$Rs might influence the intoxicating effects of alcohol in beer, a notion that warrants exploration. Our recent findings with humulone in the chimney test (Figure S6, preliminary data) suggest that the modulation of the α6β3δ subtype by humulone and possibly hop prenylflavonoids (**I, III**), could lead to alterations in motor coordination. Given the involvement of the α6 subunit in mediating ataxia induced by ethanol and benzodiazepines (Korpi et al., 1999; Hanchar et al., 2005), it is imperative to pursue further research to understand the implications of this modulation.

Furthermore, the expression of the α4β3δ subtype is pronounced in brain regions associated with sleep regulation (Pirker et al., 2000; Peng et al., 2002). This subtype mediates tonic currents in thalamic relay neurons, contributing to the cortical delta activity observed during NREM sleep (Brickley, 2018). Although benzodiazepines and Z-drugs are effective in inducing and maintaining sleep, they reduce the power of delta waves in NREM sleep, potentially diminishing sleep depth and overall restfulness (Alexandre et al., 2008; Uygun et al., 2016). Conversely, due to the region-specific localization of the extrasynaptic receptors, δ-selective ligands might provide a more targeted influence on sleep. A recent study, building upon our earlier research, administered an optimized Saaz–Saphir hop mixture (75:25 ratio) containing xanthohumol and humulone to rats for 9 days. EEG analysis revealed a significant increase in sleep time, particularly in the NREM stage, attributed to enhanced delta wave activity (Min et al., 2021). This suggests that hop compounds may offer advantages over benzodiazepines and Z-drugs. Future research could investigate their effects on the α4β3δ subtype, paving the way for a better understanding and novel therapeutic approaches in managing sleep disorders.

## 6.3 The differential effects of hop α- and β-acids on GABA$_A$ receptors and sleep

This investigation has elucidated the pharmacological basis behind the sedative and sleep-promoting properties of hops (Zanoli et al., 2005; Schiller et al., 2006). Isolated α-acids—cohumulone (F2), humulone (F3), and adhumulone (F4)—as well as isolated β-acids—colupulone (F6), lupulone (F7.1), and adlupulone (F7.2)—all positively modulated the GABA-mediated displacement of [$^3$H]EBOB binding to GABA$_A$ receptors (**II**). Given the prominence of humulone in hops and its significant potency, its activity was confirmed electrophysiologically, where it enhanced GABA-evoked currents in the sedation-mediating α1β3γ2 subtype. Consistent with humulone's effects in our pentobarbital- and ethanol-induced sleep models (**III**), an α-acid fraction containing humulone was previously found to prolong ketamine-induced sleep in mice at a dose of 21 mg/kg (Schiller et al., 2006). In a recent study, oral administration of humulone at 20 mg/kg prolonged sleep duration in a pentobarbital-induced sleep model in mice without altering sleep onset (Min et al., 2021). The subsequent introduction of picrotoxin at 4 mg/kg reversed this outcome. Importantly, picrotoxin alone did not influence sleep parameters, in line with the saline-treated control group. Additionally, after three weeks of oral administration of the humulone-containing Saaz–Saphir hop mixture, there was a pronounced increase in GABA levels in the mice brain, accompanied by upregulated expression of GABA$_A$ receptors (Min et al., 2021). These observations establish a connection between GABA$_A$ receptors and the efficacy of humulone in promoting sleep.

On the other hand, the modulatory actions of cohumulone (F2), adhumulone (F4), and β-acids appear modest in comparison to humulone. The β-acid fraction required a dose of 119 mg/kg to elicit a similar increase in sleep duration in a ketamine-induced sleep model (Schiller et al., 2006). In another study, at a lower dose of 5 mg/kg, β-acids did not influence pentobarbital-induced sleep in rats (Zanoli et al., 2007). They did, however, enhance locomotion in the open field and exhibited antidepressant-like effects in the forced swim tests. Further electrophysiological recordings from the same study showed that, at a concentration of 120 μM, β-acids negatively modulated GABA-evoked currents in cerebellar granule cells. Yet, in [$^3$H]EBOB binding assays, they exhibited positive modulation at a lower concentration of 30 μM (**II**), underscoring their multifaceted, dose-dependent effects. Nevertheless, our findings predominantly identify humulone as the key active constituent responsible for modulating GABA$_A$ receptors and the sleep-promoting attributes of *Humulus lupulus*.

## 6.4 The blood-brain barrier permeability and CNS effects of hop compounds

The ability of phytochemicals to penetrate the blood-brain barrier (BBB) is a major factor in determining their potential to modulate receptor activity in the CNS. The BBB plays a crucial role in preserving the integrity of the brain microenvironment, allowing nutrients to selectively pass to neurons while blocking the entry of harmful substances such as toxins, pathogens, and inflammatory cells (Neuhaus et al., 2008; Takeshita and Ransohoff, 2015). The BBB is mainly composed of specialized brain microvascular endothelial cells (BMEC) that are distinct from their peripheral counterparts. These cells lack fenestrations, exhibit high expression of tight junctions, and have diminished pinocytosis (microvesicle transport) (Grant et al., 1998). Additionally, the endothelium that lines the BBB expresses a complex network of efflux pumps and nutrient transporters that regulate the paracellular and transcellular passage of hydrophilic substances (Daneman and Prat, 2015).

Although there has been limited research on the BBB penetration of hop prenylflavonoids, evidence from prior *in vivo* studies suggests the potential of some of these compounds to traverse this barrier. For example, xanthohumol has demonstrated its potential to enhance cognitive performance in mice. After a five-day treatment regimen (40 mg/kg, i.p.), it was detectable in both the cerebral cortex and hippocampus (Zamzow et al., 2014). However, the concentration of xanthohumol in the brain was markedly low, measuring at only 4 nM, which is below its effective micromolar range on the GABA$_A$ receptors (**I**). In a separate study, 8-prenylnaringenin at 750 mg led to a reduction in luteinizing hormone (LH) levels in the brains of postmenopausal women (Rad et al., 2006). LH is released by the anterior pituitary gland under the control of the hypothalamic secretion of gonadotropin-releasing hormone (GnRH). The plausibility of this pathway being modulated by estrogen-like flavonoids, including 8-prenylnaringenin, underscores the potential of its BBB penetration through the hypothalamic-pituitary axis (Christoffel et al., 2006). Consistently, 8-prenylnaringenin exhibited panicolytic effects in rats, as determined by the elevated T-maze test (Bagatin et al., 2014). However, the concentration of 8-prenylnaringenin in the brain was not assessed in that study. Conversely, the BBB permeability of 6-prenylnaringenin in mice was confirmed, with no observable influence on locomotor activity, as evidenced by open-field and rotarod tests (Sekiguchi et al., 2018). The measured concentrations within the mice brain 10 and 30 minutes after i.p. injection (30 mg/kg) were recorded at 1.95 μM and 2.20 μM, respectively. Notably, both of these doses were found to be sufficient for enhancing GABA-mediated responses of native GABA$_A$ receptors *in vitro* (**II**).

As for humulone and its potential metabolites, no studies have reported their actual concentrations in the brain following beer consumption or direct administration. The isomerized form of α-acids in beer, including isohumulone, has been proven to cross the BBB upon oral administration, as confirmed by HPLC-MS/MS analysis (Ano et al., 2017). This penetration resulted in increased anti-inflammatory and amyloid beta (Aβ) phagocytosis activities by brain microglia, correlating with a decrease in cerebral plaque burden and improved cognitive performance in a mouse model of Alzheimer's disease. In conclusion, further research is essential to correlate the pharmacological activity observed in our studies on hop constituents with their physiological brain-mediated effects.

## 6.5 Methodological significance: [$^3$H]EBOB as a functional marker for GABA$_A$ receptor complex interaction

In this thesis, [$^3$H]EBOB served as the primary radioligand targeting the non-competitive blocker site of the GABA-gated chloride channel (Cole and Casida, 1992; Kume and Albin, 1994). Our investigation established a correlation between electrophysiological measurements and [$^3$H]EBOB binding results (**III**), emphasizing the reliability and relevance of [$^3$H]EBOB as a radioligand for evaluating drug modulation of GABA$_A$ receptor function. While [$^{35}$S]TBPS has been a recognized tool for labeling and studying GABA$_A$ receptors (Im and Blakeman, 1991; Akk et al., 2007), [$^3$H]EBOB, comparatively, exhibits several advantages over [$^{35}$S]TBPS, including higher affinity, greater stability, and a longer radioisotopic half-life (Cole and Casida, 1992; Casida, 1993; Yagle et al., 2003). Moreover, the binding of [$^3$H]EBOB to the chloride channel can be readily influenced by conformational alterations induced by varying levels of GABA and/or the presence of GABA$_A$ receptor modulators (Atucha et al., 2009; L'Estrade et al., 2019; **III**). This characteristic supports the notion that the [$^3$H]EBOB binding site is closely associated with the chloride channel blocker site. Remarkably, there is a strong quantitative alignment between the potency to displace [$^3$H]EBOB binding and the ability to modulate GABA-induced chloride influx (Huang and Casida, 1996; Yagle et al., 2003). Therefore, the use of [$^3$H]EBOB can be a valuable approach in providing functional insights into the interaction of novel therapeutic modulators with the GABA$_A$ receptor complex.

## 6.6    Research limitations

While novel insights were gained through the presented studies, some limitations are important to acknowledge. First, only a select few of the numerous bioactive phytochemicals in hops were screened and characterized, even though hops contain a vast array of compounds with potential additive, synergistic, or antagonistic effects on pharmacological activity. Second, the recombinant GABA$_A$ receptor systems examined do not fully capture the complexity of native receptors in neuronal cells. Moreover, the sleep architecture was not evaluated using EEG, which is central for understanding impacts in specific stages of sleep. Third, the optimal dosages in mice or humans are unclear due to the limited number of animals in the *in vivo* experiments. A dose-response study with adequate sample size, dose selection, and outcome measures can help to determine the minimum effective dose, the maximum tolerated dose, and the therapeutic window, accounting for the inter-individual variability and the environmental influence on sleep in humans. Furthermore, aspects like the permeability of the blood-brain barrier and brain concentrations of active compounds post-oral administration were not quantified. While some interactions between compounds were studied, a comprehensive analysis of all possible combinations was missing. Finally, a more detailed toxicity profile from acute and repeated doses is needed for the potential translation of these findings into human studies or drug development (International Conference on Harmonisation, 2009). Despite these limitations, the results underscore the potential of hop constituents to modulate a crucial neurotransmitter system involved in sleep regulation and sedation.

# 7    Summary and Conclusions

In the quest to address the challenges of insomnia, GABA$_A$ receptors stand out as a crucial target, with natural modulators, such as hops, offering potential solutions. The primary objective of this thesis was to uncover the key phytochemicals from hops that influence GABA$_A$ receptor activity and to establish a deeper understanding of their modes of action, subtype selectivity, and interactions within the receptor complex.

The predominant prenylflavonoid in hops, xanthohumol, along with its derivatives isoxanthohumol and 8-prenylnaringenin, has shown positive modulatory effects on GABA-mediated responses in both native and recombinant αβγ/δ GABA$_A$ receptors at low micromolar levels. This modulation is not exclusively dependent on the γ or δ subunits, suggesting potential binding sites within the α/β interfaces of the GABA$_A$ receptor pentameric complex. Potentiation of the GABA-mediated response displayed marked subtype selectivity for δ-GABA$_A$Rs, which are known to be responsible for tonic inhibition in various neuron types within the brain. While these subtypes are not sensitive to benzodiazepines, they exhibit sensitivity to ethanol, particularly at doses that induce moderate intoxication in humans.

It is important to note that xanthohumol, isoxanthohumol, and 8-prenylnaringenin are resistant to flumazenil, a classical benzodiazepine site antagonist. Given their limited displacement of [$^3$H]Ro 15-4513 binding, it is evident that their positive modulatory actions do not function through the classical benzodiazepine binding site. This particular site is associated with drug tolerance, dependence, and withdrawal symptoms. Surprisingly, the positive modulatory action of another prenylflavonoid, 6-prenylnaringenin, also remains unaffected by flumazenil antagonism. However, binding experiments using [$^3$H]flunitrazepam and [$^3$H]Ro 15-4513, indicated discernible selectivity for the classical benzodiazepine binding site by 6-prenylnaringen. Molecular modeling provided further validation for this observation.

Through *in silico* analysis, which was supported by our binding findings, the extracellular α+/β− interface emerged as a positive allosteric modulatory site for hop prenylflavonoids. Notably, 6-prenylnaringenin displayed binding energies and interactions with residues similar to those of flumazenil at the extracellular α+/γ2− interface. However, its resistance to flumazenil antagonism in [$^3$H]EBOB binding implies that 6-prenylnaringenin acts as a silent modulator at the classical benzodiazepine-binding site without influencing its positive modulatory effect via the α+/β− interface.

The unique dual functionality of 6-prenylnaringenin highlights its potential applications in addressing neuronal excitability while inhibiting benzodiazepine-mediated responses and in identifying drugs that exert their modulatory effects solely through the classical benzodiazepine binding site. Since the α+/β− interface is present across all GABA$_A$ receptor subtype combinations, modulation at this site suggests a broader effect. This is in contrast to the classical benzodiazepine site ligands, which predominantly bind to specific receptor subtypes at the α+/γ− interface. Compounds that modulate via the α+/β− interface function by augmenting the ongoing GABAergic transmission while also exhibiting selectivity for brain regions with elevated excitatory and GABAergic activity, potentially showcasing minimal toxicity (Sieghart et al., 2012; Varagic et al., 2013).

Analysis of individual hop fractions has unveiled additional constituents that positively modulate GABA-mediated responses in native GABA$_A$ receptors, although with varying potencies. The humulone fraction, the primary α-acid in hops, exhibits potent modulatory activity within a low micromolar range. Radioligand binding experiments indicate that humulone's action is not mediated via the classical benzodiazepine binding site. Electrophysiological validation confirmed the functional activity of humulone as it amplified non-saturating GABA-evoked currents in the prevalent α1β3γ2 receptor subtype, which is correlated with GABA$_A$ receptor-mediated sedative effects.

*In vivo* experiments in mice have shown that humulone induces sedative effects and enhances the hypnotic actions of pentobarbital and ethanol, thereby solidifying its role in the sleep-promoting properties of hops. Further research should explore the influence of humulone on sleep patterns and overall sleep quality using electroencephalography. Subsequent radioligand binding experiments revealed a synergistic relationship between humulone and ethanol, as well as additive interactions between humulone and hop prenylflavonoids at GABA$_A$ receptors. The prolonged sleep induced by the combination of ethanol and humulone may be attributed to their collective action on the GABA$_A$ receptor complex. These interactions could amplify the intoxicating effects of hop-enriched beers. Therefore, it is imperative to further probe its implications on human alcohol consumption habits and their associated rewards.

The findings of this study have significant implications for various stakeholders including consumers, producers, regulators, and researchers. It unveils the potential benefits and risks of ingesting hop-enriched beers or non-alcoholic beverages, especially for sleep quality and alcohol intoxication. In addition, it suggests potential strategies for product development and marketing, such as increasing hop content and decreasing alcohol content in beer, or promoting hop-based nonalcoholic beverages as sleep aids. Furthermore, it informs policy-making and public health interventions, such as setting standards and guidelines for hop content and labeling in beers and formulations, or educating and supporting consumers about the effects and interactions of hop compounds and alcohol.

Finally, the study encourages further investigations and collaborations, such as testing more hop varieties or brewing techniques to exploit the findings or elucidating the molecular mechanisms and binding sites of hop compounds on $GABA_A$ receptors. For example, different hop varieties may have varying levels and compositions of prenylflavonoids and α-acids, which could influence their modulatory effects on $GABA_A$ receptors. Therefore, selecting or breeding hop varieties with higher levels of these compounds could enhance the sleep-promoting or intoxicating properties of hop-based preparations. Alternatively, different brewing techniques may influence the extraction and conversion of hop compounds, such as the duration and temperature of boiling, timing and amount of hop addition, or the type and activity of yeast. Therefore, optimizing or modifying these parameters can increase the availability and bioactivity of these compounds in the final product. These approaches could lead to new findings and innovations in the field of hop-based remedies and their pharmacological effects.

In conclusion, from "Nature's Brewery to Bedtime", this thesis has revealed the intricate interplay between hop compounds and $GABA_A$ receptors, opening new avenues for developing natural sleep aids and optimizing interventions for insomnia. More research is needed to fully elucidate the therapeutic potential of hops and inform regulatory guidelines for the safe and responsible use of hop-based products.

# Acknowledgements

As I approach this momentous occasion, I would like to express my appreciation for the people who have illuminated my way through the twists and turns of my doctoral expedition. This dissertation, a result of years of commitment and persistence at the University of Turku's Centre of Integrative Physiology and Pharmacology, Institute of Biomedicine, is not merely an academic pursuit. It stands as a testament to the collective synergy, intellectual brilliance, and profound enlightenment shared by all those around me.

I am immensely grateful to the Vice Dean of Education, Prof. Eriika Savontaus, and the Director of the Drug Research Doctoral Programme (DRDP), Prof. Ullamari Pesonen, for providing excellent facilities and resources, and for fostering a conducive research environment. I also owe a debt of gratitude to Prof. Markku Koulu for his support throughout my academic studies. His abiding passion for drug development and his mentorship during my early research endeavors have been an unremitting source of inspiration.

My deepest gratitude extends to my supervisor, Adj. Prof. Mikko Uusi-Oukari, for his unwavering guidance and scientific counsel over the course of my doctoral journey. Mikko has been more than a supervisor; he has been a trusted mentor, a friend, and a solid pillar of support. He nurtured my intellectual growth and independent thinking, encouraging me to seek knowledge with objectivity and open mind. He has empowered me to pursue my interest, liberally offering his expertise and feedback every step of the way. I could not have asked for a better supervisor and feel honored and privileged to have worked with him.

I also express my heartfelt thankfulness to my co-supervisor, Senior Researcher Sanna Soini, who has been an instrumental figure in my research success and learning experience. Sanna's invaluable guidance provided a fresh perspective on my research topic, and her insightful suggestions helped me refine my ideas and elevate my presentation. Sanna's unwavering encouragement has been a source of strength and motivation throughout my doctoral project, and I am immensely indebted for her confidence in my abilities.

I also warmly thank Adj. Prof. Christine Engblom for serving as a member of my follow-up committee. I value her kindness and availability to always make time for insightful discussions about my progress, and I appreciate her support and affirmation during my doctoral studies.

My special recognition extends to the esteemed co-authors, Adj. Prof. Monika Stompor, Adj. Prof. Outi Salo-Ahen, Adj. Prof. Oskar Laaksonen, Asst. Prof. Maaria Kortesniemi, Prof. Baoru Yang, Senior Researcher Kim Eerola, Nora Logrén, and Tamara Sombora. Their contribution of expertise, time, and resources has been essential in advancing my dissertation work. I am especially thankful to Nora and Tamara for their hard work and active participation in the Hops project during their master's studies. I am also grateful to Kim for his unwavering support and stimulating discussions. He has always given me positive and constructive feedback that has motivated me to improve and excel. It has been a privilege and a pleasure to work alongside such a brilliant scientist and a kind person.

I sincerely appreciate my examiners, Prof. Garry Wong and Adj. Prof. Olli Kärkkäinen, for their thorough review of this dissertation. Their expertise and valuable insights into the theoretical framework, along with suggestions for further directions, helped me produce a refined piece of work. Moreover, I am deeply honored to have Prof. Esa Korpi as my opponent for my doctoral defense. His willingness to engage in my research reflects his commitment to academic excellence, and I eagerly anticipate our constructive exchange.

I acknowledge the funding provided by The Finnish Foundation for Alcohol Studies (Alkoholitutkimussaatio), Drug Research Doctoral Programme (DRDP), and Turku University Foundation (Turun Yliopistosäätiö). Their financial support has been essential for conducting this research. I value their investment in my project and their commitment to advancing knowledge in the field.

Special thanks to Eeva Valve, coordinator of DRDP and FinPharmaNet, for her indispensable assistance during my studies. She guided me through the administrative processes over the years and has played an instrumental role in organizing scientific meetings for the program.

I extend heartfelt thanks to my colleagues, both current and former, in Farmis (Medisiina C6) for making my work experience rewarding and enjoyable. This includes but is not limited to, Prof. Olli Pentikäinen and his group, Adj. Prof. Petri Vainio, Adj. Prof. Petteri Rinne, Asst. Prof. Aleksi Tornio, Kim Eerola, Jonne Laurila, Arto Liljeblad, James Kadiri, Keshav Thapa, Anni Suominen, Aya Bouazza, Mauricio Velasco, Minna Eriksson, Minttu Mattila, Liisa Ailanen, Darin Al-Ramahi, and Milka Hauta-aho. The warm and welcoming atmosphere you have cultivated has lifted my spirits. Hats off to the lab professionals, Hanna Haukkala, Sanna Bastman, Karla Saukkonen, for facilitating the efficiency of my experiments. Your patience, technical expertise, and prompt responses to my inquiries have been invaluable.

To my comrades in the quest for knowledge, Obada Al-Zghool, James Kadiri, Ashour Abdelrahman, Sherif Bayoumy, Tarek Omran, Khalil Shahramian, Muhammad Waqar, and Arafat Siddiqui, I offer sincere appreciation for the fellowship, lively discussions, and solid support during my time in the University of Turku. You have made this experience more meaningful with your varied outlooks, kind gestures, and shared adventures. I regard the friendships we have formed and treasure the memories we created together. In this exceptional assembly, Obada, James, and Ashour have carved out a unique place in my heart. Enormous thanks to my three confidants and kindred spirits, who have been a constant source of joy, laughter, and companionship. Their strong bonds, infectious positivity, and readiness to lend a shoulder and counsel have helped me in tough times and have enriched my life in countless ways.

A big shout-out to Zahir, Houssine, Sajid, Joonas, Eero, Ekko, Janne, Anton, Mostafa, Tamer, Rapson, Raj, Ziad, Dahmane, Karim, and Mouloud, whose friendship and backing have been a constant source of warmth and solidarity.

With boundless sentiment, I extend my utmost gratitude to my family for their love, reassurance, and steadfast presence throughout my odyssey. To my parents, thank you for instilling in me the values of hard work, perseverance, and the pursuit of knowledge. Your love and prayers have fueled me to pursue my goals, and I appreciate your sacrifices and devotion. Heartfelt thanks to my siblings, Moaadh, Arwa, Alaa, and Afnan, for their ability to shift my perspective and lighten my mood. Their humor has been a lifeline, pulling me out of research-induced trance and reminding me of the joy that exists beyond the lab.

To my incredible wife, Maria, whose love, belief in my aspirations, and dedication to our family have been the pillars of strength that empowered me to achieve this milestone. Thank you for shouldering the challenges that came with my studies, for understanding my late-night writing marathons, and for always being my rock. To my cherished sons, Elias, the four-year-old trailblazer, and Sami, the one-year-old ball of sunshine, you taught me the profound essence of patience, perseverance, and unconditional love. I proudly present this dissertation to you, hoping that you will grow curious, compassionate, and driven individuals. May you pursue your passions with determination, and never lose sight of your dreams.

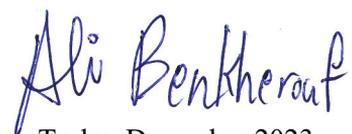

Turku, December 2023

# Appendices

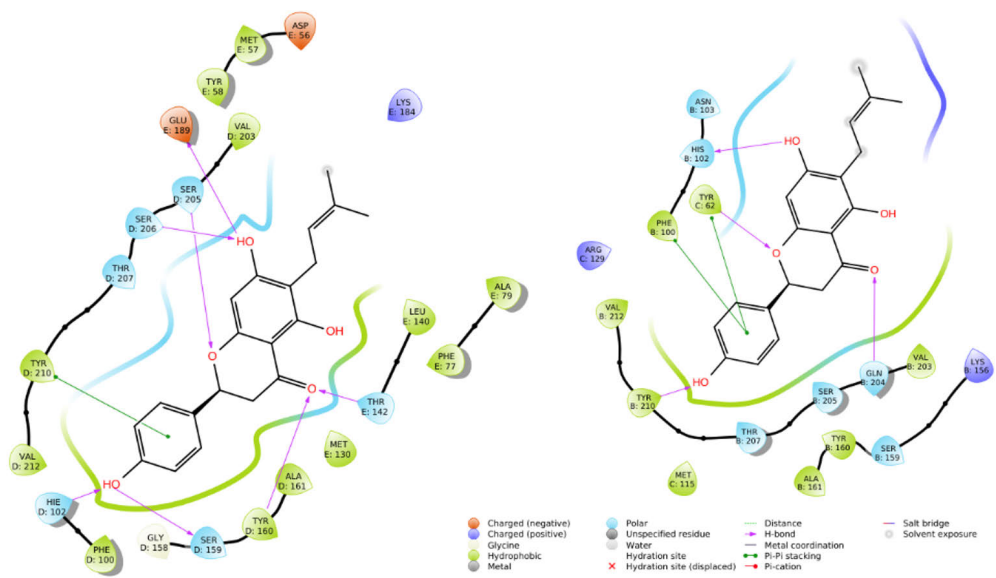

**Figure S1.** The 2D ligand interaction diagrams showcasing the docked, putative poses of 6-prenylnaringenin with the α1β2γ2 GABA$_A$ receptor isoform (PDB ID: 6D6U). The left panel represents the α+/γ2− interface, wheras the right panel captures the α+/β− interface.

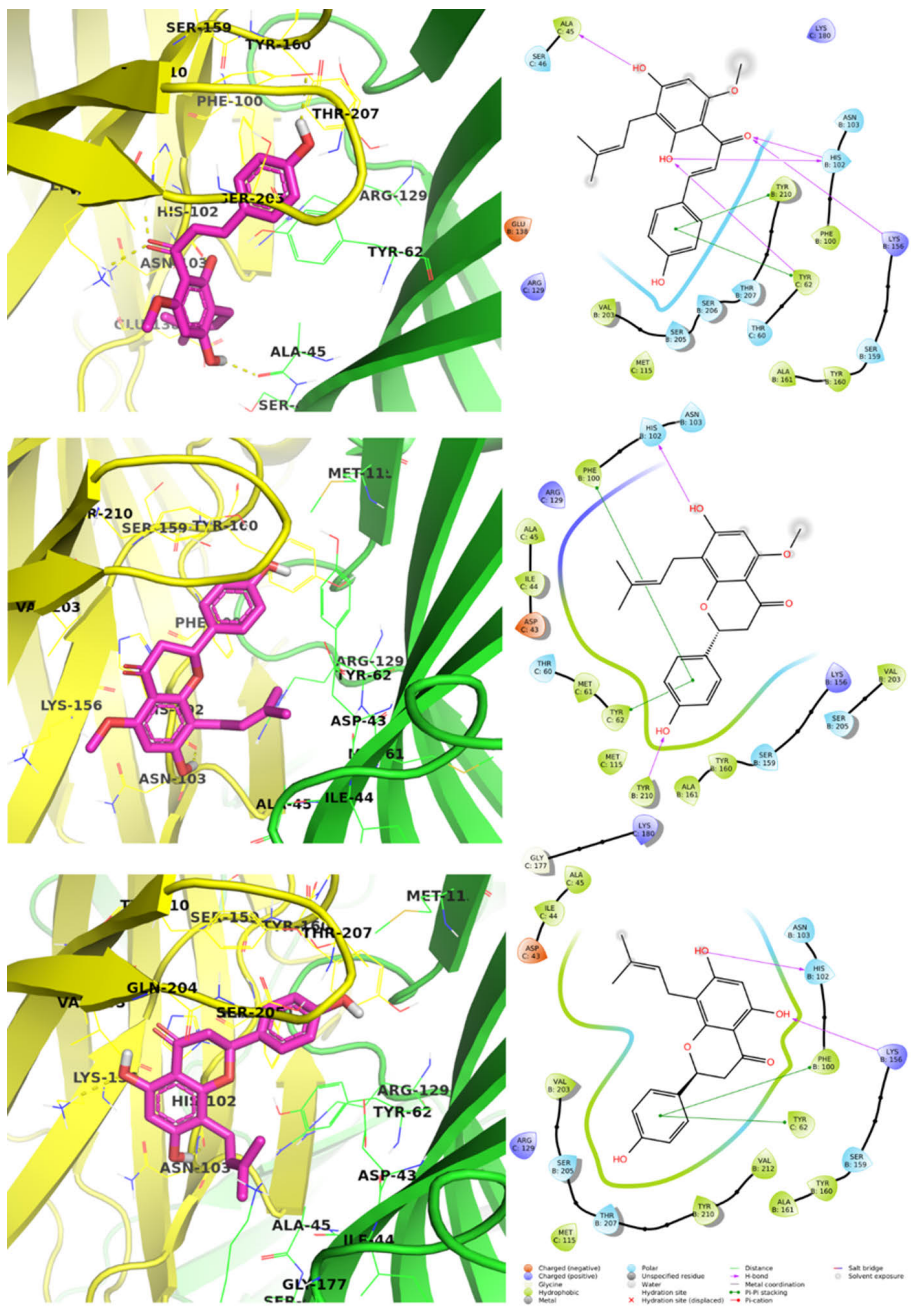

**Figure S2.** The α1β2γ2 GABA_A receptor isoform (PDB ID: 6D6U) illustrates the putative binding poses of hop prenylflavonoids at the α+/β− interface. Subunits are color-coded: α1 in bright yellow, γ2 in grey, and β2 in green. Lines and labels highlight the interacting residues, with polar interactions marked by yellow dashed lines. Non-polar hydrogens are omitted for visual clarity. Atom colors: carbon – varies by molecule, nitrogen – blue, oxygen – red, sulfur – dark yellow; hydrogen – white. From top to bottom, the flavonoids presented are: Xanthohumol, Isoxanthohumol, and 8-prenylnaringenin. The right panel details the ligand interaction for each pose.

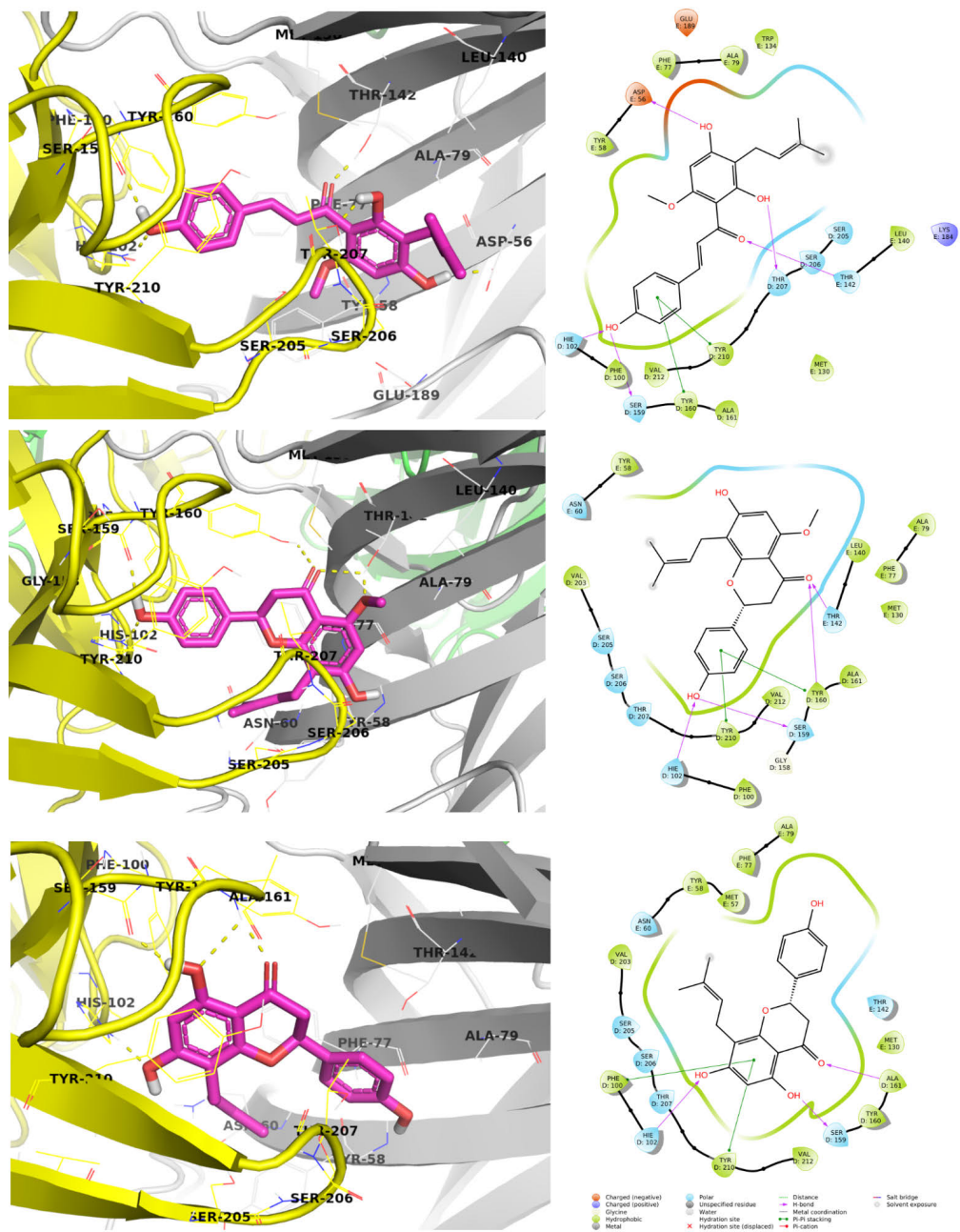

**Figure S3.** The α1β2γ2 GABA_A receptor isoform (PDB ID: 6D6U) illustrates the putative binding poses of hop prenylflavonoids at the α+/γ2− interface. Subunits are color-coded: α1 in bright yellow, γ2 in grey, and β2 in green. Lines and labels highlight the interacting residues, with polar interactions marked by yellow dashed lines. Non-polar hydrogens are omitted for visual clarity. Atom colors: carbon – varies by molecule, nitrogen – blue, oxygen – red, sulfur – dark yellow; hydrogen – white. From top to bottom, the flavonoids presented are: Xanthohumol, Isoxanthohumol, and 8-prenylnaringenin. The right panel details the ligand interaction for each pose.

**Table S1.** Analysis of hop prenylflavonoids docking at α1+/γ2- and α1+/β2- binding sites of the α1β2γ2 GABA$_A$R experimental structure (PDB ID: 6D6U)

| Ligand | Interface α1+/γ2- | | | Interface α1+/β2- | | | |
|---|---|---|---|---|---|---|---|
| | Glide XP score (kcal/mol) | MM-GBSA ΔG bind (kcal/mol) | Key interacting receptor from subunits α1/γ2[d] | Glide XP score (kcal/mol)[e] | IFD Score[f] | MM-GBSA ΔG bind (kcal/mol)[g] | Key interacting residues from subunits α1/β2[h] |
| 8-prenylnaringenin | -9.824 | -50.65 | Phe100, His102, Ser159, Ala161, Tyr210/ Met57, Tyr58 | -3.530 | -3002.40 | -21.68/-60.74 | Phe100, His102, Lys156, Tyr210/ Tyr62 |
| Xanthohumol | -7.829 | -70.34 | His102, Ser159, Tyr160, Thr207, Tyr210/ Asp56, Thr142 | -3.681 | -3000.83 | -35.71/-61.74 | His102, Lys156, Tyr210/ Ala45, Tyr62 |
| Isoxanthohumol | -8.898 | -63.00 | His102, Ser159, Tyr160, Tyr210/ Thr142 | -2.467 | -2998.18 | -24.64/-60.97 | Phe100, His102, Tyr210/ Tyr62 |
| Flumazenil | -9.001[a] / -6.494[b] | -57.95[a]/ -64.71[b] | Phe100, His102, Ala161, Thr207, Tyr210/ Thr142 | N/A | N/A | N/A | N/A |
| CGS 9895 | N/A[c] | N/A | N/A | -2.246 | -2992.31 | -23.33/-69.64 | His102, Ser206, Tyr210/ Tyr62 |

[a] Top-scoring XP pose rotated ~180° from experimental pose.

[b] Docking pose matches experimental (-57.52 kcal/mol). Only receptor's H-atoms minimized; heavy atoms fixed.

[c] Not applicable.

[d] HIS102: histidine at position 102 with proton on epsilon nitrogen.

[e] Non-optimal rigid binding site.

[f] Score for induced fit docking: GlideScore + 0.05 x PrimeEnergy.

[g] Prime/MM-GBSA energy for the Glide XP/induced fit poses.

[h] HIS102: histidine at position 102 with proton on delta nitrogen.

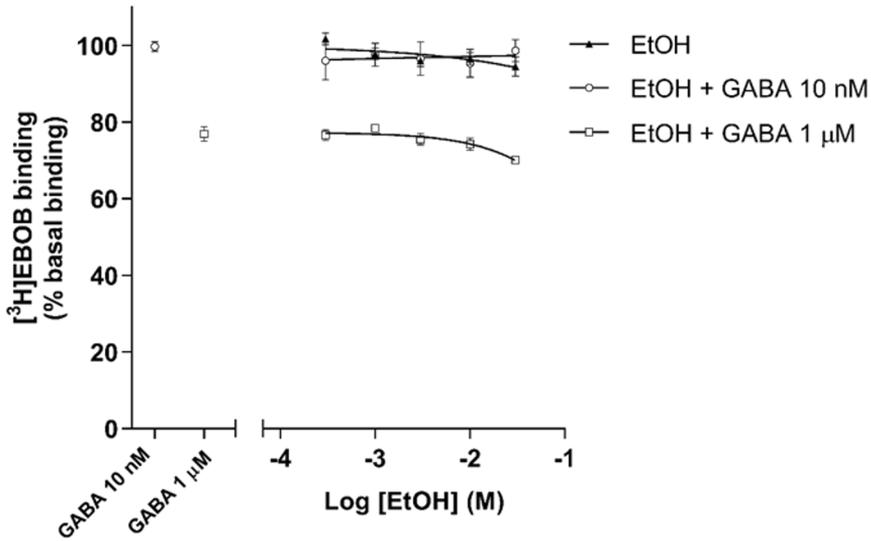

**Figure S4.** Displacement curves of ethanol (EtOH)-free [³H]EBOB (1 nM) binding to rat cerebellar membranes across various concentrations (0.3-30 mM) of EtOH, either with (A) or without (B) 10 nM/1 µM GABA. Data points denote the mean ± SEM from 3 to 4 distinct experiments, each conducted in triplicate.

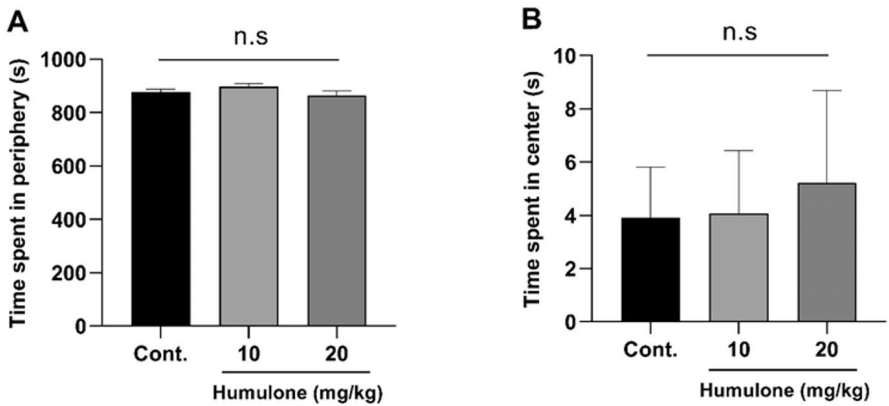

**Figure S5.** Effect of acute treatment with humulone (10 and 20 mg/kg) on mice's anxiety-related behavior in the open field paradigm. Both the time spent in the periphery (A) and time spent in the center (B) were observed over a 15-minute interval, initiated 45 minutes post-intraperitoneal injection of either humulone or the vehicle. Every bar denotes the mean ± SEM from 7 to 11 mice per group. The label "n.s", indicates no significant differences among groups, as ascertained by a one-way ANOVA paired with Tukey's *post hoc* test.

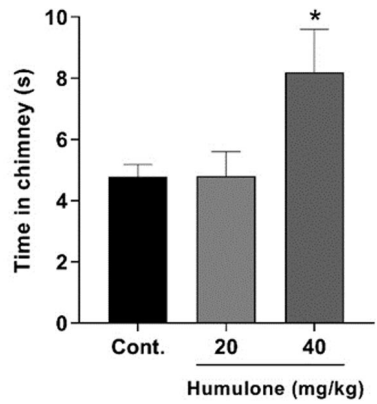

**Figure S6.** Influence of acute treatment with humulone (10 and 20 mg/kg) on mouse motor coordination in the chimney test. The time necessary to climb up the vertical tube was measured during a 1-minute interval, initiated 45 minutes post-intraperitoneal injection of either humulone or vehicle. Every bar denotes the mean SEM from 5 to 9 mice per group. Significance marker *p < 0.05, indicating differences from the group receiving the vehicle only, as ascertained by a one-way ANOVA paired with Dunnett's multiple comparison test.